\documentclass[a4paper,fleqn,amssymb,usenatbib,useAMS,usedcolumns]{mnras}
\usepackage{times} 
\usepackage{latexsym,amssymb,amsmath,amsfonts}
\usepackage{comment}
\usepackage{rotating} 
\usepackage{float}
\usepackage{graphicx}
\usepackage{natbib}
\usepackage{longtable}
\usepackage{url}
\usepackage{dcolumn}
\usepackage{textcomp}
\usepackage{multicol}
\usepackage{bm}
\usepackage{pdflscape}
\usepackage[T1]{fontenc}
\usepackage{ae,aecompl}
\usepackage[caption=false]{subfig}
\usepackage[dvipsnames]{xcolor}

\usepackage{xr}
\makeatletter
\newcommand*{\addFileDependency}[1]{
  \typeout{(#1)}
  \@addtofilelist{#1}
  \IfFileExists{#1}{}{\typeout{No file #1.}}
}
\makeatother

\newcommand*{\myexternaldocument}[1]{%
    \externaldocument{#1}%
    \addFileDependency{#1.tex}%
    \addFileDependency{#1.aux}%
}
\myexternaldocument{appendix}

\definecolor{JKcolor}{rgb}{0.8, 0.2, 0.8}
\definecolor{PPcolor}{rgb}{0.0, 0.8, 0.9}

\graphicspath{{/media/gitika/Laptop_working_disk_2TB/laptop_data_working_space/data_analysis/Ly_targets/ALL_Lya_TARGETS/ALL_PLOTS/}
 {/media/gitika/Laptop_working_disk_2TB/laptop_data_working_space/data_analysis/Ly_targets/ALL_Lya_TARGETS/PS1_radio_overlay/}
 {/media/gitika/Laptop_working_disk_2TB/laptop_data_working_space/data_analysis/Ly_targets/ALL_Lya_TARGETS/ALL_PLOTS/EMISSION_LINES_PLOT/}} 
\begin{document}

\newcommand{\cf}{{\textrm c.f.}}
\newcommand{\eg}{{\textrm e.g.}}
\newcommand{\ie}{{\textrm i.e.}}
\newcommand{\km}{\ensuremath{\mbox{~km}}}
\newcommand{\pc}{\ensuremath{\mbox{~pc}}}
\newcommand{\kpc}{\ensuremath{\mbox{~kpc}}}
\newcommand{\ebv}{\mbox{$E(B-V)$}}
\newcommand{\bvri}{\mbox{$BVRI$}}
\newcommand{\degree}{\mbox{$^\circ$}}
\newcommand{\maghundred}{\mbox{mag (100 d)$ ^{-1} $}}
\newcommand{\msun}{\mbox{M$_{\odot}$}}
\newcommand{\msol}{\mbox{M$_{\odot}$}}
\newcommand{\zsol}{\mbox{Z$_{\odot}$}}
\newcommand{\rsun}{\mbox{R$_{\odot}$}}
\newcommand{\kms}{\mbox{$\rm{\,km\,s^{-1}}$}}
\newcommand{\kkms}{\mbox{$\times10^3\rm{\,km\,s^{-1}}$}}
\newcommand{\ergs}{\mbox{$\rm{\,erg\,s^{-1}}$}}
\newcommand{\logl}{\mbox{$\log L/{\rm L_{\odot}}$}}
\newcommand{\nickel}{\mbox{$^{56}$Ni}}
\newcommand{\cobalt}{\mbox{$^{56}$Co}}
\newcommand{\iron}{\mbox{$^{56}$Fe}}
\newcommand{\mum}{\mbox{$\mu{\rm m}$}}
\newcommand{\el}{\mbox{${e}^{-}$}}
\newcommand{\ld}{\mbox{$\lambda$}}
\newcommand{\ldld}{\mbox{$\lambda\lambda$}}
\newcommand{\ergscm}{{$ \mathrm{erg\ s^{-1} cm^{-2}}$}}
\newcommand{\ergscma}{{$ \mathrm{erg\ s^{-1} cm^{-2} arcsec^{-2}}$}}
\newcommand{\fcgs}{{$ \mathrm{erg\ s^{-1} cm^{-2}}$\AA$^{-1}$}}
\newcommand{\arc}{$\mathrm{^{\prime\prime}}$}
\newcommand{\cross}{$\mathrm{\times}$}
\newcommand{\apr}{$\mathrm{\approx}$}
\newcommand{\til}{$\mathrm{\sim}$}
\newcommand{\zem}{$\mathrm{$z_{em}$}$}
\newcommand{\plm}{$\mathrm{\pm}$}
\newcommand{\sblya}{$\mathrm{SB_{Ly\alpha}}$}
\newcommand{\hi}{\mbox{H\,{\sc i}}}
\newcommand{\mgii}{\mbox{Mg\,{\sc ii}}}
\newcommand{\mgi}{\mbox{Mg\,{\sc i}}}
\newcommand{\feii}{\mbox{Fe\,{\sc ii}}}
\newcommand{\oi}{\mbox{O\,{\sc i}}}
\newcommand{\cii}{\mbox{C\,{\sc ii}}}
\newcommand{\ciii}{\mbox{C\,{\sc iii}}}
\newcommand{\ci}{\mbox{C\,{\sc i}}}
\newcommand{\sii}{\mbox{Si\,{\sc ii}}}
\newcommand{\znii}{\mbox{Zn~{\sc ii}}}
\newcommand{\caii}{\mbox{Ca\,{\sc ii}}}
\newcommand{\nai}{\mbox{Na\,{\sc i}}}
\newcommand{\civ}{\mbox{C\,{\sc iv}}}
\newcommand{\heii}{\mbox{He\,{\sc ii}}}
\newcommand{\nv}{\mbox{N\,{\sc v}}}
\newcommand{\siv}{\mbox{Si\,{\sc iv}}}
\newcommand{\flya}{$f\mathrm{_{Ly\alpha}}$}
\newcommand{\llya}{$L\mathrm{_{Ly\alpha}}$}
\newcommand{\fciv}{$f\mathrm{_{CIV}}$}
\newcommand{\fheii}{$f\mathrm{_{HeII}}$}
\newcommand{\Smjy}{$\mathrm{S_{1.4GHz}}$}
\newcommand{\mbh}{$M_{BH}$}
\newcommand{\lumcon}{$\mathrm{\lambda\ L_{1350}}$}
\newcommand{\lumbol}{$\mathrm{L_{bol}}$}
\newcommand{\lumion}{$\mathrm{L_{912}}$}
\newcommand{\lumft}{$\mathrm{L_{420}}$}
\newcommand{\lumopf}{$\mathrm{L_{1.4GHz}}$}
\newcommand{\sid}{$\alpha^{1.4}_{0.4}$}

\def\h2{$\rm H_2$}
\def\Nh2{$N$(H${_2}$)}
\def\chin{$\chi^2_{\nu}$}
\def\chiu{$\chi_{\rm UV}$}
\def\sys{J0441$-$4313~}
\def\lya{\ensuremath{{\rm Ly}\alpha}}
\def\lymana{\ensuremath{{\rm Lyman}-\alpha}}
\def\kms{km\,s$^{-1}$}
\def\cms{cm$^{-2}$}
\def\cc{cm$^{-3}$}
\def\zabs{$z_{\rm abs}$}
\def\zem{$z_{em}$}
\def\nhi{$N$($\hi$)}
\def\ln{log~$N$}
\def\nh{$n_{\rm H}$}
\def\ne{$n_{e}$}
\def\21{21-cm}
\def\ts{T$_{s}$}
\def\th{T$_{01}$}
\def\t0{T$_{0}$}
\def\ll{$\lambda\lambda$}
\def\l{$\lambda$}
\def\fc{$C_{f}$}
\def\c21{$C_{21}$}
\def\mjb{mJy~beam$^{-1}$}
\def\taudv{$\int\tau dv$}
\def\taup{$\tau_{\rm p}$}
\def\ha{H\,$\alpha$}
\def\hb{H\,$\beta$}
\def\oi{[O\,{\sc i}]}
\def\oii{[O\,{\sc ii}]}
\def\oiii{[O\,{\sc iii}]}
\def\nii{[N\,{\sc ii}]}
\def\sii{[S\,{\sc ii}]}
\def\taudvl{$\int\tau dv^{3\sigma}_{10}$}
\def\taudv{$\int\tau dv$}
\def\vshift{$v_{\rm shift}$}
\def\wmg{$W_{\mgii}$}
\def\wfe{$W_{\feii}$}
\def\dgi{$\Delta (g-i)$}
\def\ebv{$E(B-V)$}
\def\sig{$\sigma$}

\def\pathfig{images/}

%

\title[Ly$\alpha$ emission from RLQs]{
Spatially resolved Lyman-$\alpha$ emission around radio bright quasars
}
\author[Shukla et.al]{Gitika Shukla$^{1}$\thanks{E-mail: gitika@iucaa.in}, Raghunathan Srianand$^{1}$, Neeraj Gupta$^{1}$, Patrick Petitjean$^2$,
\newauthor{Andrew J. Baker$^3$, Jens-Kristian Krogager$^2$, Pasquier Noterdaeme$^2$}
{} \\
\\
 $^{1}$Inter-University Centre for Astronomy and Astrophysics (IUCAA), Post Bag 4, Pune 411007, India \\
 $^{2}$Institut dAstrophysique de Paris, UMR 7095, CNRS-SU, 98bis boulevard Arago, 75014 Paris, France\\
 $^{3}$Department of Physics and Astronomy, Rutgers, the State University of New Jersey, 136 Frelinghuysen Road, Piscataway,\\ NJ 08854-8019, USA\\
}

\date{ }
\pubyear{2021}
\maketitle
\label{firstpage}
\pagerange{\pageref{firstpage}--\pageref{lastpage}}
%
%
\begin {abstract}  
We use Southern African Large Telescope (SALT) to perform long-slit spectroscopic observations of 23 newly discovered radio-loud quasars (RLQs) at $2.7<z<3.3$.  The sample consists of powerful AGN brighter than 200 mJy at 1.4 GHz and is selected on the basis of mid-infrared colors i.e., unbiased to the presence of dust. We report 7 confirmed and 5 tentative detections of diffuse Ly$\alpha$ emission in the sample. We present the properties of diffuse Ly$\alpha$ emission and discuss in detail its relationship to different quasar properties. We find strong dependence of Ly$\alpha$ halo detection rate on the extent of radio source, spectral luminosity of RLQ at 420\,MHz ($L_{\rm 420MHz}$), presence of associated C IV absorption and nuclear He II emission line equivalent width. As seen in previous surveys, the FWHM of diffuse Ly$\alpha$ emission in the case of confirmed detections are much higher (i.e.$>$1000 km/s in all, except one). Using the samples of high-$z$ radio-loud quasars and galaxies from literature,  we confirm the correlation between the Ly$\alpha$ halo luminosity and its size with $L_{\rm 420MHz}$. The same quantities are found to be correlating weakly with the projected linear size of the radio emission. Our sample is the second largest sample of RLQs being studied for the presence of diffuse Ly$\alpha$ emission and fills in a redshift gap between previous such studies. Integral Field Spectroscopy is required to fully understand the relationship between the large scale radio emission and the overall distribution, kinematics and over density of Ly$\alpha$ emission in the field of these RLQs.
\end{abstract}

%
%
%
\section{Introduction} 
\label{sec_introduction}

Detailed investigations of the spatial distribution, kinematics, and excitation of the gas  traced  by  the  extended  Ly$\alpha$ emission can  provide  vital clues on  various  feedback  processes  that  drive  star  formation  and AGN activities in  high-$z$ galaxies.
The advent of ultra sensitive and high spatial resolution integral field spectrographs (IFS) like the Multi-Unit Spectroscopic Explorer \citep[MUSE;][]{bacon2010} and the Keck Cosmic Web Imager \citep[KCWI;][]{morrissey2012} on 8-10 metre class telescopes, have aided tremendously to study the environments of galaxies and quasars up to  redshifts as high as $z$\til 6 \citep[see,][]{Farina2019,Drake2019}. Several studies using IFS observations now routinely report the presence of large scale (few tens to hundreds of kpc) extended \lya\ halos around quasars and star forming galaxies \citep[][]{Borisova2016,Wisotzki2016,leclercq2017,Arrigoni2018,Arrigoni2019,cai2019,Osullivan2020,Mackenzie2021,Fossati2021}. The detection rate of extended \lya\ halos has gone up to 100\% in most of these studies that achieve typical surface brightness sensitivities of few$\times10^{-19}$erg  s$^{-1}$cm$^{-2}$ arcsec$^{-2}$.

The most commonly discussed origins for the source of the observed  extended \lya\ emission are:  (i) shock induced radiation, powered by radio jets or outflows \citep{mori2004,allen2008}; (ii) gravitational cooling radiation \citep{haiman2000,dijkstra2006,Rosdahl2012}; (iii) fluorescent \lya\ emission due to photoionization by UV luminous sources like an AGN or star formation activity \citep{mccarthy1993,cantalupo2005,geach2009,overzier2013}, and (iv) resonant scattering of \lya\ photons from embedded sources \citep{villar1996,dijkstra2008}. The presence of associated high-ionization lines like \civ\ and \heii\ can provide additional information on the kinematics of the gas and help disentangle the various physical processes powering the \lya\ emission \citep[e.g.][]{Prescott2015b,Arrigoni2015b}.

In the case of radio-loud AGN, the relation between the properties of the extended \lya\ emission and their radio morphology can also be investigated \citep{vanojik1997,heckman1991a,heckman1991b,villar2007}. For example, the observed morphology of the radio emission can, in principle, be used to quantify (i) the orientation of the putative ionizing cone and dusty torus, (ii) the anisotropic nature of the ambient medium  or the young nature of the radio source (based on morphological asymmetries), and (iii)  the interaction between the radio-jet and the ambient medium (based on the distorted structures and hot-spots) \citep[e.g.,][]{Fanti01, Saikia03}. Such studies are also important to probe the viewing angle based unification models \citep{Barthel1989} of  radio-loud AGN in which radio-loud quasars (RLQs) and high-z radio galaxies (HzRGs) are of the same class of objects but viewed differently, along and perpendicular to radio axis, respectively. Alternatively, RLQs may simply be the maximum quasar activity phase of a radio galaxy.

It has been found that HzRGs are hosted by massive star-forming protoclusters \citep{miley2008,Mayo2012,Galametz2012,Wylezalek2013,Dannerbauer2014} and are usually associated with luminous and often large gaseous halos \citep[][]{mccarthy1988,Chambers1989,vanojik1997,Best2000,Reuland2003}. Strong correlation is seen between the radio axis and the major axis of the diffuse/optical gas emission \citep[][]{Chambers1987,vanojik1997,villar2007a, Humphrey2006, Humphrey2007}. Using long-slit spectroscopy, \citet{vanojik1997} have found a clear correlation between the sizes of the radio sources and the diffuse \lya\ emission, along with an anti-correlation between the \lya\ velocity width and the radio size. In more than 60\% of these cases, strong associated \hi\ absorption ( with $N_\mathrm{{HI}}\geq 10^{18} \mathrm{cm^{-2}}$) was detected. The detection rate of \hi\ absorption was found to be higher in smaller radio sources (i.e.,  $\sim 90\%$ when the source size is $<50$ kpc and 25\% when $>50$ kpc). In 61$\%$ of the cases, \lya\ was more extended than the radio source itself. The inner parts of the \lya\ halo within the extent of the radio emission showed perturbed kinematics (FWHM $>1000$ \kms), whereas the more extended (\til $100$ kpc) regions were dominated by quiescent kinematics (FWHM $<700$ \kms ).

One of the first studies  of spatially resolved \lya\ emission in a large statistical sample of RLQs was by \cite{heckman1991a}.  Their sample consists of 19 RLQs  in the redshift range $1.98\le z \le 2.91$ with a median $z = 2.2$.  The 1.4 GHz flux density is in the range 0.14 to 2.2~Jy with a median of 754 mJy. They specifically considered extended radio sources -- 16 of their targets have largest angular size (LAS) measured in the range 3 - 17\arc\ and four are compact (LAS<1.5\arc). They detected extended \lya\ halos with a typical surface brightness sensitivity of few$\times 10^{-17}$erg s$^{-1}$ cm$^{-2}$ arcsec$^{-2}$ around 15 quasars all with extended radio morphology. The halos associated with these RLQs were typically \til 100 kpc large with \lya\ halo luminosity of \til\ few $\times 10^{44}$\ergs, and showed alignment between the radio axis and the \lya\ morphological axes to within 30\degree. However, unlike in the case of HzRGs the radio size do not correlate with the size of the extended \lya\ halo. Nevertheless, the brightest regions of the \lya\ emission are located on the brightest side of the radio emission. \citet{heckman1991a} concluded that the \lya\ halos were most likely ionized by the UV continuum of the quasars and the alignment between the radio and \lya\ axis was ascribed to dense gas existing along the radio axis and/or anisotropic emission from quasar escaping preferentially along the radio axis.

Based on the spectroscopic follow up of 5  RLQs in their sample, \citet{heckman1991b} have inferred that the associated \lya\ halos were kinematically perturbed with FWHM of 1000-1500 \kms. They also found that the equivalent width (EQW) of the nuclear \heii\ emission strongly correlates with the EQW of the extended \lya\ halo. They suggested that the presence of strong narrow He~{\sc ii}$\lambda$1640 nuclear emission line should flag high-z quasars with prominent \lya\ halos. \citet{Roche2014} have reobserved 6 sources from \citet{heckman1991a}, two of them being common with those studied by \citet{heckman1991b}. They claimed to have seen infall signature (with infall velocities in the range 250-460 \kms\ with respect to the \lya\ emission) in three of these cases.

\citet{Arrigoni2019} have presented MUSE observations of a sample of 61 $z \sim 3$ quasars, of which 15 are radio-loud i.e., satisfy the radio-loudness criteria, R=$f_{\nu,5 {\rm GHz}}/f_{\nu,4400} > 10 $, of \citet{Kellermann1989}. \citet{Borisova2016} have studied  17 bright quasars with MUSE, two being radio-loud. Diffuse \lya\ emission is detected around both radio-loud and radio-quiet quasars and no clear distinction has emerged. While \citet{Borisova2016} notice higher velocity dispersion (i.e., $>$1000 \kms) for the two RLQs they observed, \citet{Arrigoni2019} did not see any violent kinematics introduced by radio jet - gas interactions in their sample. However, the sample of \citet{Arrigoni2019} consists of relatively less powerful RLQs (i.e., only 5 of them have L-band flux density  $\ge$100 mJy) compared to objects in the sample of \citet{heckman1991a}. Also as of now, most of the RLQs studied with MUSE have compact morphology (only one out of 11 sources studied by \citet{Arrigoni2019}, for which arcsec scale radio images can be found, has  radio emission  extending beyond arcsec scales). In that source (TEX1033+137) the extended \lya\ emission seems to avoid the extended radio emitting regions. Thus, based on the data available at the moment from MUSE studies there is no clear picture of the influence of radio sources on the morphology and kinematics of the \lya\ halo.

We have recently completed a large spectroscopic campaign \citep[see][]{krogager2018, Gupta2021qsosurvey} using the Nordic Optical Telescope (NOT) and the Southern African Large Telescope (SALT) to confirm the nature and measure the redshifts of a subset of radio bright \citep[flux density in excess of 200 mJy at 1.4 GHz in the NRAO VLA Sky Survey (NVSS);][]{condon1998} southern (i.e $\delta < +20^\circ$) AGN candidates selected using WISE MIR-colors (i.e $\mathrm{W_1-W_2 <1.3\times (W_2 -W_3) - 3.0}$ and $\mathrm{W_1 - W_2 > 0.6}$). This has resulted in 250 spectroscopic identifications with median redshift $z =1.8$. Overall, objects in this MALS-SALT-NOT sample are optically fainter ($\Delta i$ = 0.6 mag) and redder (0.2 mag) than radio-selected quasars, and are representative of the fainter quasar population detected in deep optical surveys \citep[][]{Gupta2021qsosurvey}. These objects are being studied as part of the MeerKAT Absorption Line Survey \citep[MALS; see][for key science objectives]{Gupta2016}, an ongoing large survey at the South African precursor MeerKAT \citep[][]{Jonas2016} of the upcoming Square Kilometer Array (SKA).

There are 25 AGN in MALS-SALT-NOT sample at $z>2.7$. \citep[see][for details]{Gupta2021qsosurvey}. We select all 24 AGN i.e., 23 quasars and 1 radio galaxy at $2.7<z<3.5$ as the basic sample for the present study where we search for diffuse \lya\ emission around these objects using long-slit spectroscopy. The detailed properties of the radio galaxy and the associated \lya\ halo are already presented in \citet[][]{Shukla2021}. Here we will focus on understanding various aspects relating the  optical/radio properties of quasars to the properties of the extended \lya\ halos.

The organization of the paper is as follows. In Section \ref{sec_sample_obs}, we provide details of our sample, long-slit spectroscopic observations using SALT, radio observations using upgraded Giant Meterwave Radio Telescope (uGMRT) and data reduction. In Section \ref{sec_results}, we discuss quasar parameters extracted from the 1D spectra, 2D spectral analysis to detect diffuse line emission around quasars and radio properties of all the objects in our sample. In Section \ref{sec:results}, we discuss our results. In particular we explore the connection between the extended \lya\ emission and (i) the presence of \civ\ associated absorption, (ii) the equivalent width of the nuclear He~{\sc ii} emission, (iii) the size of the radio emission and (iv) the quasar parameters like the mass of the black-hole, the bolometric luminosity, the Eddington ratio and radio power. In this section we also compare our results with the findings from the literature. Finally, in Section \ref{sec_summary}, we summarize our results. In this paper, we have adopted a cosmology with H$_0$ = 67.4 \kms Mpc$^{-1}$, $\mathrm{\Omega_m}$ = 0.315 and  $\mathrm{\Omega_\Lambda}$ = 0.685 \citep[][]{planck2018}. 

%
\section{sample and observations}
\label{sec_sample_obs}

\begin{table*}
\scriptsize
\setlength{\tabcolsep}{2pt}
\caption{Log of RSS/SALT long-slit observations for our sample}
\centering
\begin{tabular}{lcccccccccccccccccc}
\hline
\hline
ID
&Source
&RA
&DEC
&PA
&Observing date  
&Wavelength coverage
&Air mass
&Exposure time
& Spectral PSF 
& Sky$^{a}$  \\

&
&(J2000)
&(J2000)
&(deg) 
& year/mm/dd      
&(\AA)               
&          
&(sec)                        
& (arcsec)
& Conditions\\        

(1)
&(2)
&(3)
&(4) 
&(5)      
&(6)               
&(7)       
&(8)                        
&(9)
&(10)
&(11)\\            
\hline
  
1&    M025035.54$-$262743.10	    &02:50:35.54&   -26:27:43.10        &	30	    &2018-09-15	&4203-7261                  &1.23	&	2$\times$1200	&2.9	& TN \\
&    			                    &           &                       &	120     &2019-08-20	&"                      	&1.17	&	4$\times$1200	&2.2	& CL \\
2&    M041620.54$-$333931.30	    &04:16:20.54&   -33:39:31.30        &	15	    &2016-12-22	&4486-7533                  &1.26	&	2$\times$1300	&1.6	& CL \\
&    			                    &           &                       &	111	    &2018-02-10	&"                      	&1.19	&	1$\times$1200	&2.4	& CL overhead \\
3&    M050725.04$-$362442.90	    &05:07:25.04&   -36:24:42.90        &	118 	&2019-08-23	&4203-7261                  &1.19	&	2$\times$1400	&2.2	& CL \\
&    			                    &           &                       &	178 	&2019-08-20 &"                          &1.20	&	2$\times$1400	&1.9	& CL \\
4&    M052318.55$-$261409.60	    &05:23:18.55&   -26:14:09.60        &	89	    &2019-10-19	&4486-7533                  &1.24	&	2$\times$1200	&2.1	& NA \\
&    			                    &            &                      &	167 	&2019-08-25	&"                      	&1.23	&	4$\times$1200	&2.4	& CL \\
5&    M061038.80$-$230145.60	    &06:10:38.80&   -23:01:45.60        &	60	    &2018-09-30	&4061-7124                  &1.23	&	2$\times$1200	&2.0	& Clouds/NPH \\
&    			                    &           &                       &	306 	&2017-11-18&"                           &1.25	&	5$\times$1200	&1.9	& CL \\

6&    M063613.53$-$310646.30	    &06:36:13.53&   -31:06:46.30        &	34	    &2018-10-01	&3919-6987                  &1.26	&	2$\times$1200	&2.8	& TN/NPH \\
&    			                    &           &                       &	118 	&2019-10-06	&"                      	&1.25	&	4$\times$1200	&2.0	& CL \\
7&    M080804.34$+$005708.20	    &08:08:04.34&   +00:57:08.20        &	17	    &2017-03-30	&4486-7533                  &1.27	&	2$\times$1300	&1.8	& CL \\
&    			                    &           &                       &	124	    &2017-01-21	&"                      	&1.20	&	2$\times$1200	&2.6	& CL \\ 
8&    M101313.10$-$254654.70	    &10:13:13.10&   -25:46:54.70        &	84	    &2017-12-25	&4344-7397              	&1.23	&	2$\times$1150	&1.9	& CL \\
9&    M104314.53$-$232317.50	    &10:43:14.53&   -23:23:17.50        &	10	    &2018-02-17	&4203-7261                  &1.23	&	2$\times$1150	&1.7	& Cloudy \\
10&   M114226.58$-$263313.70	    &11:42:26.58&   -26:33:13.70        &	100	    &2018-02-27	&4627-7668                  &1.27	&	2$\times$1200	&1.5	& CL \\
11&   M121514.42$-$062803.50	    &12:15:14.42&   -06:28:03.50        &	65	    &2017-01-22	&4486-7533              	&1.27	&	2$\times$1300	&2.0	& CL \\
&    			                    &           &                       &	124	    &2017-01-09	&"                          &1.22	&	2$\times$1300	&2.2	& CL/PH \\ 
12&   M123410.08$-$332638.50	    &12:34:10.08&   -33:26:38.50        &	95	    &2019-04-12	&4061-7124                  &1.38	&	1$\times$1200	&2.7	& CL \\
13&   M124448.99$-$044610.20	    &12:44:48.99&   -04:46:10.20        &	12	    &2017-03-04	&4486-7533                  &1.19	&	2$\times$1200	&1.7	& CL \\
&  			                        &           &                       &	111	    &2017-03-08	&"                          &1.22	&	2$\times$1200	&1.6	& CL$^{*}$ \\ 
14&   M125442.98$-$383356.40	    &12:54:42.98&   -38:33:56.40        &	11	    &2020-02-02	&4203-7261                  &1.23	&	4$\times$1200	&2.3	& CL \\
$15^\dag$&   M131207.86$-$202652.40	    &13:12:07.86&   -20:26:52.40        &	22B	    &2018-02-24	&6582-9530                  &1.23	&	1400,1300	    &2.3	& Cloudy/NPH \\
16&   M135131.98$-$101932.90	    &13:51:31.98&   -10:19:32.90        &	0	    &2017-04-02	&4486-7533                  &1.21	&	2$\times$1300	&1.9	& CL \\
&  			                        &           &                       &	157	    &2020-03-17	&4344-7397                  &1.23	&	2$\times$1300	&2.6	& CL \\
17&   M141327.20$-$342235.10	    &14:13:27.20&   -34:22:35.10        &	76  	&2019-06-29	&4061-7124                  &1.30	&	4$\times$1200	&2.4	& CL \\
18&   M151304.72$-$252439.70	    &15:13:04.72&   -25:24:39.70        &	72	    &2017-05-23	&4486-7533                  &1.30	&	2$\times$1200	&1.8	& CL \\
&  			                        &           &                       &	350	    &2017-05-18	&"                          &1.28	&	2$\times$1200	&1.4	& CL \\
19&   M155825.35$-$215511.50	    &15:58:25.35&   -21:55:11.50        &	15	    &2019-04-12	&3919-6987                  &1.23	&	2$\times$1400	&2.4	& TN$^{**}$ \\
&  			                        &           &                       &	97	    &2018-09-03	&"                          &1.30	&	2$\times$1300	&2.6	& CL$^{***}$ \\
20&   M161907.44$-$093953.10	    &16:19:07.44&   -09:39:53.10        &	0	    &2018-06-05	&4203-7261                  &1.29	&	2$\times$1200	&1.9	& CL \\
&  			                        &           &                       &	89	    &2018-08-15	&"                          &1.29	&	2$\times$1300	&2.5	& CL \\
21&   M204737.66$-$184141.60	    &20:47:37.66&   -18:41:41.60        &	30	    &2017-07-20	&4203-7261                  &1.23	&	2$\times$1300	&2.0	& CL/TN \\
&  			                        &           &                       &	110	    &2017-07-23	&"                          &1.21	&	1300,1125	    &1.5	& CL \\
22&   M210143.29-174759.20	        &21:01:43.29&   -17:47:59.20        &	30	    &2017-06-20	&4203-7261                  &1.22	&	2$\times$1300	&2.1	& CL \\
&	            	                &           &                       &	112	    &2017-06-25	&4203-7261                  &1.25	&	2$\times$1300	&2.2	& TN \\
23&   M215445.08$-$382632.50	    &21:54:45.08&   -38:26:32.50        &	45 	    &2017-07-26,2018-07-04	&4203-7261,4061-7124                  &1.26	&	4$\times$1200	&1.8	& PH, CL \\
&  			                        &           &                       &	135 	&2017-07-27,2018-08-18	&4203-7261,4061-7124                  &1.26	&	4$\times$1200	&1.9	& CL, CL \\
24&   M222332.81$-$310117.30	    &22:23:32.81&   -31:01:17.30        &	0	    &2017-05-21	&4486-7533                  &1.21	&	1200,407	    &1.7	& TN/NPH \\
&  			                        &           &                       &	90	    &2017-05-17	&"                          &1.23	&	2$\times$1200	&1.4	& CL \\
25&   M235722.47$-$073134.30	    &23:57:22.47&   -07:31:34.30        &	20	    &2018-08-18	&4203-7261              	&1.22	&	2$\times$1300	&2.5	& CL \\
&  			                        &           &                       &	90  	&2018-09-15	&"                          &1.23	&	2$\times$1300	&2.5	& TN \\     
\hline
\hline
\end{tabular}
\label{tab_observations}
\begin{flushleft}
Columns 1, 2, 3, and 4: Source ID, name, ra and dec, respectively; Column $5$:  PA along which long-slit were aligned; Column 6:  Date of observations \\
Column 7:  Wavelength coverage of the spectrum; 
columns 8 and 9:  Air mass and total exposure time as multiple of number of exposures, respectively \\
Column {10}:  Spectral point spread function (SPSF). 
Column {11}: Sky condition during the observations. The labels are defined as: CL-Clear night; PH-Photometric night; TN-Thin clouds; NPH-Not photometric night; NA-information not available.\\
$\dag$ Not part of the \lya\ sample properties discussed here.  \\
$^{*}:$ SCAM (a UV–Visible imaging and acquisition camera of SALT) died midway through the first exposure, took a few minutes to start up again (star seemed to still be on the slit).\\
$^{**}:$Not totally clear, but better moon and seeing conditions were reported.\\
$^{***}:$ Some issues were reported with guidance.\\

\end{flushleft}

\end{table*}

\subsection{The Sample}
\label{sub_sample}

In order to study the extended \lya\ emission, we have selected all 24 sources having emission redshift  $2.7<$\zem$<3.5$ in our MALS-NOT-SALT sample. The redshift range is chosen such that the \lya, \civ\ and He~{\sc ii}  lines are fully covered in our observations and \lya\ falls in the most sensitive part of the detector. The highest redshift source in our sample, M131207.86$-$101932.90 at \zem = 5.064, is not considered here for the statistical analysis as the observed wavelength range of \lya\ emission in this case is severely affected by skylines and fringes (Table~\ref{tab_observations} lists all 25  AGN at $z>2.7$). Based on \civ\ emission line width, we have only one radio galaxy (M151304.72$-$252439.70 at \zem = 3.1318 with $\mathrm{FWHM_{CIV}<2000}$ \kms ) and 23 quasars in our sample. As shown by \citet{Gupta2021qsosurvey}, the fraction of AGN with broad emission lines in our sample is consistent with the expectations from the SDSS. The radio loudness parameter, $R = f_{\rm 5GHz}/f_{2500}$, for 23 RLQs in the sample considered here are in the range, $\mathrm{1400\le R \le 7000}$, confirming that our objects are among the most radio bright high-z quasars known.

\begin{figure*} 
\centerline{\vbox{
\centerline{\hbox{ 
\includegraphics[height=0.9\textheight,angle=0]{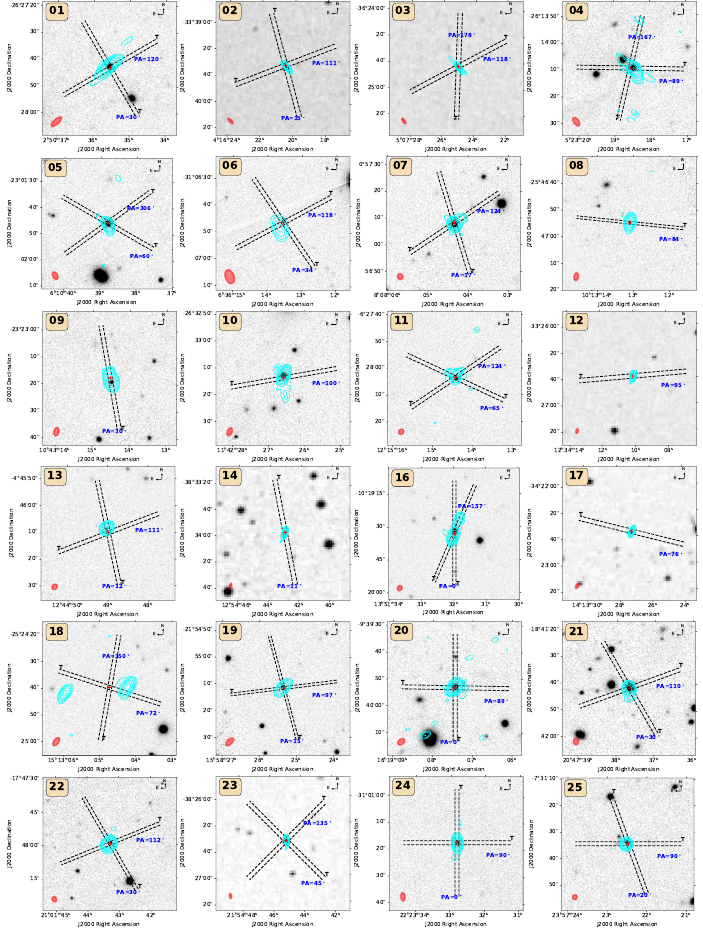} }} }}  
\caption{Pan-STARRS1 i-band or Digitized Sky Survey 
(for sources with declination $\delta<-30^\circ$) images of 24 objects (excluding $\# 15$)(DSS images were obtained from http://archive.eso.org/dss/dss).
The slit positions along which the long-slit observations were carried out are shown using dashed lines. The red plus marks the WISE position. The cyan contours correspond to Band-5 (1.4 GHz) radio emission from our uGMRT observations, and the corresponding synthesized beam is shown in red in the lower left corner. The contour levels are plotted at 4 \cross\ (rms) (-1, 1, 2, 4, 8, 16, 32, 64, ..), where rms is taken from column 5 of Table~\ref{tab_radio_props}. The symbol ``T" corresponds to top of the 2D long-slit spectra shown in Figs. \ref{fig_lya_samp1}, \ref{fig_lya_samp2} and \ref{fig_lya_samp3}. 
}
\label{fig_ps1}    
\end{figure*} 

As initial target selection is based on NVSS, their radio morphology at a few arcsec scales is a priory unknown. Therefore, we obtained broad band uGMRT observations in Band-3 \citep[presented in][]{gupta2021pband} and Band-5 (presented here). Based on the Band-5 (1.4\,GHz) uGMRT images we find that only six of our sources are clearly well resolved, 17 are compact ($<1$\arc) and one is partially resolved ($\sim$1.1\arc). We point out that though object M063613.53$-$310646.30 is compact in Band-5, it shows extended structure in Band-3 (0.42\,GHz) image, whereas M135131.92$-$101932.90 and M151304.72$-$252439.70 (radio galaxy) have extended radio morphology in both the bands.

As mentioned before, the only dedicated \lya\ survey around RLQs with similar radio luminosity is by \citet{heckman1991a}. Our sample differs from the \cite{heckman1991a} sample in the following way. \cite{heckman1991a} primarily selected {\it known} optically selected quasars with large radio sizes. On the other hand, our selection criteria picks radio bright AGN unbiased to dust through MIR colour selection \citep[][]{Gupta2021qsosurvey}. Our selection process preferentially picks $z>1.4$ quasars associated with radio emission that is compact or of the symmetric radio morphology \citep[e.g., Compact Symmetric Objects;][]{Odea21}.   The RLQs in our sample are also at a higher redshift \citep[median \zem\ \til\ 2.9 in comparison to 2.2 of][]{heckman1991a}. 

\subsection{Long-slit observations with RSS/SALT}
\label{sub_longslit}
To obtain long-slit spectra of sources in our sample, we have used Robert Stobie Spectrograph \citep[RSS,][]{burgh2003,kobul2003} on the Southern African Large Telescope \citep[SALT,][]{buckley2006}. The RSS detector is a combination of three CCD detectors with total 3172 \cross\ 2052 pixels and spatial resolution of 0.1267\arc\ per pixel. We further used 2 \cross\ 2 binning to improve SNR. For all our observations we have used long-slit with width of 1.5\arc (matched to the typical seeing in Sutherland) and grating PG0900. The typical spectral resolution obtained  is \til\ 350 \kms.  The grating angle (GR-ANGLE) and camera angle (CAMANG), which determine the wavelength range covered, were set carefully from known redshift for each source such that \lya\ line falls in the most sensitive part of the CCD, and \civ\ and \heii\ lines are also covered. In all cases, we choose the sky position angle (PA) to accommodate a comparison star in the slit. This star spectrum is used for secondary flux calibration and constructing the spectral point spread function (SPSF) determination whenever needed. The observations were carried out between December 2016 to February 2020.

The observational details are provided in Table \ref{tab_observations} (see the table foot note for the column information). In the following we identify sources using the source IDs given in the first column of this table. Each target is observed mostly along two position angles (PAs), so that we can study the spatial distribution of gas around the quasars. For extended radio sources, we also made an effort to align the slit along the radio axis.  However, due to bad weather or scheduling constraints, for six targets spectra could be obtained along only one PA. The total on-source exposure times for each PA along with the number of exposures are provided in column 9 of the Table~\ref{tab_observations}. Each observation is usually split into two exposures of \til\ 1200 s, however for 6 sources we have more than two exposures available and for 2 only single exposure was obtained (see Table \ref{tab_observations} and Fig.~\ref{fig_ps1}). Therefore, total on-source exposure time is not uniform for all the sources in the sample. The sky conditions prevailed during our observations are summarized in the last column of the Table \ref{tab_observations}.

The data were reduced using the SALT science pipeline \citep{crawford2010} and standard {\tt IRAF} tasks\footnote{IRAF is distributed by the National Optical Astronomy Observatories, which are operated by the Association of Universities for Research in Astronomy, Inc., under cooperative agreement with the National Science Foundation.}. For  cosmic ray removal, we used the IRAF task {\tt crmedian}, based on median filtering approach to identify cosmic rays and bad pixels. The algorithm creates a residual image by first subtracting median filtered image from input image and then dividing the same by the sigma image. The pixels in the residual image are then flagged depending on the \sig\ threshold chosen. We have used {\tt lsigma=10} and {\tt hsigma=3}. This means that values above {\tt hsigma} are identified as cosmic rays and below {\tt lsigma} as bad pixels. The chosen \sig\ threshold provide us satisfactory results. The wavelength calibration was performed using Xenon/Argon arc lamp. 1D spectra are extracted from 2D using IRAF task {\tt apall} choosing the aperture size manually such that it includes entire emission above background level.
The spectrophotometric standard stars observed close to the date of observation of source with same instrumental setting were used for flux calibration, where corrections for atmospheric extinction and air mass have been taken into account. The wavelengths were then shifted to vacuum wavelengths. We further apply correction to all the flux calibrated spectra using the magnitudes of  sources obtained from PanSTARRS-1  \citep[PS1;][]{ps12018} catalogue. Correction factor is estimated for each band as the ratio between average flux corresponding to PS1-band magnitude and average flux within 100 \AA\ around the effective wavelength of the filter band. The final correction factor is the mean of the correction factors estimated from all the bands. Since, the SALT standard star spectra were not obtained at the same time as the target source, this correction is critical to set the flux scale accurately.

There are 8 sources with $\delta <-30^\circ$ for which PS1 magnitudes are unavailable. We obtained their photometry from the SuperCosmos Sky Survey\citep[SSS,][]{Hambly2001}. We used R-band photometry to correct for the flux or B band if R magnitude is not present. Once all the exposures are corrected for flux, we combined them using {\tt IRAF} task {\tt scombine} with {\tt scale=median} and {\tt weight=median}. The final 1D spectrum is combination of 1D spectra obtained from individual exposures of two PAs (when available) and are shown in Fig.~\ref{fig_1dspectra}.
The same scaling factors are also applied to 2D exposures. The flux corrected exposures corresponding to individual PA of each source were combined using median weight. Here we emphasize that while correcting for flux, we have ignored the variability in quasar magnitudes between the epoch of our observations and that of the PS1 or SSS observations. As ADC was unavailable during our observations, we have not corrected for differential slitloss due to atmospheric dispersion.

\subsection{uGMRT Band-5 and complementary radio data}
\label{sub_gmrt}

The Band-5 observations (10 hrs of total observing time) of the sample were carried out on 2018 June 30 and 2018 July 14. We used uGMRT Wideband Backend (GWB) with  a  base-band bandwidth of 200 MHz covering 1260-1460\,MHz and split into  8192 frequency channels.  Each target was observed for typically $\sim$15\,mins.  3C\,48, 3C\,147 and 3C\,286 were observed for flux density and bandpass calibrations. A complex gain calibrator was also observed for each target source. Only parallel hand correlations XX and YY were recorded.   

The data were processed using the Automated Radio Telescope Imaging Pipeline (ARTIP) that has been developed to perform the end-to-end processing (i.e., from the ingestion of the raw visibility data to the spectral line imaging) of data from the uGMRT and MeerKAT absorption line surveys. The pipeline is written using standard Python libraries and the Common Astronomy Software Applications (CASA) package; details are provided in \cite{Gupta2020}. In short, following data ingestion, the pipeline automatically identified bad antennas, baselines, time ranges and radio frequency interference  (RFI), using directional and median absolute deviation (MAD) statistics. After excluding these bad data, the complex antenna gains as a function of time and frequency were determined using the standard flux/bandpass and phase calibrators. Applying these gains, a continuum map was obtained, which was then improved using four rounds of phase-only and two rounds amplitude-and-phase self calibrations. 

The uGMRT radio continuum emission overlaid on the optical images are shown in Fig. \ref{fig_ps1}.  The synthesize beams, continuum rms, peak and integrated flux densities of the radio sources are provided in columns 4 - 7 of  Table \ref{tab_radio_props}. In the last two columns of this table, we also provide the largest angular size (LAS) and largest linear size (radio size in kpc at the redshift of the quasars) of the radio emission.  The upper limits correspond to deconvolved size estimated using a single Gaussian component fit. Note that in 6 cases radio emission shows extended radio morphology ($>$1\arc) with multiple components.  In three of these, i.e., M1351-1019 (\#16), M1513-2524 (\#18) and M2101-1747(\#22), the radio emission is in the form of double lobe structure. In the case of M1142-2633 (\#10), multiple components exist but the majority of emission is associated with a single, presumably core, component. For M1043-2323 (\#9), the emission is barely resolved to ascertain the morphology. M0808+0057 (\#7) has most of the emission in a single component and a nearby weak component (3\% flux) which may also be an unrelated source.

The uGMRT Band-3 (250 - 500\,MHz) images of our sample were obtained as part of a larger survey to search for high-$z$ \hi\ 21-cm and OH 18-cm absorption lines. Specifically, relevant for this paper are the spectral indices, $\alpha_{0.4}^{1.4}$, estimated using the NVSS 1.4\,GHz and uGMRT 0.42\,GHz images. The details of these observations, the radio source properties and $\alpha_{0.4}^{1.4}$ are provided in \citet[][]{gupta2021pband}. Although M0636$-$3106 (\#6) is compact at 1.4\,GHz, it exhibits structure at arcsec scales in the Band-3 image \citet[][]{gupta2021pband}. This is an indication of weak extended emission resolved out in our 1.4\,GHz image. For the analysis presented in this paper, we will consider this object to be compact at arcsec scales.

\begin{table*}
\scriptsize
\setlength{\tabcolsep}{3pt}
\caption{Radio properties of the sample from NVSS (column\,3) and uGMRT (columns 4-8) survey}
\centering
\begin{tabular}{lcccccccccccccccccc}
\hline
\hline

ID
&Name
&$\mathrm{F_{1.4GHz}}$
&Beam$\mathrm{}$
&rms 
&F$\mathrm{_{p,1.4 GHz}}$
&F$\mathrm{_{1.4 GHz}}$
&Radio size
&Radio size\\

&
&(mJy)
&
&($\mathrm{mJy\ beam^{-1}}$)
&($\mathrm{mJy\ beam^{-1}}$)
&(mJy)
&(arcsec)  
&(kpc)\\

(1)
&(2)
&(3)
&(4) 
&(5)
&(6)
&(7) 
&(8) 
&(9) \\

\hline
    1&   M025035.54-262743.10&   389.20&   5.0$^{\prime \prime}$\cross1.9,$^{\prime \prime}$-49.1$^{\degree}$&     0.4&     208&     212&   $<$0.5             &$<$ 4 	 \\[0.0cm]
    2&   M041620.54-333931.30&   264.10&   5.6$^{\prime \prime}$\cross2.0,$^{\prime \prime}$+43.2$^{\degree}$&     0.9&     145&     149&   $<$0.3             &$<$ 2 	 \\[0.0cm]
    3&   M050725.04-362442.90&   212.40&   5.0$^{\prime \prime}$\cross1.9,$^{\prime \prime}$+34.8$^{\degree}$&     0.8&     162&     164&   $<$0.5             &$<$ 4 	 \\[0.0cm]
    4&   M052318.55-261409.60&   1354.90&   3.8$^{\prime \prime}$\cross2.0,$^{\prime \prime}$+35.7$^{\degree}$&     3.4&    1266&    1278&   $<$0.5            &$<$ 4 	 \\[0.0cm]
    5&   M061038.80-230145.60&   360.20&   3.2$^{\prime \prime}$\cross2.0,$^{\prime \prime}$+28.2$^{\degree}$&     1.3&     313&     321&   $<$0.5             &$<$ 4 	 \\[0.0cm]
    6&   M063613.53-310646.30&   208.00&   5.4$^{\prime \prime}$\cross3.3,$^{\prime \prime}$+19.8$^{\degree}$&     1.9&      95&      99&   $<$0.9$^\dag$      &$<$ 7 	 \\[0.0cm]
    7&   M080804.34+005708.20&   317.00&   2.3$^{\prime \prime}$\cross2.1,$^{\prime \prime}$+21.7$^{\degree}$&     0.4&     287&     305&   3.9 (Resolved?)    &    30  \\[0.0cm]
    8&   M101313.10-254654.70&   248.80&   3.0$^{\prime \prime}$\cross1.8,$^{\prime \prime}$-10.3$^{\degree}$&     0.7&     206&     220&   $<$0.6             &$<$ 5 	 \\[0.0cm]
    9&   M104314.53-232317.50&   212.10&   3.3$^{\prime \prime}$\cross1.9,$^{\prime \prime}$-13.0$^{\degree}$&     0.5&     105&     182&   3.7 (Resolved)     &    30  \\[0.0cm]
   10&   M114226.58-263313.70&   294.70&   3.3$^{\prime \prime}$\cross1.9,$^{\prime \prime}$-22.5$^{\degree}$&     0.6&     216&     239&   11.9 (Core+diffuse)&    92  \\[0.0cm]
   11&   M121514.42-062803.50&   360.40&   2.2$^{\prime \prime}$\cross1.9,$^{\prime \prime}$-39.4$^{\degree}$&     1.5&     202&     270&   $<$0.8$^\ddag$     &$<$ 6 	 \\[0.0cm]
   12&   M123410.08-332638.50&   297.90&   3.9$^{\prime \prime}$\cross1.9,$^{\prime \prime}$-17.0$^{\degree}$&     0.7&     182&     228&   $<$1.1             &$<$ 9 	 \\[0.0cm]
   13&   M124448.99-044610.20&   384.90&   2.3$^{\prime \prime}$\cross1.9,$^{\prime \prime}$-45.9$^{\degree}$&     0.8&     348&     366&   $<$0.4             &$<$ 3 	 \\[0.0cm]
   14&   M125442.98-383356.40&   219.20&   4.4$^{\prime \prime}$\cross1.9,$^{\prime \prime}$-18.5$^{\degree}$&     0.7&     237&     239&   $<$0.2             &$<$ 2 	 \\[0.0cm]
   16&   M135131.98-101932.90&   726.10&   2.7$^{\prime \prime}$\cross2.1,$^{\prime \prime}$-33.1$^{\degree}$&     0.9&     413&     599&   9.5 (Double)       &    75  \\[0.0cm]
   17&   M141327.20-342235.10&   274.70&   4.2$^{\prime \prime}$\cross1.9,$^{\prime \prime}$-25.8$^{\degree}$&     0.7&     229&     230&   $<$0.5             &$<$ 4 	 \\[0.0cm]
   18&   M151304.72-252439.70&   217.60&   3.7$^{\prime \prime}$\cross1.9,$^{\prime \prime}$-35.5$^{\degree}$&     0.4&     100&     152&   23.7 (Double)      &    185 \\[0.0cm]
   19&   M155825.35-215511.50&   206.90&   3.9$^{\prime \prime}$\cross2.0,$^{\prime \prime}$-45.9$^{\degree}$&     0.8&     138&     141&   $<$0.5             &$<$ 4 	 \\[0.0cm]
   20&   M161907.44-093953.10&   340.30&   3.1$^{\prime \prime}$\cross2.1,$^{\prime \prime}$-51.4$^{\degree}$&     0.4&     300&     307&   $<$0.3             &$<$ 2 	 \\[0.0cm]
   21&   M204737.66-184141.60&   241.70&   2.5$^{\prime \prime}$\cross2.1,$^{\prime \prime}$+13.8$^{\degree}$&     0.7&     198&     209&   $<$0.6             &$<$ 5 	 \\[0.0cm]
   22&   M210143.29-174759.20&   959.50&   2.5$^{\prime \prime}$\cross2.1,$^{\prime \prime}$+12.8$^{\degree}$&     1.1&     537&     948&   2.6 (Double)       &    21  \\[0.0cm]
   23&   M215445.08-382632.50&   759.80&   3.9$^{\prime \prime}$\cross1.8,$^{\prime \prime}$+6.7$^{\degree}$&     2.3&     638&     649&   $<$0.4              &$<$ 3 	 \\[0.0cm]
   24&   M222332.81-310117.30&   231.70&   3.2$^{\prime \prime}$\cross1.7,$^{\prime \prime}$+2.7$^{\degree}$&     0.6&     212&     224&   $<$0.5              &$<$ 4 	 \\[0.0cm]
   25&   M235722.47-073134.30&   235.50&   2.0$^{\prime \prime}$\cross1.8,$^{\prime \prime}$-40.1$^{\degree}$&     0.5&     185&     216&   $<$0.7             &$<$ 6 	 \\[0.0cm]

\hline

\end{tabular}
\label{tab_radio_props}
\begin{flushleft}
Column 1, 2: Source ID, name. 
Column 3: Radio flux denstity at 1.4 GHz from NVSS.
Column 4: Synthesize beams and beam PA with respect to North, for Band-5  uGMRT radio observations.
Column 5-7: Continuum rms, peak and integrated flux densities from Band-5 uGMRT observations.
Column 8,9: Largest angular size (LAS) of the radio emission and corresponding linear size at the redshift of the quasars.\\  
$\dag$  Compact at 1.4\,GHz, it shows extended structure in Band-3 image at 0.42\,GHz. $\ddag$ The uGMRT image is of poor quality.  The LAS constraint is based on the 3\,GHz VLA Sky Survey (VLASS) image. 

\end{flushleft}
\end{table*}

%

\begin{table*}
\scriptsize
\setlength{\tabcolsep}{3pt}
\caption{Measurements based on broad emission lines
}
\centering
\begin{tabular}{lcccccccccccccccccc}
\hline
\hline
ID  
&Name
&\zem
&Line
&Flux
&FWHM
&$\mathrm{\Delta V_{sys}}$
&EQW  
&\flya/\fciv
&\flya/\fheii
&\fciv/\fheii\\

&
&
&
&($\mathrm{10^{-16}\ erg\ s^{-1} cm^{-2}}$) 
&(\kms)
&(\kms)
&(\AA) 
&
&
&\\

(1)
&(2)
&(3)
&(4) 
&(5)
&(6)
&(7)
&(8)
&(9)
&(10)
&(11)\\

\hline

1&   M025035.54-262743.10&   2.9257$\pm$0.0037&    \lya&   61.54$\pm$0.23&   8547$\pm$954&    1476&   64.73&    3.44&$>$   321.73&$>$   93.63\\
&        &        &    \civ&   17.91$\pm$0.16&   5735$\pm$407&       0&   23.53&        &        &        \\
&        &        &   \heii&$<$    0.19&       -&       -&$<$    0.25&        &        &        \\[0.06cm]
2&   M041620.54-333931.30&   3.0409$\pm$0.0006&    \lya&   42.01$\pm$0.21&   5697$\pm$229&    1001&   75.22&    1.96&$>$   451.90&$>$   230.54\\
&        &        &    \civ&   21.43$\pm$0.13&   8090$\pm$136&       0&   47.88&        &        &        \\
&        &        &   \heii&$<$    0.09&       -&       -&$<$    0.19&        &        &        \\[0.06cm]
3&   M050725.04-362442.90&   2.9344$\pm$0.0016&    \lya&   71.23$\pm$0.17&   11299$\pm$2023&    -282&   70.96&    3.80&$>$   603.00&$>$   158.77\\
&        &        &    \civ&   18.75$\pm$0.09&   6190$\pm$261&       0&   23.37&        &        &        \\
&        &        &   \heii&$<$    0.12&       -&       -&$<$    0.15&        &        &        \\[0.06cm]
4&   M052318.55-261409.60&   3.1125$\pm$0.0005&    \lya&   146.01$\pm$0.22&   5748$\pm$653&    -508&   165.91&    2.62&$>$   846.94&$>$   323.48\\
&        &        &    \civ&   55.76$\pm$0.13&   5184$\pm$48&       0&   79.37&        &        &        \\
&        &        &   \heii&$<$    0.17&       -&       -&$<$    0.22&        &        &        \\[0.06cm]
5&   M061038.80-230145.60&   2.8308$\pm$0.0009&    \lya&   117.08$\pm$0.17&   8903$\pm$602&    1231&   78.14&    4.15&$>$   446.80&$>$   107.62\\
&        &        &    \civ&   28.20$\pm$0.11&   5542$\pm$174&       0&   23.55&        &        &        \\
&        &        &   \heii&$<$    0.26&       -&       -&$<$    0.22&        &        &        \\[0.06cm]
6&   M063613.53-310646.30&   2.7559$\pm$0.0030&    \lya&   13.34$\pm$0.17&   2737$\pm$944&     914&   52.98&    2.41&   61.54&   25.50\\
&        &        &    \civ&   5.53$\pm$0.11&   4508$\pm$601&       0&   27.54&        &        &        \\
&        &        &   \heii&   0.22$\pm$0.05&   755$\pm$225&     640&    1.08&        &        &        \\[0.06cm]
7&   M080804.34+005708.20&   3.1402$\pm$0.0035&    \lya&   115.36$\pm$0.19&   4776$\pm$1573&     261&   70.89&    3.69&$>$   617.93&$>$   167.24\\
&        &        &    \civ&   31.22$\pm$0.12&   6641$\pm$99&       0&   24.00&        &        &        \\
&        &        &   \heii&$<$    0.19&       -&       -&$<$    0.14&        &        &        \\[0.06cm]
8&   M101313.10-254654.70&   2.9647$\pm$0.0004&    \lya&   3.51$\pm$0.04&   2043$\pm$257&     382&   60.84&    1.59&$>$   60.87&$>$   38.30\\
&        &        &    \civ&   2.21$\pm$0.03&   2738$\pm$952&       0&   47.85&        &        &        \\
&        &        &   \heii&$<$    0.06&       -&       -&$<$    1.13&        &        &        \\[0.06cm]
9&   M104314.53-232317.50&   2.8784$\pm$0.0009&    \lya&   31.85$\pm$0.13&   2834$\pm$351&     293&   80.57&    2.67&   157.90&   59.25\\
&        &        &    \civ&   11.95$\pm$0.09&   4403$\pm$223&       0&   36.83&        &        &        \\
&        &        &   \heii&   0.20$\pm$0.04&   777$\pm$177&      54&    0.57&        &        &        \\[0.06cm]
10&   M114226.58-263313.70&   3.2372$\pm$0.0002&    \lya&   108.38$\pm$0.18&   2408$\pm$180&     450&   101.83&    2.61&   77.01&   29.47\\
&        &        &    \civ&   41.48$\pm$0.11&   3916$\pm$51&       0&   48.45&        &        &        \\
&        &        &   \heii&   1.41$\pm$0.06&   1534$\pm$168&     783&    1.56&        &        &        \\[0.06cm]
11&   M121514.42-062803.50&   3.2237$\pm$0.0020&    \lya&   50.15$\pm$0.13&   1528$\pm$28&    -468&   118.39&    2.74&   80.20&   29.29\\
&        &        &    \civ&   18.32$\pm$0.17&   2428$\pm$144&       0&   53.77&        &        &        \\
&        &        &   \heii&   0.63$\pm$0.06&   629$\pm$88&    -435&    1.72&        &        &        \\[0.06cm]
12&   M123410.08-332638.50&   2.8182$\pm$0.0013&    \lya&   15.68$\pm$0.38&   1940$\pm$236&     490&   236.10&    2.98&$>$   61.59&$>$   20.68\\
&        &        &    \civ&   5.27$\pm$0.17&   2558$\pm$361&       0&   96.86&        &        &        \\
&        &        &   \heii&$<$    0.25&       -&       -&$<$    3.85&        &        &        \\[0.06cm]
13&   M124448.99-044610.20&   3.1052$\pm$0.0066&    \lya&   10.18$\pm$0.15&   2082$\pm$131&    1281&   43.42&    2.91&$>$   52.72&$>$   18.15\\
&        &        &    \civ&   3.50$\pm$0.11&   3737$\pm$856&       0&   18.64&        &        &        \\
&        &        &   \heii&$<$    0.19&       -&       -&$<$    1.05&        &        &        \\[0.06cm]
14&   M125442.98-383356.40&   2.7793$\pm$0.0012&    \lya&   29.58$\pm$0.18&   3600$\pm$348&     311&   93.92&    3.74&$>$   128.59&$>$   34.37\\
&        &        &    \civ&   7.91$\pm$0.17&   2827$\pm$353&       0&   31.54&        &        &        \\
&        &        &   \heii&$<$    0.23&       -&       -&$<$    0.80&        &        &        \\[0.06cm]
16&   M135131.98-101932.90&   3.0006$\pm$0.0004&    \lya&   110.35$\pm$0.18&   4605$\pm$611&     668&   69.41&    2.47&   96.37&   39.01\\
&        &        &    \civ&   44.67$\pm$0.12&   5921$\pm$178&       0&   34.93&        &        &        \\
&        &        &   \heii&   1.15$\pm$0.07&   2032$\pm$421&     240&    0.88&        &        &        \\[0.06cm]
17&   M141327.20-342235.10&   2.8106$\pm$0.0018&    \lya&   278.24$\pm$0.72&   4769$\pm$189&     375&   205.45&    3.30&$>$   280.26&$>$   84.90\\
&        &        &    \civ&   84.29$\pm$0.47&   5571$\pm$443&       0&   56.86&        &        &        \\
&        &        &   \heii&$<$    0.99&       -&       -&$<$    0.55&        &        &        \\[0.06cm]
18&   M151304.72-252439.70&   3.1312$\pm$0.0004&    \lya&   14.64$\pm$0.11&   1383$\pm$27&     188&   197.10&    4.44&   15.09&    3.40\\
&        &        &    \civ&   3.29$\pm$0.08&   1810$\pm$102&       0&   53.45&        &        &        \\
&        &        &   \heii&   0.97$\pm$0.05&   897$\pm$46&      82&   15.38&        &        &        \\[0.06cm]
19&   M155825.35-215511.50&   2.7633$\pm$0.0023&    \lya&   25.56$\pm$0.22&   2745$\pm$366&    -525&   86.70&    2.41&   65.42&   27.13\\
&        &        &    \civ&   10.60$\pm$0.17&   3350$\pm$241&       0&   32.46&        &        &        \\
&        &        &   \heii&   0.39$\pm$0.09&   661$\pm$158&      -6&    1.05&        &        &        \\[0.06cm]
20&   M161907.44-093953.10&   2.9031$\pm$0.0063&    \lya&   47.72$\pm$0.23&   3491$\pm$315&     -94&   137.62&    5.09&$>$   204.53&$>$   40.20\\
&        &        &    \civ&   9.38$\pm$0.14&   5428$\pm$280&       0&   21.64&        &        &        \\
&        &        &   \heii&$<$    0.23&       -&       -&$<$    0.44&        &        &        \\[0.06cm]
21&   M204737.66-184141.60&   2.9956$\pm$0.0007&    \lya&   81.88$\pm$0.14&   5194$\pm$251&     738&   77.02&    3.60&$>$   451.03&$>$   125.41\\
&        &        &    \civ&   22.77$\pm$0.08&   5024$\pm$78&       0&   26.70&        &        &        \\
&        &        &   \heii&$<$    0.18&       -&       -&$<$    0.20&        &        &        \\[0.06cm]
22&   M210143.29-174759.20&   2.8030$\pm$0.0005&    \lya&   15.39$\pm$0.14&   942$\pm$772&     832&   109.85&    1.47&   21.92&   14.95\\
&        &        &    \civ&   10.50$\pm$0.10&   4051$\pm$232&       0&   91.48&        &        &        \\
&        &        &   \heii&   0.70$\pm$0.05&   840$\pm$162&     -10&    4.80&        &        &        \\[0.06cm]
23&   M215445.08-382632.50&   2.7913$\pm$0.0007&    \lya&   250.80$\pm$0.19&   10144$\pm$1340&    2055&   49.19&    5.57&$>$   704.18&$>$   126.35\\
&        &        &    \civ&   45.00$\pm$0.14&   7321$\pm$48&       0&   10.29&        &        &        \\
&        &        &   \heii&$<$    0.36&       -&       -&$<$    0.09&        &        &        \\[0.06cm]
24&   M222332.81-310117.30&   3.2035$\pm$0.0024&    \lya&   61.51$\pm$0.21&   10197$\pm$153&    1387&   53.99&    2.90&$>$   293.18&$>$   100.93\\
&        &        &    \civ&   21.17$\pm$0.13&   7016$\pm$542&       0&   23.04&        &        &        \\
&        &        &   \heii&$<$    0.21&       -&       -&$<$    0.24&        &        &        \\[0.06cm]
25&   M235722.47-073134.30&   2.7648$\pm$0.0046&    \lya&   28.60$\pm$0.56&   1967$\pm$315&     129&   41.88&    2.18&$>$   38.83&$>$   17.80\\
&        &        &    \civ&   13.11$\pm$0.40&   4631$\pm$205&       0&   22.16&        &        &        \\
&        &        &   \heii&$<$    0.74&       -&       -&$<$    1.24&        &        &        \\[0.06cm]

\hline
\end{tabular}
\begin{flushleft}
Column 1-2, Source ID, name.
Column 3: redshift measured using \civ\ emission line. 
Column 4-6: line ID, line flux and velocity width of the corresponding line.
Column 7-8: Velocity shift of the line with respect to measured redshift and equivalent width of the line.
Column 9-11: Line ratios based on flux provided in Column 5.
\end{flushleft}
\label{tab_fluxes}
\end{table*}
\begin{table*}
\scriptsize
\setlength{\tabcolsep}{1pt}
\caption{Radio loud quasar sample observed properties}
\centering
\begin{tabular}{lcccccccccccccccccc}
\hline
\hline
ID  
&Name
&$\lambda L_{1350}$
&$L_{bol}$
&log$_{10}[M_{BH}/M_\odot$]
&Eddington
&$L_{912}$
&$L_\mathrm{{420MHz}}$
&$L_\mathrm{{1.4GHz}}$
&$\alpha \mathrm{^{1.4}_{0.4}}$ \\
& 
&($10^{46}$\ergs)
&($10^{46}$\ergs)
&
& ratio
&$(\mathrm{10^{30} erg\ s^{-1}Hz^{-1}})$
&$(\mathrm{10^{27} W\ Hz^{-1}})$
&$(\mathrm{10^{27} W\ Hz^{-1}})$
& \\
(1)
&(2)
&(3)
&(4) 
&(5)
&(6)
&(7)
&(8)
&(9)
&(10) \\

\hline
    1&   M025035.54-262743.10&    3.47&   13.18&   9.53$\pm$0.06&    0.30&   10.31$\pm$0.05&    8.84&    8.21&   -0.06 \\[0.1cm]
    2&   M041620.54-333931.30&    2.34&    8.91&   9.73$\pm$0.01&    0.13&   6.84$\pm$0.04&    1.00&    2.24&    0.64 \\[0.1cm]
    3&   M050725.04-362442.90&    3.72&   14.13&   9.61$\pm$0.04&    0.27&   10.94$\pm$0.03&   22.26&    9.97&   -0.64 \\[0.1cm]
    4&   M052318.55-261409.60&    3.89&   14.79&   9.46$\pm$0.01&    0.39&   11.51$\pm$0.05&    2.80&    8.40&    0.88 \\[0.1cm]
    5&   M061038.80-230145.60&    5.01&   19.05&   9.58$\pm$0.03&    0.39&   14.68$\pm$0.08&    0.79&    2.20&    0.82 \\[0.1cm]
    6&   M063613.53-310646.30&    0.76&    2.88&   8.97$\pm$0.10&    0.24&   2.26$\pm$0.05&   19.98&    8.76&   -0.66 \\[0.1cm]
    7&   M080804.34+005708.20&    7.41&   28.18&   9.83$\pm$0.01&    0.32&   21.92$\pm$0.03&   14.65&   10.30&   -0.28 \\[0.1cm]
    8&   M101313.10-254654.70&    0.22&    0.83&   8.25$\pm$0.17&    0.36&   0.65$\pm$0.01&    3.78&    4.31&    0.10 \\[0.1cm]
    9&   M104314.53-232317.50&    1.41&    5.37&   9.09$\pm$0.04&    0.34&   4.00$\pm$0.15&   24.37&   10.22&   -0.69 \\[0.1cm]
   10&   M114226.58-263313.70&    5.37&   20.42&   9.29$\pm$0.01&    0.80&   15.93$\pm$0.04&   12.96&   11.21&   -0.35 \\[0.1cm]
   11&   M121514.42-062803.50&    2.24&    8.51&   8.68$\pm$0.05&    1.37&   6.22$\pm$0.05&   14.03&   10.95&   -0.20 \\[0.1cm]
   12&   M123410.08-332638.50&    0.22&    0.83&   8.19$\pm$0.12&    0.42&   0.63$\pm$0.05&   32.70&   13.71&   -0.69 \\[0.1cm]
   13&   M124448.99-044610.20&    1.05&    3.98&   8.88$\pm$0.20&    0.41&   3.09$\pm$0.04&   22.58&   14.11&   -0.38 \\[0.1cm]
   14&   M125442.98-383356.40&    0.98&    3.80&   8.62$\pm$0.10&    0.70&   2.92$\pm$0.10&    7.49&    5.45&   -0.25 \\[0.1cm]
   16&   M135131.98-101932.90&    6.31&   23.99&   9.69$\pm$0.03&    0.38&   18.76$\pm$0.04&   109.63&   42.19&   -0.76 \\[0.1cm]
   17&   M141327.20-342235.10&    5.01&   19.05&   9.59$\pm$0.07&    0.38&   8.98$\pm$0.29&    0.35&    1.28&    1.02 \\[0.1cm]
   18&   M151304.72-252439.70&    0.35&    1.32&   7.99$\pm$0.04&    1.04&   0.98$\pm$0.05&   106.36&   28.24&   -1.26 \\[0.1cm]
   19&   M155825.35-215511.50&    1.05&    3.98&   8.78$\pm$0.06&    0.51&   1.82$\pm$0.16&    4.76&    4.17&   -0.10 \\[0.1cm]
   20&   M161907.44-093953.10&    1.51&    5.75&   9.28$\pm$0.04&    0.23&   2.14$\pm$0.01&   29.00&   14.19&   -0.57 \\[0.1cm]
   21&   M204737.66-184141.60&    4.27&   16.22&   9.45$\pm$0.01&    0.44&   12.46$\pm$0.03&    6.88&    5.85&   -0.13 \\[0.1cm]
   22&   M210143.29-174759.20&    0.46&    1.74&   8.76$\pm$0.05&    0.23&   1.30$\pm$0.13&   123.74&   47.92&   -0.76 \\[0.1cm]
   23&   M215445.08-382632.50&   16.60&   64.57&   10.10$\pm$0.01&    0.40&   44.56$\pm$0.15&    9.23&    7.01&    0.50 \\[0.1cm]
   24&   M222332.81-310117.30&    5.62&   21.38&   9.81$\pm$0.07&    0.26&   16.50$\pm$0.04&   11.13&    7.81&   -0.28 \\[0.1cm]
   25&   M235722.47-073134.30&    2.19&    8.32&   9.23$\pm$0.04&    0.38&   5.71$\pm$0.70&   12.29&    7.27&   -0.42 \\[0.1cm]

\hline
\end{tabular}
\begin{flushleft}
Column 1-2: Source ID, name.
Column 3: Continuum luminosity at 1350 \AA, measured from the power law fit.
Column 4: Bolometeric luminosity estimated using $L_{1350}$ and bolometric correction factor from \cite{richards2006a}.
Column 5: Virial black hole mass estimated using Eqn. \ref{eqn_bhmss} \citep[see][]{Vestergaard_2006}.
Column 6: Eddington ratio ($L_\mathrm{{bol}}$/$L_{\mathrm{Edd}}$).
Column 7: 912 \AA\ luminosity extrapolated from  power law fit to \lya\ continuum region.
Column 8: Radio power at 420 MHz estimated using radio flux density provided by \cite{gupta2021pband} using uGMRT radio observations at 420 MHz.
Column 9: radio power at 1.4 GHz estimated from NVSS observations provided in Column 3 of Table~\ref{tab_radio_props}.
Column 10: Radio spectral index taken from \cite{gupta2021pband}.
\end{flushleft}
\label{tab_luminosities}
\end{table*}


\section{Analysis}
\label{sec_results}
In this section, we summarize line fluxes, velocity width, equivalent width and line ratios of \lya, \civ\ and \heii\ emission lines, and details of absorption systems detected towards each quasar sight line. We also describe our \lya\ halo detection technique and properties of the \lya\ halos, and properties of the radio emission.

\subsection{Analysis of 1D spectra}

\subsubsection{Emission line analysis}
\label{sec_emi}
In Table~\ref{tab_fluxes}, we present \zem, line fluxes, FWHM, velocity separation with respect to systemtic redshift ($\mathrm{\Delta V_{sys}}$), rest equivalent width (EQW) of emission lines and emission line ratios. To measure these quantities, we fit the emission line profiles using multiple Gaussians to measure velocity width and line peak. For this we use the combined spectrum from all the available spectra (including those obtained with different PA).
The method we follow is similar to that discussed in \citet{shen2011} for the analysis of SDSS spectra. Basically, we identify regions around each emission line to define the continuum. For example, in the case of \civ\ we use the rest wavelength ranges 1445$-$1465\AA\ and  1700$-$1705\AA, and for \heii\ we used 1620$-$1635\AA\ and 1650$-$1660\AA. We fitted the measured fluxes in these regions using a powerlaw of the form, $f_\lambda = A \lambda^\alpha$.
We subtracted this fitted continuum from the observed spectrum. Then, we fit the emission lines in this continuum subtracted spectrum using multiple Gaussians -- up to three Gaussians for \civ\ and five in the case of \lya\ were used. The actual number of Gaussian components needed to fit an emission line is decided by the fit that provides minimum $\chi^2$ and Akaike Information Criterion with correction (AICC) value\footnote{The Akaike information criterion with correction  is given by AICC=$\mathrm{\chi^2_{min}+2kN/(N-k-1)}$, where $\mathrm{\chi^2_{min}}$ is the minimum Chi-squared of the fit, N is the number of data points and k is the number of  fitted  parameters \citep{Akaike1974}.}. While fitting the emission line profile we masked the regions affected by narrow absorption lines, CCD gap regions and artifacts from the cosmic ray removal and sky background subtraction. The errors in the Gaussian parameters were estimated using Monte Carlo method. We randomly generated 50 mock profiles of the emission line using the error on original spectrum and fitted them using the same method as above. The best fit parameters and the errors are obtained from these 50 measurements.
This method of estimating errors is widely used for measuring emission line properties \citep[][]{shen2008,shen2011}.  

Similar to \cite{shen2011}, we use the best fitted profiles to measure the FWHM. While doing this we reject any Gaussian component that contains $<5\%$ of the total flux. Fits to the \civ\ and \lya\ emission lines are shown in Fig.~\ref{fig_civfit} and ~\ref{fig_lyafit} respectively. The gray shaded regions are excluded from the fits. The \civ\ and \lya\ fluxes and equivalent widths are measured from the observed spectra within four times of \civ\ FWHM around peak of the line determined from the fit \citep[see also][]{roettgering1997}. The \heii\ emission line is detected significantly ($>4\sigma$ level) only in eight cases. The fits to the \heii\ emission lines are shown in Fig.~\ref{fig_heiifit}. In most cases we were able to fit this line with a single Gaussian component. In the case of non-detections, we estimate the 3\sig\ upper limits on the \heii\ line flux by adopting the \civ\ line FWHM.

As \heii\ emission line is particularly weak and not detected for most of the quasars in our sample, we have used peak of the overall fit to the \civ\ line to estimate the redshift. In this paper, we will treat \civ\ based redshift measurements (tabulated in column 3 of Table \ref{tab_fluxes}) as systemic redshift. It is well known that \civ\ lines can under-predict the systemic redshift by $\sim$810\,\kms\ for RQQs and $\sim$ 360 \kms\ for RLQs \citep[][]{Richards2011}. Also this shift is known to be correlated with the quasar luminosity \citep[e.g.,][]{Paris2012, shen2016}. As we do not apply luminosity based redshift corrections,  we will treat the \zem\ provided here with caution.

\subsubsection{Virial Black hole mass from \civ\ emission lines}
\label{sub_BHmass}

To measure the black hole mass ($M_{BH}$), we use \civ\ emission line which is readily observed for all the sources in our sample and has also been calibrated as virial black hole  mass  estimator \citep{Vestergaard_2006, park2013}. We measure the FWHM of \civ\ using the fitted profile (also listed in Table~\ref{tab_fluxes}). We estimated the continuum luminosity at rest frame 1350\AA\ using our flux calibrated spectra. These are also summarized in the third column of the Table~\ref{tab_luminosities}. The bolometric luminosity ($L_\mathrm{{bol}}$) estimated using $L_\mathrm{{1350}}$ and the bolometric correction factor ($\mathrm{BC_{1350}}=3.81$) from \cite{richards2006a} are provided in column 4 of Table~\ref{tab_luminosities}. The virial BH mass calibration equation used is from \cite{Vestergaard_2006} and is given by,
%
\begin{align}
\Big(\frac{\text{log}M_{BH}}{M_\odot}\Big)&=0.660+0.53~ \text{log} \Big(\frac{\lambda L_{1350}}{10^{44} \mathrm{erg\ s^{-1}}}\Big)\notag\\
&\phantom{spcspace}+ 2\text{log}\Big(\mathrm{\frac{FWHM_{CIV}}{km\ s^{-1}}}\Big).
\label{eqn_bhmss} 
\end{align}
The  BH mass estimates are summarized in column 5 of Table~\ref{tab_luminosities}. Considering only RLQs, i.e., excluding the radio galaxy (object \#18), the $M_{BH}$ are in the range, $(0.15 - 12.6)\times10^9$ M$_\odot$, with a median of $1.9\times10^9$  M$_\odot$. We also estimated the Eddington ratios ($L_\mathrm{{bol}}$/$L_{\mathrm{Edd}}$\footnote{$L_{\mathrm{Edd}}=4\pi GMcm_H \sigma_T$= 1.25$\times 10^{38} (M/M_\odot)$ \ergs, where $\sigma_T$ is Thomson scattering cross section, $m_H$ is mass of hydrogen atom and $M$ used is estimated virial $M_{BH}$ \citep[see][for details]{Rybicki1986}}.) and computed the luminosity of the H~{\sc i} ionizing Lyman continuum photon by extrapolating the power-law fit to the UV wavelengths. These are provided in columns 6 and 7 of Table~\ref{tab_luminosities}. The Eddington ratios are in the range: 0.16-1.37 with a median of 0.39. The ratio of line widths of \civ\ with respect to H$\beta$ and Mg~{\sc ii}  have shown large scatter (\til\ 0.5 dex), therefore \civ\ based virial $M_{BH}$ estimates are uncertain and should be used with caution \citep[e.g.][]{Shen2012,Trakhtenbrot2012,Ho2012}.

We note that in the $L_{bol}-M_{BH}$ plane, in comparison to the core-dominated RLQs in SDSS, the objects in our SALT-NOT sample (i.e., 86 RLQs at $1.9<z<3.5$) with \civ-based BH masses are fainter and have slightly lower $M_{BH}$ \citep[see section 5.5.2 of][for details]{Gupta2021qsosurvey}. However, the median $L_\mathrm{{bol}}$ and $M_{BH}$ of the 23 RLQs ($2.7<z<$3.5) considered here are slightly higher (0.25\,dex) compared to the MALS-SALT-NOT sample. This is simply because the \lya\ sample has selected higher redshift, and hence intrinsically more luminous, RLQs.  

\subsubsection{Absorption systems}
\label{sec_abs}

We identified all the absorption lines that are present in the 1D spectra. The list of absorption systems identified based on the presence of \civ\ doublets and/or strong H~{\sc i} \lya\ absorption are presented in Table~\ref{tab:associated_abs}. Here, we mainly focus on associated (or proximate) absorption systems, defined to be within 5000\,\kms\ of the systemic redshift of the quasar. In column 3 of Table~\ref{tab:associated_abs}, we identify these systems by suffix ``A" appended to the absorption redshift. The main idea is to quantify the nature of gas flows close to the AGNs in our sample (i.e infall/outflow signatures and/or high density environments) and to connect the properties of diffuse \lya\ halos to these.

As is typical of high-$z$ radio galaxies \citep[][]{vanojik1997}, in the case of the \zem = 2.7602 radio galaxy M1513-2524, we do see \lya\ absorption signatures superimposed on the \lya\ emission profile. There are also indications that the absorbing region is spatially extended. However, we do not detect associated \civ\ absorption. A detailed analysis of this object is presented by \citet{Shukla2021}.

We find 6 damped \lya\ absorption systems (DLAs), \citep[i.e., \lya\ absorbers with H~{\sc i} column density in excess of $2\times 10^{20}$\cms,][]{Wolfe2005}, in our sample. One of these is within 3000 \kms to the quasar redshift. Details of these systems along with the results of H~{\sc i} 21-cm absorption searches are presented in \citet{gupta2021pband}. \citet{gupta2021pband} have also identified a potential proximate sub-DLA towards  M0507$-$3624 (\#3). This system has \zabs$\sim$\zem\ and shows signatures of damping wing in the \lya\ absorption profile and absorption from singly ionized species. However, we do detect some residual flux in the core of the \lya\ absorption. In our recent spectrum we also notice similar associated absorption with \lya\ showing damping wings with some residual flux in the core in case of M0523-2614 (\#4). 
In this case also the absorption redshift is slightly more than the emission redshift (see Table.~\ref{tab:associated_abs}). DLAs at the redshift of the quasars with non-zero residual flux are detected in SDSS spectra in a few cases \citep[see for example,][]{Fathivavsari2018,Noterdaeme2019}. Given the low resolution of our spectra it is difficult to confirm whether the residual flux is indeed related to the partial coverage and not an artifact of poor spectral resolution. In summary, in total there are three high column density \lya\ absorbers (i.e two potential sub-DLAs and a DLA)  with detectable low ion absorption very close (i.e relative velocity within 3000 \kms\ to the quasar) to the quasar. In all these three cases the radio source is compact ($<$1\arc).

\begin{figure} 
\centerline{\hbox{ 
\includegraphics[viewport=30 0 670 670,height=0.36\textheight,angle=0,clip=true]{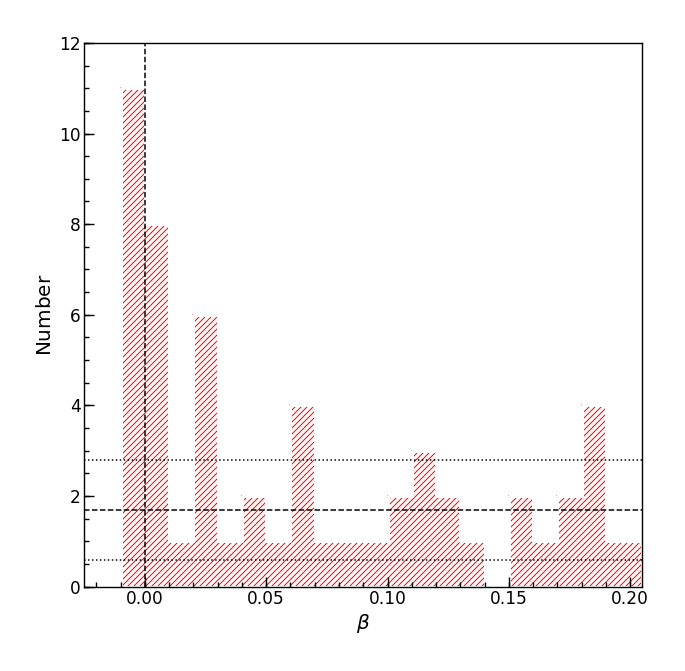}
}} 
\caption{Distribution of number of \civ\ absorbers at different $\beta$. The dashed and dotted horizontal lines respectively give the mean and standard deviation of number of absorbers per bin for $\beta\ge 0.05$. The excess absorption is clearly visible in the low $\beta$ bins.} 
\label{fig_beta_plot}   
\end{figure} 
%
\begin{table}
    \centering
    \caption{Absorption systems detected in our SALT spectra.}
    \resizebox{8cm}{!}{
    \begin{tabular}{lcll}
    \hline\hline
    \multicolumn{1}{c}{Quasar} & \zem &\multicolumn{1}{c}{\zabs} & \multicolumn{1}{c}{Species}\\

\multicolumn{1}{c}{(1)} & (2) &\multicolumn{1}{c}{(3)} & \multicolumn{1}{c}{(4)}\\

   \hline
M025035.54-262743.10  & 2.9257 & 2.4134 & \civ\\
                      &        & 2.8419 & \lya, C~{\sc iv}, Si~{\sc iv}\\
                      &        & 2.8774A & \lya, C~{\sc iv}, Si~{\sc iv}\\
                      &        & 2.9393A& \lya, N~{\sc v}, C~{\sc iv}\\
M041620.54-333931.30  & 3.0409 & 2.3927 & \civ \\ 
                      &        & 2.6525 & \civ, Si~{\sc iv}\\
                      &        & 2.8558 & DLA \\
                      &        & 3.0395A& \lya, \civ \\
M050725.04-362442.90  & 2.9344 & 2.3554 & C~{\sc iv}\\
                      &        & 2.9544A & pDLA \\
M052318.55-261409.60  & 3.1125 & 2.8375 & \lya, \civ \\
                      &        & 3.0076A& C~{\sc iv}, \lya\\
                      &        & 3.1129A& pDLA\\
M061038.80-230145.60  & 2.8308 & 2.3975 & DLA \\ 
                      &        & 2.4400 & \civ, Si~{\sc iv} \\ 
                      &        & 2.6441 & \civ, \lya \\
                      &        & 2.7519 & \lya, \civ, Si~{\sc iv}  \\
                      &        & 2.8138A&  \lya, N~{\sc v}, \civ, Si~{\sc iv} \\
M063613.53-310646.30  & 2.7559 & 2.3449L& \lya, C~{\sc ii}, Si~{\sc ii}, \civ \\ 
                      &        & 2.7641A& \lya, N~{\sc v}, Si~{\sc iv}, \civ\\ 
M080804.34+005708.20  & 3.1402 & 2.4990 & \civ, Si~{\sc ii}, Si~{\sc iv}, \\
&&&Al~{\sc iii}, Al~{\sc ii}\\
                      &        & 2.6784 & \civ, Si~{\sc iv} \\
M101313.10-254654.70  & 2.9647 & 2.6838 & DLA \\ 
M104314.53-232317.50  & 2.8784 & 2.2276 & \civ, Si~{\sc ii}, Si~{\sc iv}, \\
& & &Al~{\sc iii}, Al~{\sc ii}\\
                      &        & 2.2358 & \civ, Si~{\sc iv}\\
                      &        & 2.4210 & \civ, Si~{\sc iv}\\
                      &        & 2.5656 & \lya, \civ, Si~{\sc iv} \\
                      &        & 2.8960A& \lya, \civ, Si~{\sc iv}\\
M114226.58-263313.70  & 3.2372 & 3.1271 & \civ, \lya \\ 
M121514.42-062803.50  & 3.2237 & 2.4869 & \civ \\
                      &        & 3.2414 & \lya(?) \\
M123410.08-332638.50  & 2.8182 & 2.8089A & \lya, \civ \\ 
M124448.99-044610.20  & 3.1052 &2.4060 &\civ   \\ 

                      &        & 3.0731A & \civ\ \\
                               &        & 3.1280A & \lya, \civ\ \\
M125442.98-383356.40  & 2.7793 & 2.7921A& \lya, N~{\sc v},\civ \\ 
M135131.98-101932.90  & 3.0006 & 2.7705 & DLA \\ 
                      &        & 3.0139A& \lya, N~{\sc v}, Si~{\sc iv}, C~{\sc iv} \\
M141327.20-342235.10  & 2.8106 & 2.1581& Si~{\sc ii}, Al~{\sc ii}, Al~{\sc iii} \\
                      &        & 2.5828 & \civ, \lya \\
M155825.35-215511.50  & 2.7633 & 2.6380 & \civ, \lya \\ 
                      &        & 2.7350 & \civ, \lya \\
                      &        & 2.7665A& \lya, \civ, N~{\sc v}\\
M161907.44-093953.10  & 2.9031 & 2.1894 & \civ\\
                      &        & 2.3382 & \civ \\
                      &        & 2.6570 & \civ, \lya \\
                      &        & 2.7923 & DLA\\
M204737.66-184141.60  & 2.9956 & 2.2309 & \civ\\ 
                      &        & 2.5226 & \civ\\
                      &        & 2.7303 & \civ, \lya\\
M210143.29-174759.20  & 2.8030 & 2.7990A& \civ, \lya\\
                      &        & 2.8065A& \civ, \lya\\
M215445.08-382632.50  & 2.7913 & 2.7611A & pDLA\\ 
M222332.81-310117.30  & 3.2035 & 2.4919 & \civ \\ 
                      &        & 2.8064 & \civ, \lya\\
M235722.47-073134.30  & 2.7648 & 2.6696 & \civ \\
                      &        & 2.7375 & \civ, \lya\\
\hline
    \end{tabular}
    }
\begin{flushleft}
Column 1: Source name. Column 2: Quasar redshift from Table~\ref{tab_fluxes}. Column 3: Redshift of the absorption systems listed in column 4. ``A" stands for associated absorption systems defined to be within 5000 \kms\ of the systemic redshift.
\end{flushleft}
    \label{tab:associated_abs}
\end{table}

Next we focus on \civ\ absorption associated with 23 RLQs.  In two cases, i.e., M1215-0628 (\#11) and M2223-3101 (\#24), the red part of the \civ\ emission line falls in the ccd gap. So we will not consider these cases for the discussion below. In the remaining 13/21 cases, we see \civ\ and associated \lya\  absorption within 5000 \kms. This implies associated absorption detection rate of $62\pm17$\%.  Even if we restrict ourselves to strong \civ\ lines i.e., with rest equivalent width $\ge$ 0.5\AA, we find associated absorption in 43$\pm$14\% (i.e., 9/21) of the quasars.  This can be interpreted as $\sim$ 40\% of the solid angle to the central source being covered by \civ\ absorbers with equivalent width greater than 0.5\AA. In 29$\pm$12\% of the cases, the strong \civ\ absorption is detected in the red wing of the \civ\ emission line, which may imply infalling gas. However this scenario needs further confirmation using accurate systemic redshift measurements obtained from rest frame optical emission lines such as H$\alpha$, H$\beta$ and [O~{\sc iii}].
%

Further, in Fig.~\ref{fig_beta_plot} we plot the distribution of \civ\ absorbers  as a function of relative ejection velocity ($\beta$) with respect to the quasar. The dashed and dotted horizontal lines in this figure correspond to the mean expected number  and  1$\sigma$ range in each $\beta$ bin computed from the observed values $\beta\ge0.05$ bins. As has been found in the earlier studies \citep[e.g.,][]{Wild2008}, we see a statistical excess of associated \civ\ absorption even in our sample. Also, the fraction of quasars showing associated \civ\ absorption is higher than that reported in the literature \citep{Vestergaard_2006, Wild2008, Nestor2008, Perrotta2016,ChenZ2017}. 

\citet{Gupta2021qsosurvey} constructed a comparison sample of quasars from SDSS with similar WISE color-cuts and radio flux density cut-off. There are 189 quasars in that sample at \zem$>2.0$. We searched for the associated \civ\ absorption in this and find the detection rate to be 31$\pm$4\% as compared to $62\pm17$\% in the \lya\ sample. 
However, due to small number of systems in our sample the difference is statistically insignificant. 
As mentioned above, Fig.~\ref{fig_beta_plot} also confirms slight excess of systems with negative $\beta$ in our sample. We note that the redshift differences are larger (i.e $\ge 1000$ \kms in more than 70\% cases) than typical error in the systemic redshift measurements using \civ\ line. Nevertheless, \civ-based measurements usually under predict systemic redshifts. Therefore, more accurate redshifts are needed to understand the nature of excessive \civ\ absorption observed in our sample.

\subsection{Analysis of 2D spectra}

We use the 2D spectra and SPSF subtraction method to detect and quantify the spatial distribution of the extended \lya\ halo. Here, we present the details of the method  \citep[see also][]{Shukla2021}, the systems with \lya\ halo detections and the measurement of various parameters of the \lya\ halo.

\subsubsection{SPSF subtraction and detection of extended Ly$\alpha$ Halos}
\label{sub_detection}

\begin{figure*} 
\centerline{\vbox{
\centerline{\hbox{ 
\includegraphics[viewport=340 400 8500 9100, height=0.95\textwidth,angle=0,clip=true]{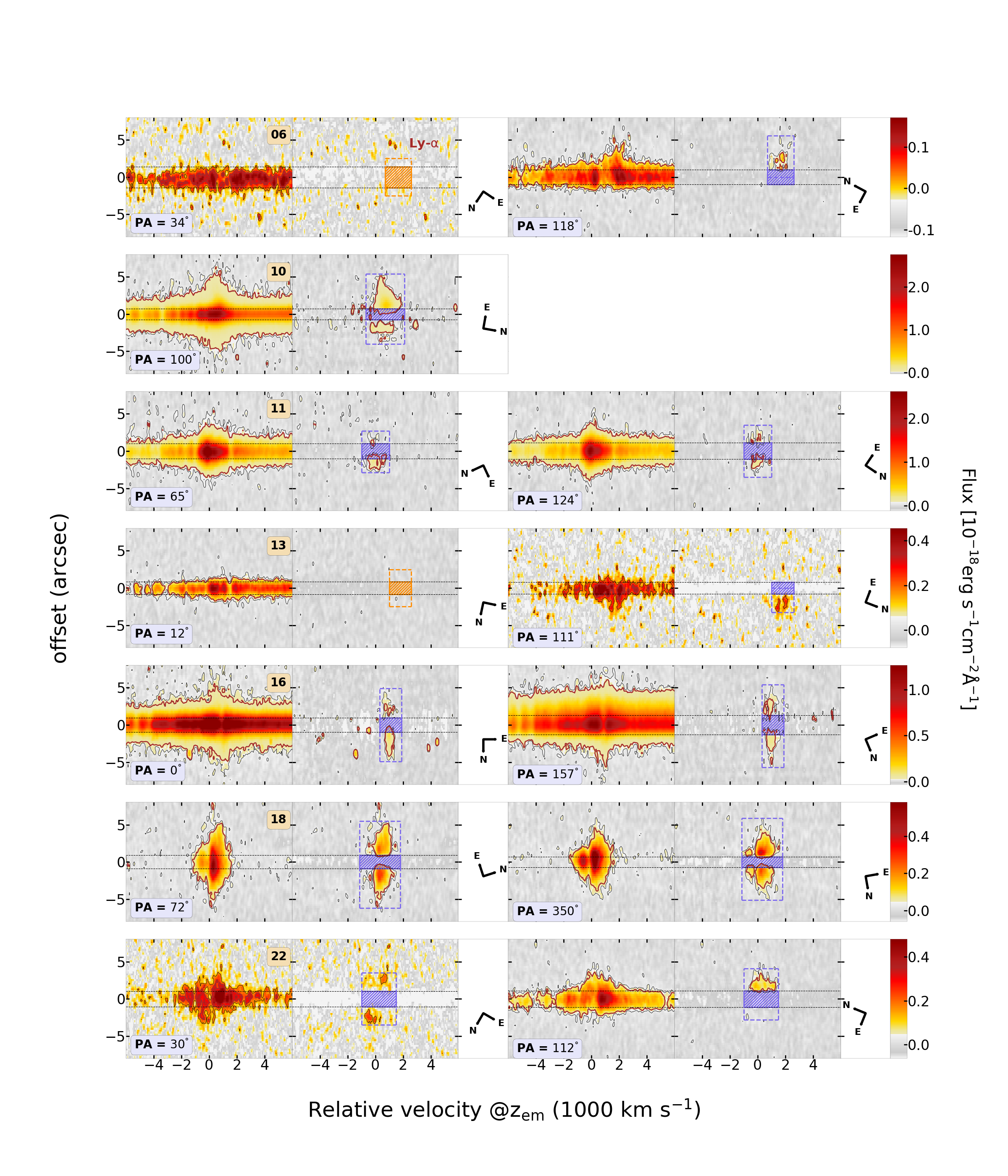}
}} 
}}  
\vskip+0.0cm  
\caption{Long-slit 2D spectra of the 7 RLQs in our sample showing presence of extended emission. In each row we have shown the 2D spectra along available PAs before and after SPSF subtraction. Black dashed horizontal lines mark  the SPSF FWHM. The clear detections are shown by blue boxes and the non-detections by orange. These boxes are used for measuring the \lya\ halo size (at 3\sig\ flux levels) for clear detections and the total flux (flux limit) for detections (non-detections). 
All the 2D spectra are smoothed by 3\cross3 pixels to improve the SNR and to remove the pixel-to-pixel correlation. The 3\sig\ and 5\sig\ flux contour level for individual PAs spectra for all the quasars are shown in black and brown, respectively.
The North and East directions are also indicated. 
} 
\label{fig_lya_samp1}   
\end{figure*} 

\begin{table*}
\scriptsize
\setlength{\tabcolsep}{2pt}
\caption{Properties of extended \lya\ halos}
\centering
\begin{tabular}{lcccccccccccccccccc}
\hline
\hline

ID  
&Name
&PA
&$z^\mathrm{{halo}}_\mathrm{{Ly\alpha}}$
&$\mathrm{L_{Ly\alpha}}$
&FWHM
&$\mathrm{f_{3\sigma}}$
&Halo Extent
&$\mathrm{\Delta V_{halo}}$\\

&
&(degree)
&
&($10^{43}$ \ergs)
&(\kms)
&($10^{-18}$ \fcgs)
&(kpc)
&(\kms)\\

(1)
&(2)
&(3)
&(4) 
&(5)
&(6)
&(7) 
&(8) 
&(9) \\

\hline
1&   J025035.54-262743.10&      30&       -&$<$    0.09&       -&    0.90&       -&       -\\
&        &     120&       -&$<$    0.13&       -&    1.22&       -&       -\\[0.0cm]
2&   J041620.54-333931.30&      15&       -&$<$    0.14&       -&    0.81&       -&       -\\
&        &     111&       -&$<$    0.16&       -&    1.50&       -&       -\\[0.0cm]
3&   J050725.04-362442.90&     118&       -&$<$    0.09&       -&    0.85&       -&       -\\
&        &     178&       -&$<$    0.13&       -&    0.77&       -&       -\\[0.0cm]
4&   J052318.55-261409.60&      89&       -&$<$    0.14&       -&    1.43&       -&       -\\
&        &     167&       -&$<$    0.16&       -&    1.58&       -&       -\\[0.0cm]
5&   J061038.80-230145.60&      60&       -&$<$    0.18&       -&    1.93&       -&       -\\
&        &     306&       -&$<$    0.12&       -&    0.71&       -&       -\\[0.0cm]
6&   J063613.53-310646.30&      34&       -&$<$    0.19&       -&    1.40&       -&       -\\
&        &     118&   2.7712&    0.68&    1144&    0.45&      41&    1222\\[0.0cm]
7&   J080804.34+005708.20&      17&       -&$<$    0.17&       -&    0.62&       -&       -\\
&        &     124&       -&$<$    0.12&       -&    0.92&       -&       -\\[0.0cm]
8&   J101313.10-254654.70&      84&   2.9672&    0.03&    1022&    0.26&      36&     187\\
&        &        &        &        &        &        &        &        \\[0.0cm]
9&   J104314.53-232317.50&      10&   2.8781&    0.23&     678&    0.59&      34&     -27\\
&        &        &        &        &        &        &        &        \\[0.0cm]
10&   J114226.58-263313.70&     100&   3.2473&   17.11&    1007&    0.54&      68&     713\\
&        &        &        &        &        &        &        &        \\[0.0cm]
11&   J121514.42-062803.50&      65&   3.2172&    4.13&    1016&    1.27&      39&    -462\\
&        &     124&   3.2161&    3.91&     906&    1.38&      49&    -539\\[0.0cm]
12&   J123410.08-332638.50&      95&       -&$<$    0.23&       -&    2.48&       -&       -\\
&        &        &        &        &        &        &        &        \\[0.0cm]
13&   J124448.99-044610.20&      12&       -&$<$    0.16&       -&    0.94&       -&       -\\
&        &     111&   3.1291&    1.80&    1154&    3.92&      24&    1745\\[0.0cm]
14&   J125442.98-383356.40&      11&       -&$<$    0.17&       -&    1.25&       -&       -\\
&        &        &        &        &        &        &        &        \\[0.0cm]
16&   J135131.98-101932.90&       0&   3.0144&    3.18&     474&    0.70&      72&    1035\\
&        &     157&   3.0137&    2.87&     476&    0.60&      84&     983\\[0.0cm]
17&   J141327.20-342235.10&      76&       -&$<$    0.28&       -&    3.55&       -&       -\\
&        &        &        &        &        &        &        &        \\[0.0cm]
18&   J151304.72-252439.70&      72&   3.1375&   14.91&     921&    0.72&      87&     461\\
&        &     350&   3.1366&   15.06&    1150&    0.81&      83&     392\\[0.0cm]
19&   J155825.35-215511.50&      15&       -&$<$    0.17&       -&    1.89&       -&       -\\
&        &      97&   2.7576&    0.37&     510&    1.87&      25&    -451\\[0.0cm]
20&   J161907.44-093953.10&       0&       -&$<$    0.16&       -&    1.52&       -&       -\\
&        &      89&       -&$<$    0.16&       -&    1.41&       -&       -\\[0.0cm]
21&   J204737.66-184141.60&      30&       -&$<$    0.11&       -&    1.02&       -&       -\\
&        &     110&   3.0197&    0.11&     357&    0.40&      32&    1810\\[0.0cm]
22&   J210143.29-174759.20&      30&   2.7967&    2.19&    1574&    3.17&      51&    -495\\
&        &     112&   2.8046&    2.35&    1259&    0.75&      53&     127\\[0.0cm]
24&   J222332.81-310117.30&       0&       -&$<$    0.09&       -&    0.84&       -&       -\\
&        &      90&   3.2264&    0.63&    3099&    1.46&      21&    1637\\[0.0cm]
25&   J235722.47-073134.30&      20&       -&$<$    0.33&       -&    2.17&       -&       -\\
&        &      90&       -&$<$    0.85&       -&   11.62&       -&       -\\[0.0cm]

\hline

\end{tabular}
\begin{flushleft}
Column 1-2: Same as Table~\ref{tab_observations}.
Column 3: Observed PAs for long-slit observations.
Column 4: Redshift of the \lya\ when detected, measured from Gaussian fits to the extracted spectral profile of the halos shown in Figs.~\ref{fig_halo_velprof} and \ref{fig_halo_velprof_tentative}.
Column 5: \lya\ halo luminosity (or luminoisty limit) estimated from measured fluxes or 3\sig\ flux limits.
Column 6: \lya\ halo velocity FWHM from Gaussian fits.
Column 7: 3\sig\ flux level per pixel reached in our observations.
Column 8: \lya\ halo extent estimated at corresponding 3\sig\ flux level.
Column 9: Velocity shift of the \lya\ halo with respect to the quasar redshift, positive velocities indicate redshifted halos.\\
Note: The $\mathrm{f_{3\sigma}}$ varies among each observations despite having similar exposure times. This is due to varying weather conditions and non-uniform aperture of SALT telescope during the course of observations as target moves across the sky. 

\end{flushleft}
\label{tab_halo_props}
\end{table*}

We carry out SPSF subtraction at each pixel along the wavelength axis over 300\,\AA\  
centered at the peak of the \lya\ emission line. To construct a model SPSF, we extract spatial profiles by collapsing continuum emission over the wavelength interval of \til 40 \AA\ wide in the quasar spectrum close to \lya\ line and free of contamination from strong emission and/or absorption lines. Recall that in all our observations, we also have a reference star included in the long slit. In the cases with weak quasar continuum\footnote{For objects \#8, \#12, \#13, \#18, \#19, \#22 and \#25 (PA=90\degree).}, we have used the reference star spectrum for modeling SPSF. 

At each wavelength, we construct the model profile by matching the observed spatial profile (over $\pm2$ pixels around the peak of the quasar trace) with SPSF by varying its amplitude and centroid location. The best fitted values of the amplitude and location are obtained using $\chi^2$ minimization. We then refine the centroid positions by fitting a line to centroid vs wavelength scatter plot using a lower order polymonimal. We use the fitted values to center the scaled SPSF while subtracting the 2D SPSF model from the quasar spectrum. The original and SPSF subtracted 2D spectra around the \lya\ emission lines are shown  in Figs.~\ref{fig_lya_samp1}, \ref{fig_lya_samp2} and \ref{fig_lya_samp4}. 

In the  column 10 of Table~\ref{tab_observations} we provide the measured FWHM  of the SPSF for each observations. As can be seen from the table, the FWHM of SPSF ranges from 1.4-2.9\arc, which allows us to detect \lya\ halos extending beyond central \til\ 2\arc\ region. For the cosmological parameter assumed here 1\arc\ corresponds to 7.9 proper kpc at the median redshift of our sample. This together with the inherent nature of SPSF method to over-subtract around the quasar trace will mean our spectra have poor sensitivity to detect \lya\ halos of smaller sizes (i.e., $<$10-15 kpc).  

We first smoothed the SPSF subtracted spectra by 3\cross 3 pixels and estimate the background \sig\ level per pixel reached in the data. We take a segment of \til\ 15000 \kms \cross 10\arc\ of this spectra around \lya\ peak and quasar spatial center (but avoiding the central over subtracted regions) and use that for the purpose of \lya\ halo detection. We use the connected component labelling algorithm with union finding of classical binary image analysis for identifying the \lya\ halos \cite[similar to the procedure used by][]{Borisova2016, Arrigoni2019}. In this method, pixels above a defined \sig\ threshold are selected and connected. Once the pixels are thresholded the components with total number of pixels less than 10 are rejected as suspects of cosmic rays and other artifacts. We put another threshold on the minimum size of the connected component and its total significance for it to be considered as a candidate halo, i.e. the halo sizes must be $>$350 \kms \cross 2.5\arc\  along the dispersion and spatial axis respectively with a total significance $>4$. The condition on the minimum size of the halo is based on the spectral PSF (ranging from 1.4-2.9\arc) and the spectral resolution (ranging from 200-300 \kms) reached in our data. For thresholding of pixels we have used 3\sig\ level to find the halos.

Considering 3\sig\ thresholding, we find 8 quasars possessing extended halo (i.e. quasars with IDs \#6, \#10, \#11, \#13, \#16, \#18, \#22 and \#23 in Table~\ref{tab_observations}). The extended \lya\ emission for these objects are presented in Fig.~\ref{fig_lya_samp1}. One of these objects is a radio-galaxy (M1513-2524; \#18) showing the largest \lya\ extent in our sample, for which we have presented a detailed analysis in \cite{Shukla2021}. The quasar M215445.08-382632.5 (i.e \#23) has a faint companion foreground source that is blended with the quasar light for typical seeing prevailed during our observations. Therefore, the spectrum of this source is not spatially resolved in any of our observations and its contamination results in poor SPSF subtraction which prevents us from detecting faint diffuse emission. So we cannot confirm presence/absence of \lya\ halo and will not consider this object in our analysis any further. Thus we have  \lya\ halo detections in 7 out of 23 AGNs  when we use $3\sigma$ threshold. Apart from 1 case (i.e, object \# 10), we have spectra along two PAs for the remaining 6 \lya\ halo detections. 

Thresholding pixels with 3\sig\ cut reduces our chance of detecting underlying fainter halos, as for some objects we do see clear residual \lya\ emission, but below 3\sig\ cut. So just to be sure that the method prescribed above does not miss any quasar with extended emission, we visually search for halos around the \lya\ emission peak in the smoothed 2D spectra. Once we identify the possible halo (having sizes $> 350$ \kms\ and 2.5\arc\ along the dispersion and spatial axes and having consistent signals in the two available exposures) we estimate the significance of the features by measuring the rms in random locations in our spectrum over a box of size similar to the identified \lya\ emitting region. We confirm the extended \lya\ halos in all the above mentioned 7 sources using this method as well. We also find excess \lya\ emission in additional five sources viz. \#8, \#9, \#19, \#21 and \#24 (see \ref{fig_lya_samp2}) with total significance of the emission within the box ranging from 3-5 \sig. For three sources (\#19, \#21 and \#24) we have obtained spectra along two PAs. It is evident from Figs.~\ref{fig_lya_samp2} and \ref{fig_halo_velprof_tentative} that we detect narrow \lya\ emission features in at least one of these spectra. The extended emission is seen along single PA for objects \#19 (PA=97\degree), \#21 (PA=110\degree) and \#24 (PA=90\degree). We notice that the object \#24 has large  emission close to the quasar trace in the SPSF subtracted spectra of both the PAs, which we believe to be residuals from poor subtraction. However, the emission \til\ 2200 \kms\ in PA=90\degree\ is real. In our discussions we will treat these 5 cases as tentative detections.  For the 11 remaining cases we do not have any clear signatures of extended \lya\ emission in the SPSF subtracted images (see Fig.~\ref{fig_lya_samp3}).

In summary, in the sample of 23 AGNs we confirm the presence of clear extended \lya\ emission in 7 cases (objects \#6, \#10, \#11, \#13, \#16, \#18 and \#22). Six of these are quasars and one (i.e \#18) is a radio galaxy. In  five additional cases (objects \#8, \#9, \#19, \#21 and \#24), we rather see small scale and faint halos with total significance of the emission around \lya\ in the range 3-5 $\sigma$. We consider these as tentative detections, since high SNR and better spatial resolution observations are required to  confirm the presence of extended \lya\ emission at a higher significant level.

We summarise the parameters of the extended \lya\ halos in Table~\ref{tab_halo_props} for each source along available PAs. For both clear and tentative detections, we measure the sizes of the halo from the 3\sig\ contour level (values provided in column 7) shown in Figs. \ref{fig_lya_samp1} and \ref{fig_lya_samp2}.
Here \sig\ is the standard deviation per pixel obtained in the background region. The halo extent provided in column 8 of Table~\ref{tab_halo_props} is the distance between farthest spatial points of the 3\sig\ contour and is shown by height of the box in Fig~\ref{fig_lya_samp1} and \ref{fig_lya_samp2}. To estimate the total halo flux, we mask the central SPSF FWHM+0.5 \arc\ region in the 2D spectra and sum the fluxes of all the pixels within the box and multiply with the pixel width $d \lambda$. For non-detections, we estimate the standard deviation ($\sigma_{box}$) from a region within -500 to 500 \kms\ and -2.5\arc\ to 2.5\arc\ around the \lya\ region (also shown as box in Figs. \ref{fig_lya_samp2} and \ref{fig_lya_samp3}). The lower limit on the total flux is then given by 3$\sigma_{box} \sqrt N$, where N is the total number of pixels used for estimating $\sigma_{box}$. The corresponding luminosities are provided in column 5 of Table~\ref{tab_halo_props}. Upper limits in \lya\ luminosities in the case of non-detections were obtained using the above mentioned 3$\sigma$ limits on the \lya\ flux.

\begin{figure} 
\centerline{\vbox{
\centerline{\hbox{ 
\includegraphics[width=0.5\textwidth,angle=0]{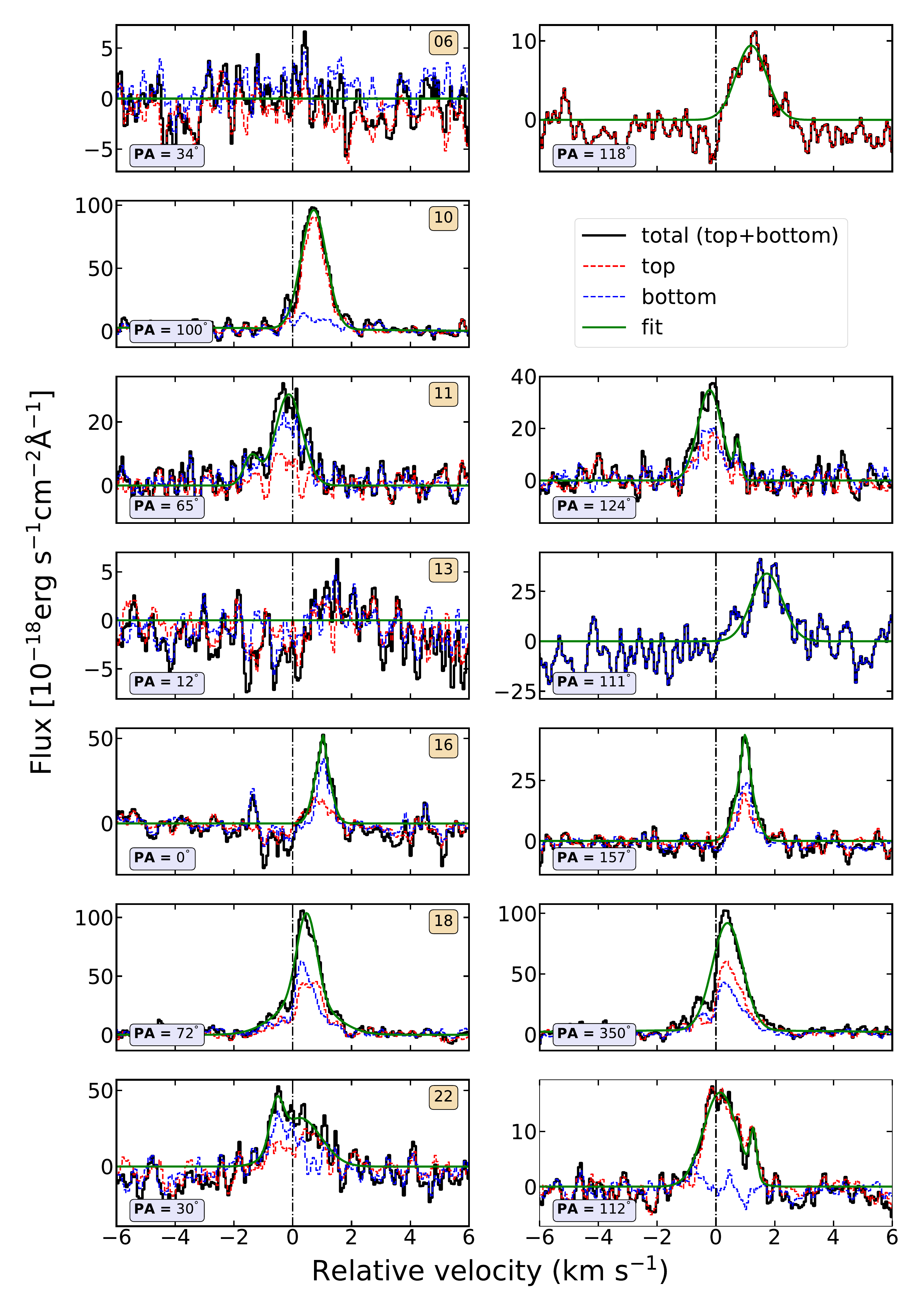}
}} 
}}  
\vskip+0.0cm  
\caption{Velocity profiles (with respect to \zem) of the extended \lya\ halos for clear detections. 
The blue and red colors are for spectra extracted from the `top' and `bottom' region around the quasar trace obtained collapsing the spatial region within the box (see Fig. \ref{fig_lya_samp1}). The black solid lines are sum of the `top' and `bottom' profiles and the green lines are Gaussian fits (see Sec. \ref{sub_detection}) to the total profile.} 
\label{fig_halo_velprof}   
\end{figure}

To measure the width and peak of the \lya\ halo emission, we extract velocity profiles summing the spatial region within the box after masking the central SPSF FWHM+0.5 \arc\ region for the 7 cases where we detect extended emission (see Fig. \ref{fig_halo_velprof}). In this figure we show the profile (i.e.,  1D spectrum extracted above and below the quasar trace separately in addition to the total profile). Derived parameters using these 1D spectra are summarised in Table~\ref{tab_halo_props}. We fit these profiles with one or two Gaussian components, and measure the redshift from the peak of the fit (column 4), width at half maximum of the fit (column 6), and the velocity separation between the \lya\ halo and quasar redshift (column 9). Positive velocities indicate redshifted halos with respect to the systemic redshift. The gaussian fits to the tentative detections are shown in Fig~\ref{fig_halo_velprof_tentative}.

In the case of confirmed detections, the measured extent of the \lya\ emission ranges from \til\ 24-87 kpc and \lya\ halo luminosities of (0.79-17.11)$\times 10^{43}$\ergs. For two cases (\#6 and \#13), the extended \lya\ emission is seen along one PA only, i.e. along 118\degree\ and 111\degree, respectively. Even in the spectra where we see extended emission we do see the emission being asymmetric with respect to the quasar trace. In the case of \#10 we have obtained spectra along only one PA. In this case also the distribution of \lya\ emission is asymmetric with respect to the quasar trace (see Fig.~\ref{fig_lya_samp1}). These results are consistent with the non-spherical nature of the extended \lya\ emission. In the remaining 4 cases the measured sizes along two PAs are consistent with each other within 20\% (see Table~\ref{tab_halo_props}).

From the column 6 of Table~\ref{tab_halo_props}, the measured FWHM of the \lya\ halo emission is more than 900 \kms\ (covering a range of 906-1574 \kms) in 6 out of 7 cases. In only one case (i.e object \#16) we have its value in the range 534-596 \kms. Typically profiles with FWHM more than 1000 \kms\  are considered systems with perturbed kinematics \citep[as per the definitions used in][]{vanojik1997} and possible jet-gas interaction. Thus it appears that most of our detections may have perturbed kinematics. We discuss this for individual sources in more detail in the Appendix.

In Fig. \ref{fig_spsf}, we plot the FWHM of SPSF against the 3\sig\ flux level (see column 7 of Table \ref{tab_halo_props}) for all the sources along each available PA (i.e., a total of 40 2D images for 23 objects). The filled circles mark the PA along which extended \lya\ emission is seen and the empty circles are for non-detections. The stars are tentative detections. The two dashed lines mark the median values of the FWHM of SPSF (2.1\arc) and f$_{3\sigma}$ flux ($1.21\times 10^{-18}$\fcgs). It is obvious that our survey is limited by seeing, sensitivity and covering factor, with maximum number of \lya\ halos (including tentative ones) detected for sources having both seeing and flux sensitivity below the respective medians. 

Our detection rate is considerably lower than 77$-$100\%  claimed in the very recent studies \citep{Borisova2016,Arrigoni2019,cai2019,Osullivan2020,Fossati2021} using IFU-spectroscopy with 8-10m class telescopes and by \citet{heckman1991a} towards extended radio sources using narrow band imaging. The average 2\sig\ surface brightness (SB$_{\rm Ly\alpha}$) limit reached in 1 arcsec$^2$ aperture in a single wavelength channel (1.25\AA) of MUSE observations is $8.8\times 10^{-19}$ \ergscma. If we simply convert this number to per pixel 3\sig\ flux level corresponding to our observations then the 3\sig\ flux limit would be $4.0\times 10^{-19}$\fcgs, which is three times deeper than median 3\sig\ flux level reached in our SALT observations. If we further split our sample into two; (i) objects with only single PA spectra observed (total 6), and (ii) objects with two PA observations available (total 17), then only 1/6 and 6/17 are sources with confirmed extended \lya\ emission in the two sets, respectively. Thus we attribute the non-detection of \lya\ halos in few cases to poor sensitivity and inherent difficulties associated with slit spectroscopy of our observations.

\begin{figure} 
\centerline{\vbox{
\centerline{\hbox{ 
\includegraphics[clip=true,height=0.25\textheight,angle=0]{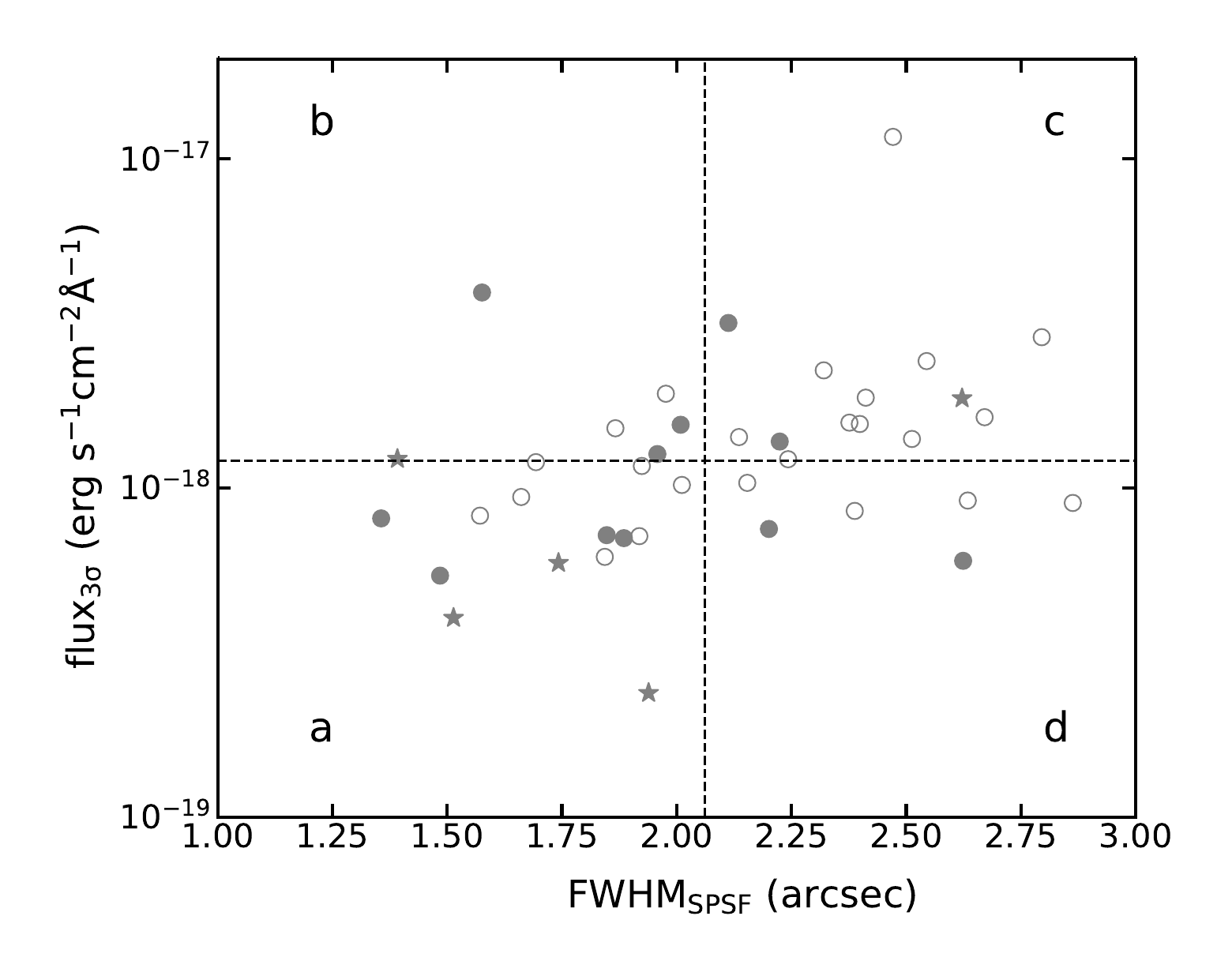}
}} 
}}  
\vskip+0.0cm  
\caption{Comparison of SPSF FWHM and f$_{3\sigma}$  for the sources with (filled cirles) and without (empty circles) extended \lya\ emission. The tentative detections are shown by stars.
} 
\label{fig_spsf}   
\end{figure} 

\subsubsection{Detection of extended \civ\ and \heii\ lines }
\begin{figure} 
\centerline{
\includegraphics[viewport=5 20 4200 2800,width=0.5\textwidth,angle=0,clip=true]{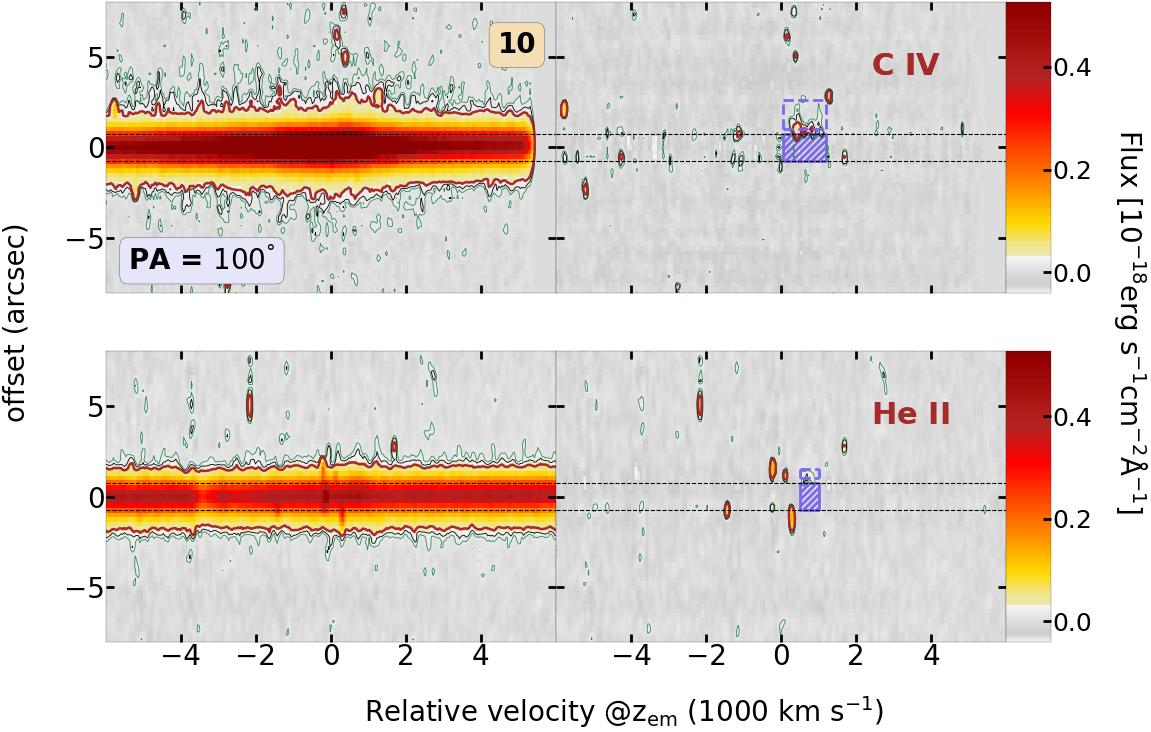}}
\vskip+0.0cm  
\caption{ 2D spectra covering
the \civ\ (top panel) and \heii\ (bottom panel) emission in source \#10. {\it Left panels} show the observed quasar spectra  and {\it right panels} show spectra after SPSF subtraction. The horizontal lines mark the SPSF FWHM.
} 
\label{fig_civ_heii_M1142}   
\end{figure}

In addition to searching for extended \lya\ halos, we have also searched for the presence of extended emission in \civ\ and \heii\ lines using the same procedure as used for \lya\ line. Extended \civ\ and \heii\ emission is seen clearly in  the radio galaxy M151304.72$-$252439.7 \citep[i.e., object \#18; see][]{Shukla2021}. In addition, M114226.58$-$2633137 (object \#10) shows clear extended emission in \civ\ (\til\ 12 kpc) and a relatively faint and low significant extension in \heii\ line as well (see Fig. \ref{fig_civ_heii_M1142}). An accurate measurement of the size of \heii\ extension is not possible for the source \#10 as the flux level is below 3\sig. The line ratios of \fciv/\flya, \fheii/\flya\ are 0.010\plm0.006 and 0.001\plm0.001, respectively. 

%
\cite{Guo2020}, compiled the sample of \cite{Borisova2016,Arrigoni2019}, which has a total of 80 AGNs, with 17 RLQs, rest RQQs and 6 unknown types, to study the extended emission in UV emission lines \civ, \heii\ and \ciii. The overall detection rate of extended \civ\ and \heii\ in their full sample is 19\% and 13\%, respectively.  For RLQs, the detection rates of extended \civ\ and \heii\ are  23\% and  \til18\%, respectively. But note that these studies have almost 100\% detection rate of extended \lya. If we consider only 7 objects with extended \lya\ emission from our sample, we get two cases with extended \civ\ and one with extended \heii. This observed detection rate for \civ\ (28\%) and \heii\ (14\%) is consistent with these results from the literature. Note that the uncertainties on these rates are also large due to small number statistics of our sample.

\subsection{Radio properties}
\label{sub_radio}

The spectral luminosities at 0.42 GHz and 1.4 GHz are listed in columns 8 and 9 of Table \ref{tab_luminosities}.  These have been estimated using flux density measurements from uGMRT Band-3 \citep[0.42\,GHz;][]{gupta2021pband} and NVSS \citep[1.4\,GHz;][]{condon1998}, and the spectral indices, $\alpha_{0.4}^{1.4}$, provided in column 10 of Table \ref{tab_luminosities}. We use higher spatial resolution uGMRT Band-5  images to estimate LAS of the radio emission (see column 8 of Table~\ref{tab_radio_props}). For objects with single radio component, we define LAS as the beam deconvolved size.  For objects with well-defined double lobed structure it is defined as the separation between the farthest hot spots.  The corresponding largest linear sizes are provided in column~9 of Table \ref{tab_radio_props}.

We note that in the case of  M1558-2155 (object \#19), in addition to a radio source coincident with the AGN identified in WISE and PS1, we detect two radio sources symmetrically placed on the opposite sides of the AGN.  If we consider these two components, hereafter referred to as North-East (NE) and South-West (SW), separated by 58\arc\ as two radio lobes then the projected linear size will be  470\,kpc at \zem, making this the largest radio source in our sample. However, in PS1 the NW component may be associated with a optical source (separted by 2.4\arc).  The nearest optical source to the SW component is 8\arc\ away, and it also has a counterpart in WISE.  Thus, NE and SW components may not be associated with M1558-2155 at all. Therefore, for M1558-2155 we have adopted LAS corresponding to the compact radio component to estimate the extent of the radio emission.

\section{Results}
\label{sec:results}
\subsection{\lya\ halos of extended radio sources}
\label{sub_lya_halos}

Here we examine any possible connection between the extent of radio emission and the presence of extended \lya\ halo. Our sample has 5 objects with extended radio emission at 1.4\,GHz.  These are objects \#9, \#10, \#16, \#18 and \#22. We do not consider object \#7 (see Section~\ref{sub_gmrt} for details).  We find 4 of these, i.e., 80\%, also exhibit extended \lya\ emission.  Note that one of these (object \#9) is a tentative detection. The typical sizes of \lya\ halos are in excess of $2.6$\arc. Interestingly, all 4 extended radio sources with the confirmed \lya\ halos also show nuclear He~{\sc ii} emission in their spectra. Thus, if we only consider objects with extended radio emission i.e., projected linear sizes $>10$ kpc from our sample \citep[similar to the criteria used in,][]{heckman1991a}, then $\sim$80\% (100\% if we consider \#9 as a detection) of these also exhibit extended \lya\ emission. The detailed comparsion of \lya\ and radio morphologies of the RLQs \#10, \#16 and \#22 is provided in Appendix~\ref{sec_appendix}.  The same for the radio galaxy (\#18) is provided by \citet[][]{Shukla2021}.

Despite the previously discussed deficiencies of our spectra, our sample indicates a strong connection between the presence of kpc-scale radio emission and the extended \lya\ halo. Already, a strong correlation between the size of the radio source and the extent of \lya\ halos has been noticed in the case of radio galaxies \citep[see][and their Fig.~7]{vanojik1997}.  However, \citet{heckman1991a} did not find any such correlation in their sample of RLQs. We discuss this  in Section~\ref{sub_comp_litertaure} in more detail using the larger sample of AGN with \lya\ halos compiled from the literature.

\subsection{Connection to the presence of associated absorption}
\label{sub_associated_abs}

In Table~\ref{tab:associated_abs}, we have summarized the \civ\ and \lya\ absorbers detected in our spectra. Out of the 21 objects listed in this table where \civ\ absorption can be searched, no associated \civ\ absorption is detected in 8 cases. We do not find any statistically significant difference in SPSF or sensitivity reached between the two sub-samples (i.e., with and without the associated \civ\ absorption). Only one of the 8 RLQs (i.e., 12.5\%) without associated absorption (i.e., M1142-2633, object \#10) shows extended \lya\ emission. While we detect a strong associated \lya\ absorption in the case of the radio galaxy (i.e., \#18) no associated \civ\ absorption was detected in that case \citep{Shukla2021}. As previously mentioned, there are 13 cases with clear detection of associated \civ\ absorption. In 9 of these, the rest equivalent width of \civ\ absorption is in excess of 0.5\AA.  There are confirmed and tentative detections of \lya\ halo in 4 and 2 (i.e., total 6/9) cases, respectively, among these. Thus, there is a slight excess of \lya\ halo detection among objects showing \civ\ associated absorption.

Note in the two cases we could not ascertain the presence of \civ\ absorption in the red wing of the \civ\ emission. One of these, i.e., M1215-0628 (object \#11) shows detectable extended \lya\ emission. In our SALT spectrum we detect a narrow \lya\ absorption at $z>$\zem. We also see a dip in the expected position of \civ\ in our NOT spectrum \citep{krogager2018}. As the spectral resolution of NOT spectrum is low, confirming the \civ\ absorption in this case will further consolidate our finding of a correlation between the detection of  \lya\ halos and presence of \civ\ associated absorption.

In general, the RLQs in our sample show excess associated \civ\ absorption (section~\ref{sec_abs}). \citet{Wild2008} performed large scale clustering analysis of absorbers around quasars. They concluded that the \civ\ absorption excess within $\pm$ 3000 \kms\ to the quasar is mainly due to the gas in the environment and not due to the large scale clustering. In such cases, the \civ\ detection rate can be linked to the covering factor of gas around quasars.  The same can also be linked to the presence of bright and possibly large \lya\ halos. {\it As deep observations usually detect Ly$\alpha$ halos in all the cases, our result means a possible correlation between the Ly$\alpha$ luminosity and/or size with the presence of associated absorption.}  It will be interesting to verify this correlation in samples with deep IFS observations.

The gas distribution around quasars can also be probed through \hi\ 21-cm absorption, an excellent tracer of cold neutral medium (CNM; T$\sim$100\,K).  Since radio emission is often extended the gas properties can be probed towards multiple sight lines \citep[see e.g.,][]{Srianand15}.  \citet{gupta2021pband} have searched for \hi\ 21-cm absorption towards all $z\ge2$ quasars in the MALS-SALT-NOT sample. Based on the lack of \hi\ 21-cm absorption (detection rate = 1.6$^{+3.8}_{-1.4}$\%) they concluded that the powerful RLQs in our sample have low CNM covering factor.  
Unfortunately the spectral range of \hi\ 21-cm absorption from 18 quasars in our sample are affected by radio frequency interference. This prevents us from directly investigating the connection between the \lya\ halos and cold \hi\ gas.
For remaining 5 objects (\#3, \#6, \#8, \#16 and \#21) \hi\ 21-cm absorption is observable but no absorption is detected.  However, for object \#3 with no \lya\ halo detection, we detect an associated pDLA at the redshift of the quasar \citep[see Table~\ref{tab:associated_abs} and section 6 of][for discussions]{gupta2021pband}. Using the lack of \hi\ 21-cm absorption in this case, we constrain the spin temperature, $\mathrm{T_S}$ $>$ 216\,K.

Recall that objects \#6 and \#16 have confirmed \lya\ halo detections.  They also exhibit associated C~{\sc iv} absorption (see Table~\ref{tab:associated_abs}). The \hi\ 21-cm absorption non-detection (assuming $\mathrm{T_S} = 100$ K) for these correspond to $N$(\hi) less than $5.6\times10^{20}$ cm$^{-2}$ and  $2.4\times10^{19}$ cm$^{-2}$, respectively. 
The objects \#8 and \#21 have tentative \lya\ halo detections. No associated \civ\ absorption is detected in these cases. The non-detection of 21-cm absorption in these cases correspond to an upper limit of $5.6\times10^{20}$ cm$^{-2}$ and $5.6\times10^{20}$ cm$^{-2}$, respectively. 
The H~{\sc i} 21-cm absorption and \lya\ halos in our observations are probing gas at different physical scales. Specifically, the radio emission in the former case probes gas at scales smaller than 10\,kpc whereas the detected \lya\ halos are much larger than this. Thus, unlike \civ\ absorption which may originate from gas with a wide range of physical conditions, there is no connection between the detection of extended \lya\ halos and the presence of large reservoirs of cold atomic gas at smaller ($<$10\,kpc) scales.

\subsection{Nuclear He~{\sc ii} emission and Ly$\alpha$ halo detection}
\label{sub_heii_line}

\begin{figure} 
\centerline{
\includegraphics[viewport=35 30 1000 1280, width=0.45\textwidth,angle=0,clip=true]{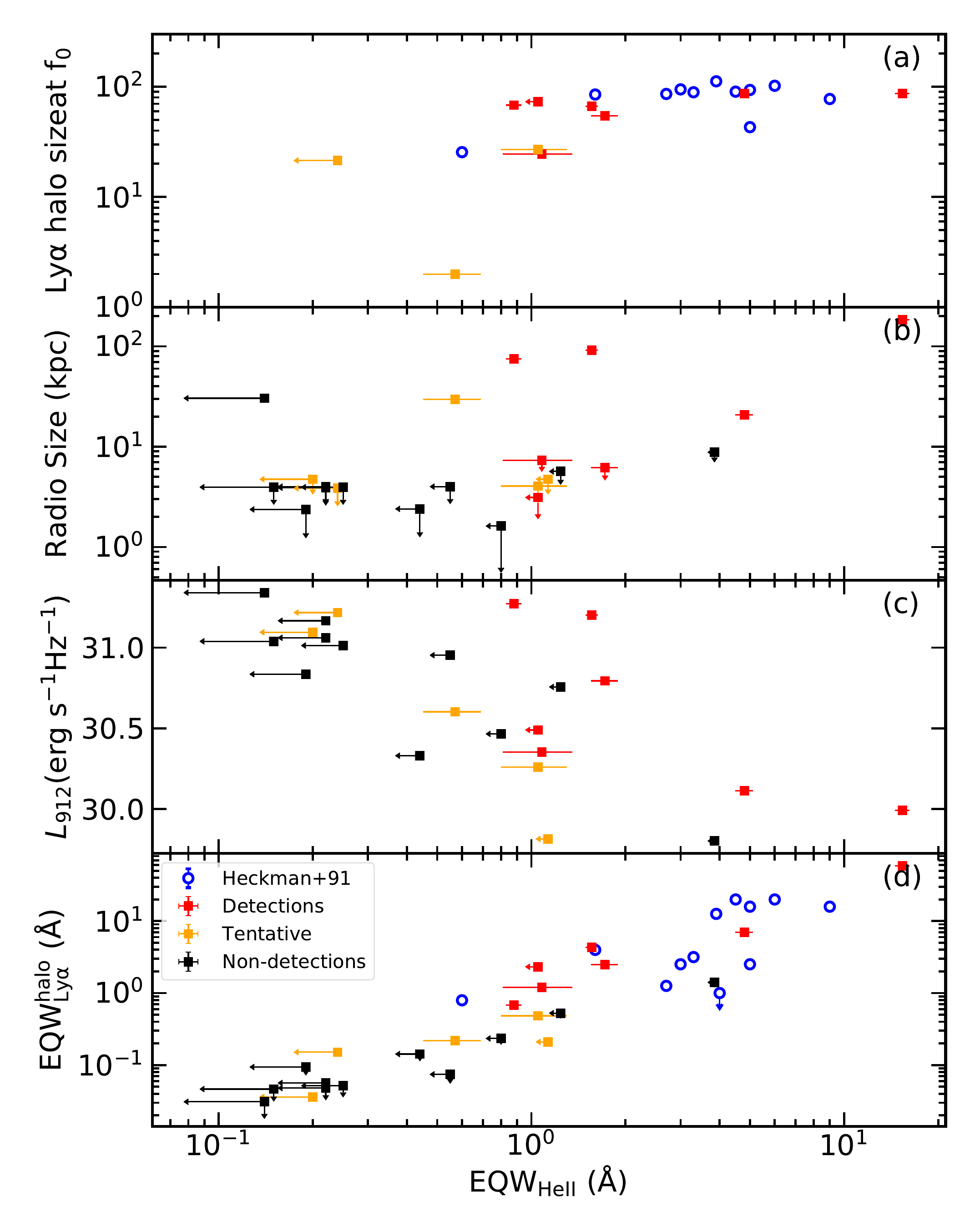}
} 
\caption{Comparison of the \heii\ equivalent width with \lya\ halo size (measured at $\mathrm{f_0}$=$1.0\times 10^{-18}$ \fcgs, Lyman continuum luminosity ($L_{912}$), Radio size and the equivalent width of the \lya\ halos ($\mathrm{EQW^{halo}_{Ly\alpha}}$) for our sources with clear extended \lya\ halo (red), tentative (orange) and non-detections (black). In the bottom panel we have included data from the  sample of \citep{heckman1991a,heckman1991b}.
}
\label{fig_comp3}
\end{figure} 

The nuclear He~{\sc ii} emission is a good probe of the quasar spectral energy distribution in the extreme-UV to soft X-ray regime. In our sample, out of 23 RLQs searched for the extended \lya\ halos, 8 show detectable nuclear He~{\sc ii} emission (see Table~\ref{tab_fluxes} and Fig~\ref{fig_heiifit}).  For 20 objects, our spectra reach a 4$\sigma$ equivalent width sensitivity of 1\AA\ or better. We detect \heii\ emission with rest equivalent width in excess of 1\AA\ in 6 cases (i.e objects \#6, \#10, \#11, \#18, \#19 and \#22). In 5 of these sources we have the firm detection of \lya\ halos and for one (i.e object \#19) we have tentative detection. In the remaining 14 cases, we have two firm detection (i.e in objects \#13 and \#16) and 2 cases show tentative \lya\ halo detection. Thus it appears that in our sample there is a clear trend of increase in the \lya\ detection rate with the rest equivalent width of the nuclear \heii\ emission line. This is consistent with the finding of \citet{heckman1991a}. 

In Fig.~\ref{fig_comp3}, we compare the rest frame \heii\ equivalent width (EQW$_{\mathrm{HeII}}$) of the nuclear emission with the \lya\ halo size, Lyman continuum luminosity ($L_{912}$), Radio size and the equivalent width of the \lya\ halos ($\mathrm{EQW^{halo}_{Ly\alpha}}$).
To estimate the equivalent width of the extended \lya\ halo emission, we have taken the ratio of the halo luminosity provided in Table \ref{tab_halo_props} and the specific luminosity at 1350\AA\  (see Table \ref{tab_luminosities}). In these plots, the confirmed and tentative \lya\ detections, and the non-detections are represented by red, orange and black filled squares, respectively.
In the bottom panel we also show the data from \citet{heckman1991a,heckman1991b} as blue circles.

We estimate, Kendall rank correlation coefficient (a.k.a. Kendall's $\tau$) between EQW$^\mathrm{halo}_\mathrm{Ly\alpha}$ and EQW$_{\mathrm{HeII}}$ (see panel (d)), using all the data points from our sample and \cite{heckman1991a}, including upper limits. We use python package `pymccorrelation' \citep{Curran2014,Privon2020} for estimating the coefficient (r) and p-value. Since the number of data points are small, we use $N_{boot}=10-30$ and find that the coefficients do not change drastically. For $N_{boot}=10$ and 100 realizations, we obtain, r=0.61 and  p= $7\times10^{-5}$, for our sample, and r=0.50 and  p=0.03 for \cite{heckman1991a} sample. For the combined data set, we get r = 0.52 and p = $3\times10^{-6}$. 
The same between \lya\ halo size and EQW$_{\mathrm{HeII}}$ (see panel (a)) gives r=0.50 and p=0.003. The \lya\ halo sizes plotted here  are measured at constant flux threshold of $1.0\times 10^{-18}$ \fcgs\ for all the sources to eliminate the effect of different sensitivities reached for individual sources. The threshold flux level is approximately the average flux level reached for clear detections+tentative detections (see Table~\ref{tab_halo_props}).
These results strengthen the suggestion that the presence of narrow \heii\ nuclear emission line is a strong indicator of finding extended gas around AGNs, especially around RLQs. It will be interesting to explore this in more detail for the objects in IFS samples.

Panel (b) of Fig.~\ref{fig_comp3} shows radio size vs EQW$_{\mathrm{HeII}}$. It is evident that the frequency of \lya\ halo detection is enhanced among sources having large radio size and high equivalent width of He~{\sc ii} nebular emission. The low EQW$_{\mathrm{HeII}}$ observed for compact sources could also be due to the additional contribution 
to the optical continuum from the radio jets. 

In the panel (c) of Fig.~\ref{fig_comp3}, we plot the \heii\ rest equivalent width vs. Lyman continuum luminosity ($L_{912}$). We do find a possible anti-correlation between $L_{912}$ and EQW$_{\mathrm{HeII}}$ (r=$-$0.23) albeit with less statistical significance (p=0.14). This is similar to the already known anti-correlation between ${L_{2500}}$ and  EQW$_{\mathrm{HeII}}$ (i.e ``Baldwin effect")  \citep[][]{Zheng1993, Laor1995, Green1996, Korista1998, Dietrich2002, Timlin2021a}. Note, we inferred $L_{912}$ using a simple power-law extrapolation from the rest UV spectrum. For RQQs, a strong correlation has been found between $\alpha_{OX}$ and EQW$_{\mathrm{HeII}}$, and anti-correlation between $\alpha_{OX}$ and $L_{2500}$ \citep{Timlin2021a}. Thus, we expect $L_{2500}$ to anti-correlate with EQW$_{\mathrm{HeII}}$ as well. Therefore, the trend observed in the top panel of  Fig.~\ref{fig_comp3}, is consistent with the general trends seen in RQQs. Also presence of beamed continuum from the compact radio core can reduce the measured equivalent widths in some of the sources in our sample. In these cases, the actual UV continuum seen by the line emitting gas will be less luminous compared to what we infer.

\subsection{Are RLQs intrinsically different?}

\begin{figure} 
\centerline{\vbox{
\centerline{\hbox{ 
\includegraphics[viewport=30 20 1300 1300,width=0.5\textwidth,angle=0, clip=true]{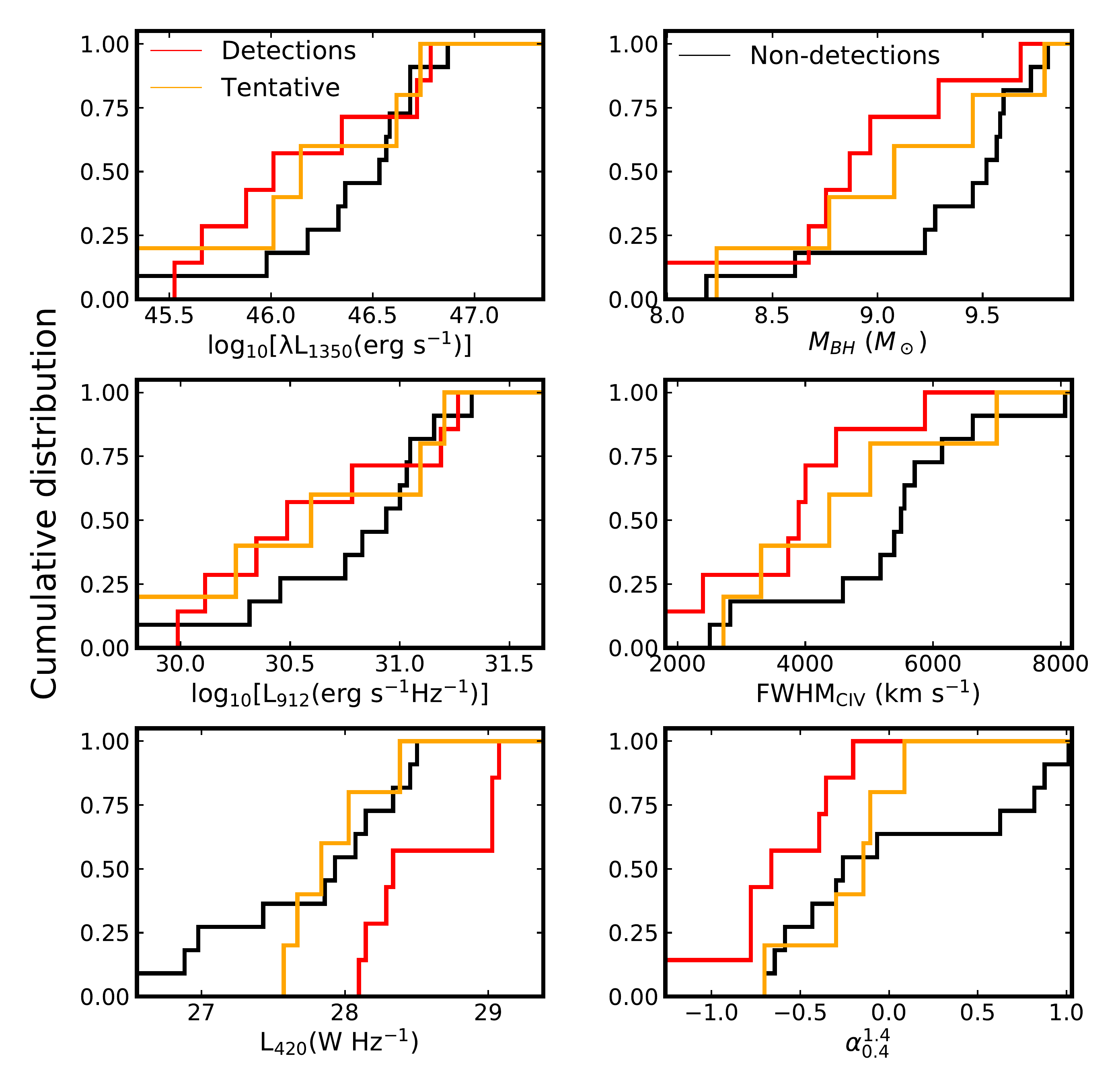}
}} 
}}  
\vskip+0.0cm  
\caption{Cummulative distributions of various quasar properties measured for our sample. The sources showing clear extended \lya\ are shown by red. The tentative detections and non-detections are shown by orange and black, respectively.
}
\label{fig_compa_det_nondet}   
\end{figure} 

\begin{table}
\scriptsize
\setlength{\tabcolsep}{2pt}
\caption{ Results of Kolmogorov-Smirnov statistics
}
\centering
\begin{tabular}{lcccccccccccccccccc}
\hline
\hline

Parameter  
&ks statistics p-value\\

&
&\\

(1)
&(2)\\

\hline
\multicolumn{2}{c}{\bf Detections vs Non-detections}\\[0.1cm]
$\lambda L_{1350}$&   4.31e-01 \\[0.0cm]
$M_{BH}$&   1.16e-01 \\[0.0cm]
$L_{912}$&   5.67e-01 \\[0.0cm]
FWHM(CIV)&   6.46e-02 \\[0.0cm]
$L_{420 \mathrm{MHz}}$&   3.42e-02 \\[0.0cm]
Radio spectral index ($\alpha^{1.4}_{0.4}$)&   1.73e-01 \\[0.0cm]
$L_{1420 \mathrm{MHz}}$&   2.05e-02 \\[0.0cm]

\hline
\hline
\multicolumn{2}{c}{\bf Detections+tentative vs Non-detections}\\[0.1cm]
$\lambda L_{1350}$&   2.41e-01 \\[0.0cm]
$M_{BH}$&   9.15e-02 \\[0.0cm]
$L_{912}$&   5.51e-01 \\[0.0cm]
FWHM(CIV)&   3.23e-02 \\[0.0cm]
$L_{420 \mathrm{MHz}}$&   3.51e-01 \\[0.0cm]
Radio spectral index ($\alpha^{1.4}_{0.4}$)&   3.26e-01 \\[0.0cm]
$L_{1420 \mathrm{MHz}}$&   5.20e-01 \\[0.0cm]

\hline
\end{tabular}
\begin{flushleft}

Kolmogorov statistics results for various parameters (Column 1) and the resulting p-values (Column 2) between the sources with clear detections and non-detections (upper half of the table) and clear detections+tentative versus non-detections (lower half of the Table).
\end{flushleft}
\label{tab_ks_stats}
\end{table}

In this section, we compare the intrinsic properties of RLQs with and without extended \lya\ halo detection. In particular, we compare the cumulative distribution functions (CDFs) of various parameters for the following three subsets: confirmed detections, tentative detections and non-detections. The CDFs of quasar continuum luminosity ($\lambda L_{1350}$), blackhole mass ($M_{BH}$), \lya\ continuum luminosity ($L_{912}$), line width of \civ\ (FWHM$\mathrm{_{CIV}}$), spectral luminosity at 420 MHz ($L_{420MHz}$) and radio spectral index ($\alpha^{1.4}_{0.4}$) are shown in Fig.~\ref{fig_compa_det_nondet}. The Kolomogorov Smirnov test (KS-test) results are provided in Table~\ref{tab_ks_stats}. Visual inspection of the Fig. \ref{fig_compa_det_nondet} shows that the detections are apparently different from the non-detections in all the variables with the median of the non-detections always being higher than those of detections except for $L_{420MHz}$, where the median of the non-detection is lower. Interestingly we also see that the distribution of tentative detections tend to be more similar to those of the detections than to the non-detections, an exception again being $L_{420MHz}$ where tentative detections are more similar to non-detections.
 However, the table shows that the KS-test p-values (see Table \ref{tab_ks_stats}) are typically not very significant.

We find the difference in the radio luminosity (both $L_{420MHz}$ and $L_{1420MHz}$) between clear detections and non-detections with p-values less than 0.03. However, the significance gets diluted when we combine the tentative detections with confirmed detections. This indicates the \lya\ halo properties to be correlated with the radio power. In the following section we explore this further using the literature sample. In the case of FWHM$_\mathrm{CIV}$ we find the p-values to be 0.06 and 0.03 when we compare  detections vs. non-detections and detections+tentative detections vs. non-detections, respectively. This suggests that objects with narrower \civ\ emission line width may have luminous and/or larger \lya\ halos.

\subsection{Comparison with sample from Literature}
\label{sub_comp_litertaure}

\begin{figure} 
\centerline{\vbox{
\centerline{\hbox{ 
\includegraphics[height=0.37\textheight,angle=0]{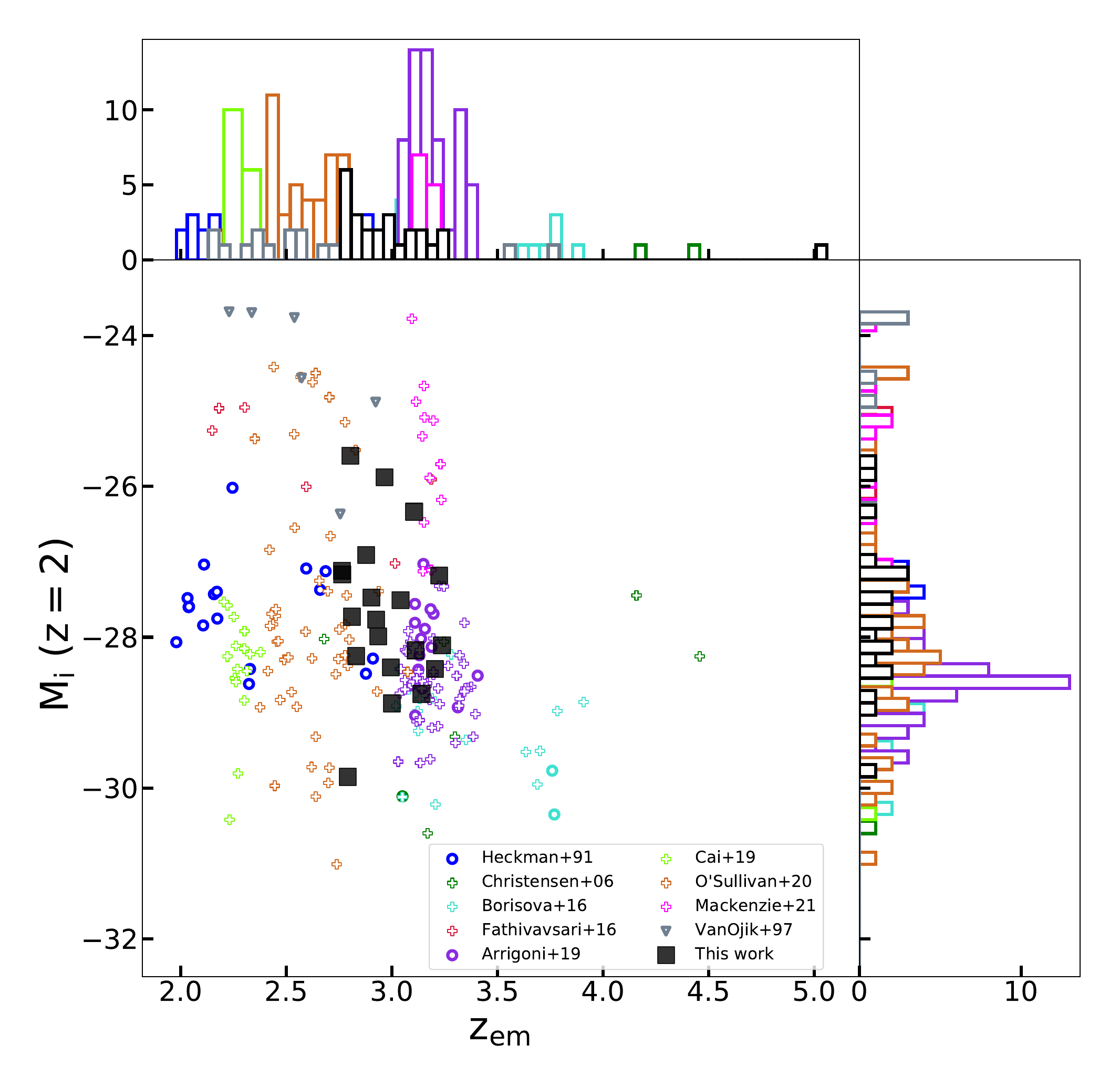}
}} 
}}  
\vskip+0.0cm  
\caption{Scatter plot of 
 absolute magnitude in the SDSS i-band  and \zem\ 
  for our sample and samples from literature studies of the extended \lya\ emission.  Circles stands for radio loud sources and + for radio quiet sources or sources with unknown type. The lower triangle is for HzRGs.  Our sample is shown by black squares.
}
\label{fig_comp1}   
\end{figure} 

\begin{figure*} 
\includegraphics[width=1.0\textwidth,angle=0]{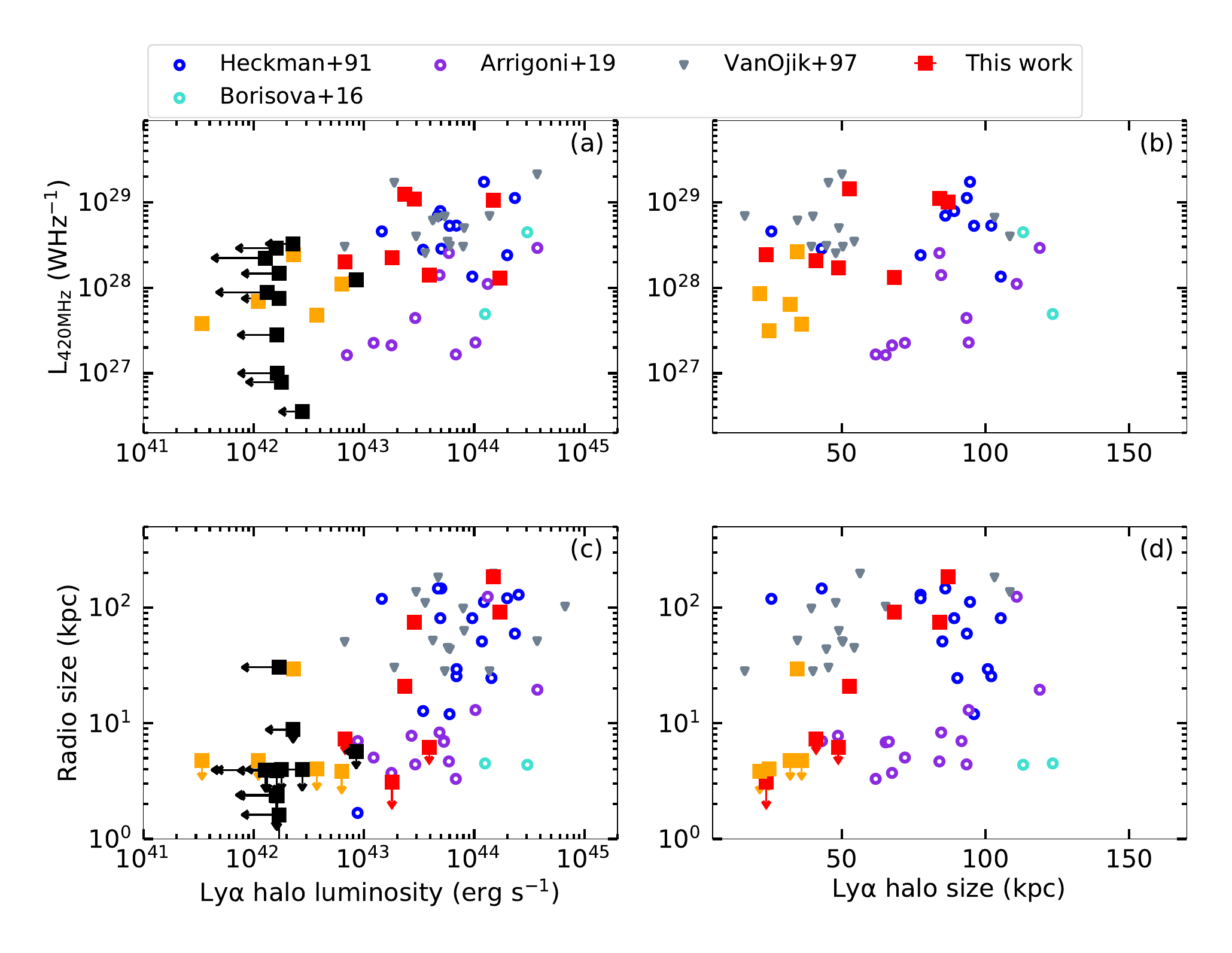}
\caption{Comparison of $L_{420MHz}$ and radio size with \lya\ halo luminosity and sizes for objects in our sample and from the literature. The squares correspond to our sample of RLQs, the red ones indicate objects with extended \lya\ halo detection and the orange and black are for tentative and non-detections, respectively. RLQs from literature sample are shown with circles and HzRGs by lower triangle. The samples from \citet{Borisova2016} (B16) and \citet{Arrigoni2019} (A19) includes 2 RLQs and 17 RLQs, respectively. Note that the number of points are not same in each subplot, as not all quantities are simultaneously available for all the objects. Note that the \lya\ halo sizes used for our sample are measured at 3\sig\ flux level.}
\label{fig_comp2}   
\end{figure*} 

\begin{table*}
\scriptsize
\setlength{\tabcolsep}{3pt}
\caption{Measured correlation Coefficients and p-values}
\centering
\begin{tabular}{lcccccccccccccccccc}
\hline
\hline

Sample
&$L_{420MHz}$-\lya\ luminosity
&$L_{420MHz}$-\lya\ halo size
&Radio size -\lya\ luminosity
&Radio size -\lya\ halo size\\

(1)
&(2)
&(3)
&(4)
&(5)\\

\hline
IFS (B16+A19 RLQs)&	0.56$\pm$0.06 (0.0158$\pm$0.0135)&	0.54$\pm$0.07 (0.0222$\pm$0.0217)&	0.25$\pm$0.07 (0.1832$\pm$0.1156)&	0.27$\pm$0.07 (0.1605$\pm$0.0961)\\[0cm]
Heckman 1991&	0.18$\pm$0.08 (0.3606$\pm$0.1514)&	0.01$\pm$0.09 (0.5026$\pm$0.1291)&	0.17$\pm$0.09 (0.3229$\pm$0.1653)&	-0.42$\pm$0.04 (0.0415$\pm$0.0215)\\[0cm]
VanOjik 1997&	0.23$\pm$0.10 (0.2498$\pm$0.1426)&	-0.08$\pm$0.06 (0.5320$\pm$0.1373)&	0.08$\pm$0.08 (0.4643$\pm$0.1267)&	0.43$\pm$0.06 (0.0333$\pm$0.0276)\\[0cm]
This work& 0.32$\pm$0.05 (0.0540$\pm$0.0430)&	0.42$\pm$0.10 (0.0797$\pm$0.0806)&	0.33$\pm$0.07 (0.0462$\pm$0.0661)&	0.61$\pm$0.07 (0.0079$\pm$0.0076)\\[0cm]
IFS+Heckman&	0.28$\pm$0.05 (0.0656$\pm$0.0522)&	0.16$\pm$0.06 (0.2903$\pm$0.1363)&	0.27$\pm$0.05 (0.0482$\pm$0.0433)&	0.06$\pm$0.05 (0.4129$\pm$0.1399)\\[0cm]
\hline

\end{tabular}
\begin{flushleft}
Kendall rank correlation coefficient between two parameters (Columns 2-5) as shown in Fig.~\ref{fig_comp2} for IFS observations, \cite{heckman1991a,heckman1991b}, \cite{vanojik1997}, our sample and IFS+\cite{heckman1991a} sample as provided in Column 1. The \lya\ halo sizes used are measured at 3\sig\ flux level.
\end{flushleft}
\label{tab_correlation_coef}
\end{table*}

\begin{figure} 
\centerline{
\includegraphics[viewport=15 25 700 560,width=0.5\textwidth,angle=0,clip=true]{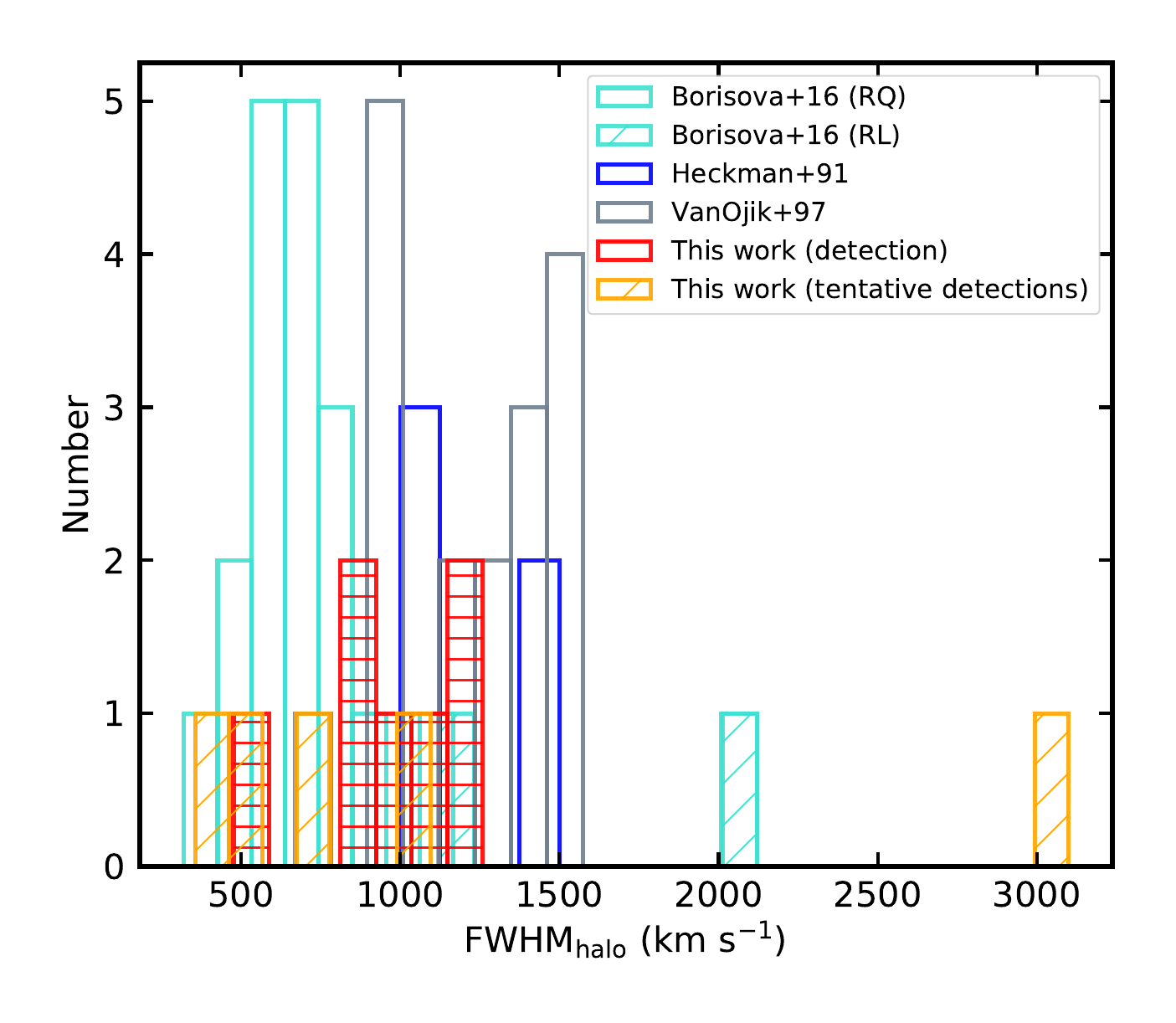}}
\vskip+0.0cm  
\caption{Distribution of FWHM of the \lya\ halo emission. We compare the distribution from different samples discussed here.
} 
\label{fig:Lyafwhm}   
\end{figure} 

In this section, we combine our sample with those from the literature to understand some of the above discussed trends. To start with, we investigate the differences between our sample and those from previous studies of extended \lya\ emission around high-z quasars and radio galaxies. 

In Fig. \ref{fig_comp1}, we show the distribution of absolute i-band magnitude vs. \zem\ for objects in our sample and various samples from the literature. For comparison, we have selected major studies on diffuse \lya\ emission from the literature: \cite{heckman1991a,heckman1991b} (H91), \cite{vanojik1997} (VO97), \cite{Christensen2006}, \cite{Borisova2016} (B19), \cite{Fathivavsari2016}, \cite{Arrigoni2019} (A19), \cite{cai2019,Osullivan2020,Mackenzie2021}. The absolute magnitudes are taken from the respective studies if provided, otherwise are estimated using the SDSS apparent i-band magnitude after k-correcting it using the method by \cite{Ross2013}. If the SDSS magnitude is also not available then we use the PS1 magnitudes \citep[see table 6 in][]{Tonry2012}/ SSS magnitudes \citep[see eqn 2 in][]{Peacock2016} to calibrate the apparent magnitude in SDSS i-band. Since the i band bandpass filters are very similar for PS1 and SDSS, calibrating SDSS i-band magnitude from PS1 has smaller error (magnitude difference, $\Delta \mathrm{m=0.002^{+0.14}_{-0.17}}$) compared to using SSS. While the $\mathrm{M_i}$ distribution of our sources are similar to those in the literature, our sample fills the gap in the distribution around $z=2.7-3.0$.

In Fig. \ref{fig_comp2}, we compare radio power $L_{420MHz}$ and radio size with the luminosity and size of \lya\ halos for our sample (squares) and sample of 17 RLQs from B16+A19, 19 from H91 (open circles) and HzRGs from VO97 (downward triangle). To estimate $L_{420MHz}$ of the sources from literature, we compute spectral index ($\alpha^{1.4}_{0.2}$) between 200 MHz flux density obtained from GaLactic and Extragalactic All-sky MWA Survey (GLEAM) Extragalactic Catalog \citep[GLEAMEGCAT;][]{Hurley-Walker2017} and 1.4\,GHz flux density measurements from NVSS/FIRST. The radio sizes for H19 and VO97 are available from literature. For RLQs from B16+A19, we use VLASS 3\,GHz images to estimate the deconvolved radio sizes in the same way as our sample. We point out that for a few sources one or both radio size and $L_{420MHz}$ measurements are unavailable, so number of points shown in each subplot of Fig. \ref{fig_comp2} are different. 
The upper limits on the quantities  are shown by downward/leftward arrows. 

We note that both \lya\ luminosity and halo size for our sample may be underestimated as we use slit-based spectra (same is the case for VO97). In the case of \citet{heckman1991a}, the size measurements are based on narrow band images and are more reliable compared to the slit based measurements.  
We compute Kendall rank correlation coefficient between the two parameters shown in each subplot, accounting for the upper limits using $N_{boot}=10$. We find the coefficient and p-values change with each realization, even when $N_{boot}=10$ is kept constant. So, we compute the correlation coefficients 100 times and report the mean and standard deviation on these. We compute these correlation coefficients separately for the sample of RLQs with IFS measurements (B16+A19), RLQs of H19, HzRGs and our sample. We also study these correlations combining the IFS and H91 samples together. The correlation coefficient r and associated p-values (in brackets) for all the cases are summarized in  Table \ref{tab_correlation_coef}.

For RLQs observed with MUSE (i.e., B16+A19), we find a strong correlation between the \lya\ luminosity and $L_{420MHZ}$  (Kendall $r=0.56$; p-value = 0.02).  Our sample also shows significant correlation. However, correlations are relatively less significant in the RLQ sample of \citet{heckman1991a} and HzRG sample of \citet{vanojik1997}. This could be related to the smaller range in $L_{420MHz}$ probed in these samples. When we combine the IFS sample with that of \citet{heckman1991a}, we find  $r = 0.28$ and a p-value of 0.07.

Similarly the IFS sample also shows strong correlation between  $L_{420MHZ}$ and \lya\ halo size (Kendall $r = 0.53$ and a p-value of 0.03). However, when we consider only the sample of \citet{heckman1991a} or \citet{vanojik1997}, we do not find any significant correlation. Correlation is also not evident when we combine the IFS sample with that of \citet{heckman1991a}.
Interestingly, in the \lya\ halo luminosity vs $L_{420MHZ}$ plane both our and \citet{heckman1991a}'s RLQs occupy similar regions. However, as expected the \lya\ halo sizes are much larger for the \citet{heckman1991a} sample as they use narrow band images to measure the sizes.

Next we consider the correlation between radio size and \lya\ halo luminosity. In the MUSE IFS sample, we find an insignificant correlation (i.e r = 0.25 and p =0.18) of the \lya\ halo properties with the radio size. However, it is evident from the Fig.~\ref{fig_comp2} that the MUSE IFS samples probe a narrow range of radio sizes with very few large radio sources (i.e $\sim$ 80 percent of the sources have sizes less than 10 kpc). 
Both samples of  \citet{heckman1991a} and \citet{vanojik1997} do not show any significant correlation between radio size and \lya\ luminosity. As discussed before both the sample lack compact radio sources. However, when we combine MUSE IFS and the \citet{heckman1991a} samples, we find r = 0.27 and p=0.048. This suggest a possible correlation between radio size and \lya\ luminosity. Presence of such a correlation will explain frequent detection of \lya\ halos in the extended radio source in our sample.

A strong anti-correlation is seen between radio source size and \lya\ halo size in the case of \citet{heckman1991a} (predominantly have large radio sources) sample and strong correlation in the case of HzRG sample of \citet{vanojik1997}. In the MUSE IFS sample, this correlation is not very significant. As expected when we combine IFS sample with that of \citet{heckman1991a}, we do not find any significant correlation. Our sample alone shows a strong correlation between these two quantities, where the halo sizes are measured at 3\sig\ flux level (see Table~\ref{tab_halo_props}). However if we find the correlation between radio size and \lya\ halo sizes measured at constant flux $\mathrm{f_0}$, then this correlation reduces significantly. It is important to probe the radio-size vs. \lya\ halo size relationship for a sample spanning large radio-sizes with IFS spectroscopy.

In Fig.~\ref{fig:Lyafwhm}, we plot the distribution of measured \lya\ FWHM from our sample (firm as well as tentative detections) and those from the literature. As we mentioned before 6 out of 7 firm detections show FWHM in excess of 900 \kms\ in our sample. Among the 5 tentative detections two have FWHM in excess of 1000 \kms.  In the sample of \citet{Borisova2016}, the two RLQs have FWHM in excess of 1000 \kms. But the radio quiet quasars have FWHM in the range 320$-$930 \kms with a median of 640 \kms. \citet{Arrigoni2019} found FWHM to be $\le$ 940 \kms\ for all the objects in their sample including RLQs. They suggested that the gas kinematics are consistent with that  expected for gas motions in the gravitational potentials of galaxies hosting these quasars. For the HzRG in the sample of \citet{vanojik1997}, the measured FWHM are in the range 670$-$1575 \kms with a median of 1237 \kms. There are 5 sources in the sample of \citet{heckman1991a} for which long-slit spectroscopy was presented in \citet{heckman1991b}. All 5 of them have FWHM in excess of 1000 \kms. Thus our detections are consistent with what is typically seen in HzRGs and RLQs.

\section{Summary}

In this work, we present a detailed analysis of long-slit spectroscopic observations with SALT of 23 newly discovered RLQs at $2.7\le z\le3.3$. These objects are part of a complete sample of 25 RLQs (brighter than $>200$ mJy at 1.4 GHz) at $z>2.7$ found in our dust-unbiased MALS-SALT-NOT survey \citep{Gupta2021qsosurvey}. We present measurements of various quasar properties based on optical spectra and L-band images from our uGMRT observations. The redshift range covered in our work fills a gap between previous studies focusing on diffuse \lya\ emission from HzRGs, RLQs and RQQs. We arrive at the following conclusions based on our detailed study and comparison with previous studies.

\begin{enumerate}
     \item We report 7 clear detection and 5 tentative detection of extended \lya\ halos. Our detection rate is much lower than \til\ 83-100\% detection rate of \lya\ halos in \cite{heckman1991a} and MUSE studies \citep[][]{Borisova2016,Arrigoni2019}. The reason for low number of \lya\ halo detection is mostly due to poor seeing (\til\ 2\arc) and flux sensitivity achieved in our case compared to the recent IFS studies. 
     
    \item{} We find the \lya\ detection to be more frequent among the extended radio sources. If we select only sources with radio sizes $>10 $ kpc, we find 4 confirmed and 1 tentative detection among the 6 extended radio sources. Note the senstivity and SPSF in these cases are typical of what we achieve for the full sample. This finding is consistent with the high detection rate found by 
     \cite{heckman1991a} for extended radio sources in their sample.

    \item  Among objects with a confirmed \lya\ halo detections we find extended \civ\ emission in only two cases and extended He~{\sc ii} emission in only one case. The extended \civ\ and He~{\sc ii} emission in the case of the radio galaxy M1513-2524 (object \#18) is discussed in detail in \citet{Shukla2021}. The frequency of detection of extended \civ\ and He~{\sc ii} emission among sources showing \lya\ halos is  consistent with very recent study by \cite{Guo2020} using MUSE data of \cite{Borisova2016,Arrigoni2019}.

    \item We find a possible connection between the detection of extended \lya\ emission and the presence of associated \civ\ absorption. Our sample, shows clear excess of associated \civ\  absorption, suggesting high covering factor of ionized gas around our quasars. We also find the \lya\ halo detection among the sources showing associated \civ\ systems are much higher than those without associated absorption. However, H~{\sc i} 21-cm searches \citep[see][]{gupta2021pband} towards 18/23 RLQs in our sample are affected by RFI. The non-detection in remaining 5 cases suggests a low covering factor of cold neutral gas even in the cases where \lya\ halos are detected. 

    \item Nuclear \heii\ emission is detected in 8 out of 23 cases. It has been shown that \heii\ emission is a good indicator of the far-UV to soft-X-ray spectral energy distribution of quasars.
    We have 6 confirmed and 2 tentative detection of \lya\ halos among the 8 sources with detectable He~{\sc ii} nuclear emission. We see a clear trend of increase in diffuse \lya\ emission equivalent width with rest frame equivalent width of \heii\ emission line. This is consistent with the findings of \cite{heckman1991a,heckman1991b}.
    
    \item We compare several quasar properties (such as $L_{1350}$, $M_{BH}$, $L_{912}$, FWHM$_\mathrm{CIV}$, $L_{420MHz}$ and $\alpha_{0.4}^{1.4}$) among detections, tentative detections and non-detections in our sample.
    Detections are found to be different from non-detections in all the parameters with the median of the non-detections being higher than those of detections except for $L_{420MHz}$. However, due to small number of objects involved the statistical significance of these differences are not high for most cases. Nevertheless  spectral luminosity ($L_{420MHz}$) and \civ\ line width ($\mathrm{FWHM_{CIV}}$) have p-values less than 5\%.

    \item{} We find all our confirmed detections have FWHM of the diffuse \lya\ emission in excess of 900 \kms, except one with FWHM$<600$ \kms. Among tentative detections 2 out of 5 have FWHM in excess of 1000 \kms. In the remaining cases the FWHM spans the range 357$-$678 \kms. This is consistent with a perturbed kinematics of the halo gas as seen in the radio-loud quasar sample of \cite{heckman1991a,heckman1991b} and radio galaxy sample of \citep{vanojik1997}. The FWHM in the case of radio quiet quasars are found to be much less than what we find in the case of radio-loud objects \citep{Borisova2016, Arrigoni2019}. Based on the low FWHM it is argued that the gas kinematics is governed by the gravitational potential of the host galaxies. In the case of radio-loud objects either the jet-gas interactions or the ability of the host galaxy to sustain large scale winds over a long period of time are invoked to understand the perturbed kinematics. It is important to have deep IFS observations of our targets to fully quantify the velocity field and its connection to the radio jet orientation. Such a study is important to understand the origin of gas kinematics.    

  \item{} Finally, we probe the correlation between the \lya\ luminosity and size with the radio power and size using data from the literature. We measure the radio size and $L_{420MHz}$ for radio loud objects studied in MUSE IFS samples. These data clearly show a strong correlation of $L_{420MHz}$ with \lya\ luminosity and halo size. A relatively weaker correlation is seen between the radio size and \lya\ luminosity and halo size. These above mentioned correlations could also be the reason for us detecting \lya\ emission  more frequently amongst the objects having higher $L_{420MHZ}$ and radio size in our sample.

\end{enumerate}
\label{sec_summary}

 \section*{Acknowledgments}
 We thank the referee for useful comments.
Most of the observations reported in this paper were obtained with the Southern African Large Telescope (SALT). We thank the staff of the GMRT for wide band observations. GMRT is run by the National Centre for Radio Astrophysics of the Tata Institute of Fundamental Research. This work 
utilized the open source software packages \textsc{Astropy} \citep{astropy2}, \textsc{Numpy} \citep{numpy}, \textsc{Scipy} \citep{scipy}, \textsc{Matplotlib} \citep{matplotlib} and \textsc{Ipython} \citep{ipython}.
\section*{Data Availability}
Data used in this work are obtained using SALT. Raw data will become available for public use 1.5 years after the observing date at https://ssda.saao.ac.za/.
%
\def\aj{AJ}%
\def\actaa{Acta Astron.}%
\def\araa{ARA\&A}%
\def\apj{ApJ}%
\def\apjl{ApJ}%
\def\apjs{ApJS}%
\def\ao{Appl.~Opt.}%
\def\apss{Ap\&SS}%
\def\aap{A\&A}%
\def\aapr{A\&A~Rev.}%
\def\aaps{A\&AS}%
\def\azh{A$Z$h}%
\def\baas{BAAS}%
\def\bac{Bull. astr. Inst. Czechosl.}%
\def\caa{Chinese Astron. Astrophys.}%
\def\cjaa{Chinese J. Astron. Astrophys.}%
\def\icarus{Icarus}%
\def\jcap{J. Cosmology Astropart. Phys.}%
\def\jrasc{JRASC}%
\def\mnras{MNRAS}%
\def\memras{MmRAS}%
\def\na{New A}%
\def\nar{New A Rev.}%
\def\pasa{PASA}%
\def\pra{Phys.~Rev.~A}%
\def\prb{Phys.~Rev.~B}%
\def\prc{Phys.~Rev.~C}%
\def\prd{Phys.~Rev.~D}%
\def\pre{Phys.~Rev.~E}%
\def\prl{Phys.~Rev.~Lett.}%
\def\pasp{PASP}%
\def\pasj{PASJ}%
\def\qjras{QJRAS}%
\def\rmxaa{Rev. Mexicana Astron. Astrofis.}%
\def\skytel{S\&T}%
\def\solphys{Sol.~Phys.}%
\def\sovast{Soviet~Ast.}%
\def\ssr{Space~Sci.~Rev.}%
\def\zap{$Z$Ap}%
\def\nat{Nature}%
\def\iaucirc{IAU~Circ.}%
\def\aplett{Astrophys.~Lett.}%
\def\apspr{Astrophys.~Space~Phys.~Res.}%
\def\bain{Bull.~Astron.~Inst.~Netherlands}%
\def\fcp{Fund.~Cosmic~Phys.}%
\def\gca{Geochim.~Cosmochim.~Acta}%
\def\grl{Geophys.~Res.~Lett.}%
\def\jcp{J.~Chem.~Phys.}%
\def\jgr{J.~Geophys.~Res.}%
\def\jqsrt{J.~Quant.~Spec.~Radiat.~Transf.}%
\def\memsai{Mem.~Soc.~Astron.~Italiana}%
\def\nphysa{Nucl.~Phys.~A}%
\def\physrep{Phys.~Rep.}%
\def\physscr{Phys.~Scr}%
\def\planss{Planet.~Space~Sci.}%
\def\procspie{Proc.~SPIE}%
\let\astap=\aap
\let\apjlett=\apjl
\let\apjsupp=\apjs
\let\applopt=\ao

\defcitealias{gaikwad2016}{Paper-I}
\defcitealias{puchwein2015}{P15}	
\defcitealias{haardt2012}{HM12}	
\defcitealias{khaire2015a}{KS15}
\defcitealias{khaire2018a}{KS18}
\defcitealias{gaikwad2018}{G18}
\bibliographystyle{mnras}
\bibliography{mybib}

\begin{thebibliography}{}
\makeatletter
\relax
\def\mn@urlcharsother{\let\do\@makeother \do\$\do\&\do\#\do\^\do\_\do\%\do\~}
\def\mn@doi{\begingroup\mn@urlcharsother \@ifnextchar [ {\mn@doi@}
  {\mn@doi@[]}}
\def\mn@doi@[#1]#2{\def\@tempa{#1}\ifx\@tempa\@empty \href
  {http://dx.doi.org/#2} {doi:#2}\else \href {http://dx.doi.org/#2} {#1}\fi
  \endgroup}
\def\mn@eprint#1#2{\mn@eprint@#1:#2::\@nil}
\def\mn@eprint@arXiv#1{\href {http://arxiv.org/abs/#1} {{\tt arXiv:#1}}}
\def\mn@eprint@dblp#1{\href {http://dblp.uni-trier.de/rec/bibtex/#1.xml}
  {dblp:#1}}
\def\mn@eprint@#1:#2:#3:#4\@nil{\def\@tempa {#1}\def\@tempb {#2}\def\@tempc
  {#3}\ifx \@tempc \@empty \let \@tempc \@tempb \let \@tempb \@tempa \fi \ifx
  \@tempb \@empty \def\@tempb {arXiv}\fi \@ifundefined
  {mn@eprint@\@tempb}{\@tempb:\@tempc}{\expandafter \expandafter \csname
  mn@eprint@\@tempb\endcsname \expandafter{\@tempc}}}

\bibitem[\protect\citeauthoryear{{Akaike}}{{Akaike}}{1974}]{Akaike1974}
{Akaike} H.,  1974, IEEE Transactions on Automatic Control, \href
  {https://ui.adsabs.harvard.edu/abs/1974ITAC...19..716A} {19, 716}

\bibitem[\protect\citeauthoryear{{Allen}, {Groves}, {Dopita}, {Sutherland}  \&
  {Kewley}}{{Allen} et~al.}{2008}]{allen2008}
{Allen} M.~G.,  {Groves} B.~A.,  {Dopita} M.~A.,  {Sutherland} R.~S.,
  {Kewley} L.~J.,  2008, \mn@doi [\apjs] {10.1086/589652}, \href
  {https://ui.adsabs.harvard.edu/abs/2008ApJS..178...20A} {178, 20}

\bibitem[\protect\citeauthoryear{{Arrigoni Battaia}, {Yang}, {Hennawi},
  {Prochaska}, {Matsuda}, {Yamada}  \& {Hayashino}}{{Arrigoni Battaia}
  et~al.}{2015}]{Arrigoni2015b}
{Arrigoni Battaia} F.,  {Yang} Y.,  {Hennawi} J.~F.,  {Prochaska} J.~X.,
  {Matsuda} Y.,  {Yamada} T.,   {Hayashino} T.,  2015, \mn@doi [\apj]
  {10.1088/0004-637X/804/1/26}, \href
  {https://ui.adsabs.harvard.edu/abs/2015ApJ...804...26A} {804, 26}

\bibitem[\protect\citeauthoryear{{Arrigoni Battaia}, {Prochaska}, {Hennawi},
  {Obreja}, {Buck}, {Cantalupo}, {Dutton}  \& {Macci{\`o}}}{{Arrigoni Battaia}
  et~al.}{2018}]{Arrigoni2018}
{Arrigoni Battaia} F.,  {Prochaska} J.~X.,  {Hennawi} J.~F.,  {Obreja} A.,
  {Buck} T.,  {Cantalupo} S.,  {Dutton} A.~A.,   {Macci{\`o}} A.~V.,  2018,
  \mn@doi [\mnras] {10.1093/mnras/stx2465}, \href
  {https://ui.adsabs.harvard.edu/abs/2018MNRAS.473.3907A} {473, 3907}

\bibitem[\protect\citeauthoryear{{Arrigoni Battaia}, {Hennawi}, {Prochaska},
  {O{\~n}orbe}, {Farina}, {Cantalupo}  \& {Lusso}}{{Arrigoni Battaia}
  et~al.}{2019}]{Arrigoni2019}
{Arrigoni Battaia} F.,  {Hennawi} J.~F.,  {Prochaska} J.~X.,  {O{\~n}orbe} J.,
  {Farina} E.~P.,  {Cantalupo} S.,   {Lusso} E.,  2019, \mn@doi [\mnras]
  {10.1093/mnras/sty2827}, \href
  {https://ui.adsabs.harvard.edu/abs/2019MNRAS.482.3162A} {482, 3162}

\bibitem[\protect\citeauthoryear{{Astropy Collaboration} et~al.,}{{Astropy
  Collaboration} et~al.}{2018}]{astropy2}
{Astropy Collaboration} et~al., 2018, \mn@doi [\aj] {10.3847/1538-3881/aabc4f},
  \href {https://ui.adsabs.harvard.edu/abs/2018AJ....156..123A} {156, 123}

\bibitem[\protect\citeauthoryear{{Bacon} et~al.,}{{Bacon}
  et~al.}{2010}]{bacon2010}
{Bacon} R.,  et~al., 2010, in \procspie. p. 773508, \mn@doi{10.1117/12.856027}

\bibitem[\protect\citeauthoryear{{Barthel}}{{Barthel}}{1989}]{Barthel1989}
{Barthel} P.~D.,  1989, \mn@doi [\apj] {10.1086/167038}, \href
  {https://ui.adsabs.harvard.edu/abs/1989ApJ...336..606B} {336, 606}

\bibitem[\protect\citeauthoryear{{Best}, {R{\"o}ttgering}  \& {Longair}}{{Best}
  et~al.}{2000}]{Best2000}
{Best} P.~N.,  {R{\"o}ttgering} H.~J.~A.,   {Longair} M.~S.,  2000, \mn@doi
  [\mnras] {10.1046/j.1365-8711.2000.03028.x}, \href
  {https://ui.adsabs.harvard.edu/abs/2000MNRAS.311...23B} {311, 23}

\bibitem[\protect\citeauthoryear{{Borisova} et~al.,}{{Borisova}
  et~al.}{2016}]{Borisova2016}
{Borisova} E.,  et~al., 2016, \mn@doi [\apj] {10.3847/0004-637X/831/1/39},
  \href {https://ui.adsabs.harvard.edu/abs/2016ApJ...831...39B} {831, 39}

\bibitem[\protect\citeauthoryear{{Buckley}, {Charles}, {Nordsieck}  \&
  {O'Donoghue}}{{Buckley} et~al.}{2006}]{buckley2006}
{Buckley} D. A.~H.,  {Charles} P.~A.,  {Nordsieck} K.~H.,   {O'Donoghue} D.,
  2006, in {Whitelock} P.,  {Dennefeld} M.,   {Leibundgut} B.,  eds,  IAU
  Symposium Vol. 232, The Scientific Requirements for Extremely Large
  Telescopes. pp 1--12, \mn@doi{10.1017/S1743921306000202}

\bibitem[\protect\citeauthoryear{{Burgh}, {Nordsieck}, {Kobulnicky},
  {Williams}, {O'Donoghue}, {Smith}  \& {Percival}}{{Burgh}
  et~al.}{2003}]{burgh2003}
{Burgh} E.~B.,  {Nordsieck} K.~H.,  {Kobulnicky} H.~A.,  {Williams} T.~B.,
  {O'Donoghue} D.,  {Smith} M.~P.,   {Percival} J.~W.,  2003, in {Iye} M.,
  {Moorwood} A. F.~M.,  eds,  Society of Photo-Optical Instrumentation
  Engineers (SPIE) Conference Series Vol. 4841, \procspie. pp 1463--1471,
  \mn@doi{10.1117/12.460312}

\bibitem[\protect\citeauthoryear{{Cai} et~al.,}{{Cai} et~al.}{2019}]{cai2019}
{Cai} Z.,  et~al., 2019, arXiv e-prints, \href
  {https://ui.adsabs.harvard.edu/abs/2019arXiv190911098C} {p. arXiv:1909.11098}

\bibitem[\protect\citeauthoryear{{Cantalupo}, {Porciani}, {Lilly}  \&
  {Miniati}}{{Cantalupo} et~al.}{2005}]{cantalupo2005}
{Cantalupo} S.,  {Porciani} C.,  {Lilly} S.~J.,   {Miniati} F.,  2005, \mn@doi
  [\apj] {10.1086/430758}, \href
  {http://adsabs.harvard.edu/abs/2005ApJ...628...61C} {628, 61}

\bibitem[\protect\citeauthoryear{{Chambers}}{{Chambers}}{1989}]{Chambers1989}
{Chambers} A.~V.,  1989, PhD thesis, UNIVERSITY OF OXFORD (UNITED KINGDOM).

\bibitem[\protect\citeauthoryear{{Chambers} \& {Pan-STARRS Team}}{{Chambers} \&
  {Pan-STARRS Team}}{2018}]{ps12018}
{Chambers} K.,  {Pan-STARRS Team} 2018, in American Astronomical Society
  Meeting Abstracts \#231. p. 102.01

\bibitem[\protect\citeauthoryear{{Chambers}, {Miley}  \& {van
  Breugel}}{{Chambers} et~al.}{1987}]{Chambers1987}
{Chambers} K.~C.,  {Miley} G.~K.,   {van Breugel} W.,  1987, \mn@doi [\nat]
  {10.1038/329604a0}, \href
  {https://ui.adsabs.harvard.edu/abs/1987Natur.329..604C} {329, 604}

\bibitem[\protect\citeauthoryear{{Chen} \& {Pan}}{{Chen} \&
  {Pan}}{2017}]{ChenZ2017}
{Chen} Z.-F.,  {Pan} D.-S.,  2017, \mn@doi [\apj] {10.3847/1538-4357/aa8d66},
  \href {https://ui.adsabs.harvard.edu/abs/2017ApJ...848...79C} {848, 79}

\bibitem[\protect\citeauthoryear{{Christensen}, {Jahnke}, {Wisotzki}  \&
  {S{\'a}nchez}}{{Christensen} et~al.}{2006}]{Christensen2006}
{Christensen} L.,  {Jahnke} K.,  {Wisotzki} L.,   {S{\'a}nchez} S.~F.,  2006,
  \mn@doi [\aap] {10.1051/0004-6361:20065318}, \href
  {https://ui.adsabs.harvard.edu/abs/2006A&A...459..717C} {459, 717}

\bibitem[\protect\citeauthoryear{{Condon}, {Cotton}, {Greisen}, {Yin},
  {Perley}, {Taylor}  \& {Broderick}}{{Condon} et~al.}{1998}]{condon1998}
{Condon} J.~J.,  {Cotton} W.~D.,  {Greisen} E.~W.,  {Yin} Q.~F.,  {Perley}
  R.~A.,  {Taylor} G.~B.,   {Broderick} J.~J.,  1998, \mn@doi [\aj]
  {10.1086/300337}, \href
  {https://ui.adsabs.harvard.edu/abs/1998AJ....115.1693C} {115, 1693}

\bibitem[\protect\citeauthoryear{{Crawford} et~al.,}{{Crawford}
  et~al.}{2010}]{crawford2010}
{Crawford} S.~M.,  et~al., 2010, in Observatory Operations: Strategies,
  Processes, and Systems III. p. 773725, \mn@doi{10.1117/12.857000}

\bibitem[\protect\citeauthoryear{{Curran}}{{Curran}}{2014}]{Curran2014}
{Curran} P.~A.,  2014, arXiv e-prints, \href
  {https://ui.adsabs.harvard.edu/abs/2014arXiv1411.3816C} {p. arXiv:1411.3816}

\bibitem[\protect\citeauthoryear{{Dannerbauer} et~al.,}{{Dannerbauer}
  et~al.}{2014}]{Dannerbauer2014}
{Dannerbauer} H.,  et~al., 2014, \mn@doi [\aap] {10.1051/0004-6361/201423771},
  \href {https://ui.adsabs.harvard.edu/abs/2014A&A...570A..55D} {570, A55}

\bibitem[\protect\citeauthoryear{{Dietrich}, {Hamann}, {Shields}, {Constantin},
  {Vestergaard}, {Chaffee}, {Foltz}  \& {Junkkarinen}}{{Dietrich}
  et~al.}{2002}]{Dietrich2002}
{Dietrich} M.,  {Hamann} F.,  {Shields} J.~C.,  {Constantin} A.,  {Vestergaard}
  M.,  {Chaffee} F.,  {Foltz} C.~B.,   {Junkkarinen} V.~T.,  2002, \mn@doi
  [\apj] {10.1086/344410}, \href
  {https://ui.adsabs.harvard.edu/abs/2002ApJ...581..912D} {581, 912}

\bibitem[\protect\citeauthoryear{{Dijkstra} \& {Loeb}}{{Dijkstra} \&
  {Loeb}}{2008}]{dijkstra2008}
{Dijkstra} M.,  {Loeb} A.,  2008, \mn@doi [\mnras]
  {10.1111/j.1365-2966.2008.13066.x}, \href
  {https://ui.adsabs.harvard.edu/abs/2008MNRAS.386..492D} {386, 492}

\bibitem[\protect\citeauthoryear{{Dijkstra}, {Haiman}  \& {Spaans}}{{Dijkstra}
  et~al.}{2006}]{dijkstra2006}
{Dijkstra} M.,  {Haiman} Z.,   {Spaans} M.,  2006, \mn@doi [\apj]
  {10.1086/506243}, \href
  {https://ui.adsabs.harvard.edu/abs/2006ApJ...649...14D} {649, 14}

\bibitem[\protect\citeauthoryear{{Drake}, {Farina}, {Neeleman}, {Walter},
  {Venemans}, {Banados}, {Mazzucchelli}  \& {Decarli}}{{Drake}
  et~al.}{2019}]{Drake2019}
{Drake} A.~B.,  {Farina} E.~P.,  {Neeleman} M.,  {Walter} F.,  {Venemans} B.,
  {Banados} E.,  {Mazzucchelli} C.,   {Decarli} R.,  2019, \mn@doi [\apj]
  {10.3847/1538-4357/ab2984}, \href
  {https://ui.adsabs.harvard.edu/abs/2019ApJ...881..131D} {881, 131}

\bibitem[\protect\citeauthoryear{{Fanti}, {Pozzi}, {Dallacasa}, {Fanti},
  {Gregorini}, {Stanghellini}  \& {Vigotti}}{{Fanti} et~al.}{2001}]{Fanti01}
{Fanti} C.,  {Pozzi} F.,  {Dallacasa} D.,  {Fanti} R.,  {Gregorini} L.,
  {Stanghellini} C.,   {Vigotti} M.,  2001, \mn@doi [\aap]
  {10.1051/0004-6361:20010051}, \href
  {https://ui.adsabs.harvard.edu/abs/2001A&A...369..380F} {369, 380}

\bibitem[\protect\citeauthoryear{{Farina} et~al.,}{{Farina}
  et~al.}{2019}]{Farina2019}
{Farina} E.~P.,  et~al., 2019, \mn@doi [\apj] {10.3847/1538-4357/ab5847}, \href
  {https://ui.adsabs.harvard.edu/abs/2019ApJ...887..196F} {887, 196}

\bibitem[\protect\citeauthoryear{{Fathivavsari}, {Petitjean}, {Noterdaeme},
  {P{\^a}ris}, {Finley}, {L{\'o}pez}  \& {Srianand}}{{Fathivavsari}
  et~al.}{2016}]{Fathivavsari2016}
{Fathivavsari} H.,  {Petitjean} P.,  {Noterdaeme} P.,  {P{\^a}ris} I.,
  {Finley} H.,  {L{\'o}pez} S.,   {Srianand} R.,  2016, \mn@doi [\mnras]
  {10.1093/mnras/stw1411}, \href
  {http://adsabs.harvard.edu/abs/2016MNRAS.461.1816F} {461, 1816}

\bibitem[\protect\citeauthoryear{{Fathivavsari} et~al.,}{{Fathivavsari}
  et~al.}{2018}]{Fathivavsari2018}
{Fathivavsari} H.,  et~al., 2018, \mn@doi [\mnras] {10.1093/mnras/sty1023},
  \href {http://adsabs.harvard.edu/abs/2018MNRAS.477.5625F} {477, 5625}

\bibitem[\protect\citeauthoryear{{Fossati} et~al.,}{{Fossati}
  et~al.}{2021}]{Fossati2021}
{Fossati} M.,  et~al., 2021, \mn@doi [\mnras] {10.1093/mnras/stab660}, \href
  {https://ui.adsabs.harvard.edu/abs/2021MNRAS.503.3044F} {503, 3044}

\bibitem[\protect\citeauthoryear{{Galametz} et~al.,}{{Galametz}
  et~al.}{2012}]{Galametz2012}
{Galametz} A.,  et~al., 2012, \mn@doi [\apj] {10.1088/0004-637X/749/2/169},
  \href {https://ui.adsabs.harvard.edu/abs/2012ApJ...749..169G} {749, 169}

\bibitem[\protect\citeauthoryear{{Geach} et~al.,}{{Geach}
  et~al.}{2009}]{geach2009}
{Geach} J.~E.,  et~al., 2009, \mn@doi [\apj] {10.1088/0004-637X/700/1/1}, \href
  {https://ui.adsabs.harvard.edu/abs/2009ApJ...700....1G} {700, 1}

\bibitem[\protect\citeauthoryear{{Green}}{{Green}}{1996}]{Green1996}
{Green} P.~J.,  1996, \mn@doi [\apj] {10.1086/177584}, \href
  {https://ui.adsabs.harvard.edu/abs/1996ApJ...467...61G} {467, 61}

\bibitem[\protect\citeauthoryear{{Guo} et~al.,}{{Guo} et~al.}{2020}]{Guo2020}
{Guo} Y.,  et~al., 2020, \mn@doi [\apj] {10.3847/1538-4357/ab9b7f}, \href
  {https://ui.adsabs.harvard.edu/abs/2020ApJ...898...26G} {898, 26}

\bibitem[\protect\citeauthoryear{{Gupta} et~al.,}{{Gupta}
  et~al.}{2016}]{Gupta2016}
{Gupta} N.,  et~al., 2016, in Proceedings of MeerKAT Science: On the Pathway to
  the SKA. 25-27 May. p.~14 (\mn@eprint {arXiv} {1708.07371})

\bibitem[\protect\citeauthoryear{{Gupta} et~al.,}{{Gupta}
  et~al.}{2020}]{Gupta2020}
{Gupta} N.,  et~al., 2020, arXiv e-prints, \href
  {https://ui.adsabs.harvard.edu/abs/2020arXiv200704347G} {p. arXiv:2007.04347}

\bibitem[\protect\citeauthoryear{{Gupta} et~al.,}{{Gupta}
  et~al.}{2021a}]{gupta2021pband}
{Gupta} N.,  et~al., 2021a, arXiv e-prints, \href
  {https://ui.adsabs.harvard.edu/abs/2021arXiv210309437G} {p. arXiv:2103.09437}

\bibitem[\protect\citeauthoryear{{Gupta} et~al.,}{{Gupta}
  et~al.}{2021b}]{Gupta2021qsosurvey}
{Gupta} N.,  et~al., 2021b, arXiv e-prints, \href
  {https://ui.adsabs.harvard.edu/abs/2021arXiv210709705G} {p. arXiv:2107.09705}

\bibitem[\protect\citeauthoryear{{Haiman}, {Spaans}  \& {Quataert}}{{Haiman}
  et~al.}{2000}]{haiman2000}
{Haiman} Z.,  {Spaans} M.,   {Quataert} E.,  2000, \mn@doi [\apjl]
  {10.1086/312754}, \href
  {https://ui.adsabs.harvard.edu/abs/2000ApJ...537L...5H} {537, L5}

\bibitem[\protect\citeauthoryear{{Hambly} et~al.,}{{Hambly}
  et~al.}{2001}]{Hambly2001}
{Hambly} N.~C.,  et~al., 2001, \mn@doi [\mnras]
  {10.1111/j.1365-2966.2001.04660.x}, \href
  {https://ui.adsabs.harvard.edu/abs/2001MNRAS.326.1279H} {326, 1279}

\bibitem[\protect\citeauthoryear{{Heckman}, {Lehnert}, {van Breugel}  \&
  {Miley}}{{Heckman} et~al.}{1991a}]{heckman1991a}
{Heckman} T.~M.,  {Lehnert} M.~D.,  {van Breugel} W.,   {Miley} G.~K.,  1991a,
  \mn@doi [\apj] {10.1086/169794}, \href
  {http://adsabs.harvard.edu/abs/1991ApJ...370...78H} {370, 78}

\bibitem[\protect\citeauthoryear{{Heckman}, {Lehnert}, {Miley}  \& {van
  Breugel}}{{Heckman} et~al.}{1991b}]{heckman1991b}
{Heckman} T.~M.,  {Lehnert} M.~D.,  {Miley} G.~K.,   {van Breugel} W.,  1991b,
  \mn@doi [\apj] {10.1086/170660}, \href
  {http://adsabs.harvard.edu/abs/1991ApJ...381..373H} {381, 373}

\bibitem[\protect\citeauthoryear{{Ho}, {Goldoni}, {Dong}, {Greene}  \&
  {Ponti}}{{Ho} et~al.}{2012}]{Ho2012}
{Ho} L.~C.,  {Goldoni} P.,  {Dong} X.-B.,  {Greene} J.~E.,   {Ponti} G.,  2012,
  \mn@doi [\apj] {10.1088/0004-637X/754/1/11}, \href
  {https://ui.adsabs.harvard.edu/abs/2012ApJ...754...11H} {754, 11}

\bibitem[\protect\citeauthoryear{{Humphrey}, {Villar-Mart{\'\i}n}, {Fosbury},
  {Vernet}  \& {di Serego Alighieri}}{{Humphrey} et~al.}{2006}]{Humphrey2006}
{Humphrey} A.,  {Villar-Mart{\'\i}n} M.,  {Fosbury} R.,  {Vernet} J.,   {di
  Serego Alighieri} S.,  2006, \mn@doi [\mnras]
  {10.1111/j.1365-2966.2006.10224.x}, \href
  {https://ui.adsabs.harvard.edu/abs/2006MNRAS.369.1103H} {369, 1103}

\bibitem[\protect\citeauthoryear{{Humphrey}, {Villar-Mart{\'\i}n}, {Fosbury},
  {Binette}, {Vernet}, {De Breuck}  \& {di Serego Alighieri}}{{Humphrey}
  et~al.}{2007}]{Humphrey2007}
{Humphrey} A.,  {Villar-Mart{\'\i}n} M.,  {Fosbury} R.,  {Binette} L.,
  {Vernet} J.,  {De Breuck} C.,   {di Serego Alighieri} S.,  2007, \mn@doi
  [\mnras] {10.1111/j.1365-2966.2006.11344.x}, \href
  {https://ui.adsabs.harvard.edu/abs/2007MNRAS.375..705H} {375, 705}

\bibitem[\protect\citeauthoryear{{Hunter}}{{Hunter}}{2007}]{matplotlib}
{Hunter} J.~D.,  2007, Computing in Science Engineering, 9, 90

\bibitem[\protect\citeauthoryear{{Hurley-Walker} et~al.,}{{Hurley-Walker}
  et~al.}{2017}]{Hurley-Walker2017}
{Hurley-Walker} N.,  et~al., 2017, \mn@doi [\mnras] {10.1093/mnras/stw2337},
  \href {https://ui.adsabs.harvard.edu/abs/2017MNRAS.464.1146H} {464, 1146}

\bibitem[\protect\citeauthoryear{{Jonas} \& {MeerKAT Team}}{{Jonas} \& {MeerKAT
  Team}}{2016}]{Jonas2016}
{Jonas} J.,  {MeerKAT Team} 2016, in MeerKAT Science: On the Pathway to the
  SKA. p.~1

\bibitem[\protect\citeauthoryear{{Kellermann}, {Sramek}, {Schmidt}, {Shaffer}
  \& {Green}}{{Kellermann} et~al.}{1989}]{Kellermann1989}
{Kellermann} K.~I.,  {Sramek} R.,  {Schmidt} M.,  {Shaffer} D.~B.,   {Green}
  R.,  1989, \mn@doi [\aj] {10.1086/115207}, \href
  {https://ui.adsabs.harvard.edu/abs/1989AJ.....98.1195K} {98, 1195}

\bibitem[\protect\citeauthoryear{{Kobulnicky}, {Nordsieck}, {Burgh}, {Smith},
  {Percival}, {Williams}  \& {O'Donoghue}}{{Kobulnicky}
  et~al.}{2003}]{kobul2003}
{Kobulnicky} H.~A.,  {Nordsieck} K.~H.,  {Burgh} E.~B.,  {Smith} M.~P.,
  {Percival} J.~W.,  {Williams} T.~B.,   {O'Donoghue} D.,  2003, in {Iye} M.,
  {Moorwood} A. F.~M.,  eds,  Society of Photo-Optical Instrumentation
  Engineers (SPIE) Conference Series Vol. 4841, \procspie. pp 1634--1644,
  \mn@doi{10.1117/12.460315}

\bibitem[\protect\citeauthoryear{{Korista}, {Baldwin}  \& {Ferland}}{{Korista}
  et~al.}{1998}]{Korista1998}
{Korista} K.,  {Baldwin} J.,   {Ferland} G.,  1998, \mn@doi [\apj]
  {10.1086/306321}, \href
  {https://ui.adsabs.harvard.edu/abs/1998ApJ...507...24K} {507, 24}

\bibitem[\protect\citeauthoryear{{Krogager} et~al.,}{{Krogager}
  et~al.}{2018}]{krogager2018}
{Krogager} J.~K.,  et~al., 2018, \mn@doi [\apjs] {10.3847/1538-4365/aaab51},
  \href {https://ui.adsabs.harvard.edu/abs/2018ApJS..235...10K} {235, 10}

\bibitem[\protect\citeauthoryear{{Laor}, {Bahcall}, {Jannuzi}, {Schneider}  \&
  {Green}}{{Laor} et~al.}{1995}]{Laor1995}
{Laor} A.,  {Bahcall} J.~N.,  {Jannuzi} B.~T.,  {Schneider} D.~P.,   {Green}
  R.~F.,  1995, \mn@doi [\apjs] {10.1086/192177}, \href
  {https://ui.adsabs.harvard.edu/abs/1995ApJS...99....1L} {99, 1}

\bibitem[\protect\citeauthoryear{{Leclercq} et~al.,}{{Leclercq}
  et~al.}{2017}]{leclercq2017}
{Leclercq} F.,  et~al., 2017, \mn@doi [\aap] {10.1051/0004-6361/201731480},
  \href {https://ui.adsabs.harvard.edu/abs/2017A&A...608A...8L} {608, A8}

\bibitem[\protect\citeauthoryear{{Mackenzie} et~al.,}{{Mackenzie}
  et~al.}{2021}]{Mackenzie2021}
{Mackenzie} R.,  et~al., 2021, \mn@doi [\mnras] {10.1093/mnras/staa3277}, \href
  {https://ui.adsabs.harvard.edu/abs/2021MNRAS.502..494M} {502, 494}

\bibitem[\protect\citeauthoryear{{Mayo}, {Vernet}, {De Breuck}, {Galametz},
  {Seymour}  \& {Stern}}{{Mayo} et~al.}{2012}]{Mayo2012}
{Mayo} J.~H.,  {Vernet} J.,  {De Breuck} C.,  {Galametz} A.,  {Seymour} N.,
  {Stern} D.,  2012, \mn@doi [\aap] {10.1051/0004-6361/201118254}, \href
  {https://ui.adsabs.harvard.edu/abs/2012A&A...539A..33M} {539, A33}

\bibitem[\protect\citeauthoryear{{McCarthy}}{{McCarthy}}{1988}]{mccarthy1988}
{McCarthy} P.~J.,  1988, {Extended optical emission in 3CR radio galaxies}

\bibitem[\protect\citeauthoryear{{McCarthy}}{{McCarthy}}{1993}]{mccarthy1993}
{McCarthy} P.~J.,  1993, \mn@doi [\araa] {10.1146/annurev.aa.31.090193.003231},
  \href {https://ui.adsabs.harvard.edu/abs/1993ARA&A..31..639M} {31, 639}

\bibitem[\protect\citeauthoryear{{Miley} \& {De Breuck}}{{Miley} \& {De
  Breuck}}{2008}]{miley2008}
{Miley} G.,  {De Breuck} C.,  2008, \mn@doi [\aapr]
  {10.1007/s00159-007-0008-z}, \href
  {https://ui.adsabs.harvard.edu/abs/2008A&ARv..15...67M} {15, 67}

\bibitem[\protect\citeauthoryear{{Mori}, {Umemura}  \& {Ferrara}}{{Mori}
  et~al.}{2004}]{mori2004}
{Mori} M.,  {Umemura} M.,   {Ferrara} A.,  2004, \mn@doi [\apjl]
  {10.1086/425255}, \href
  {https://ui.adsabs.harvard.edu/abs/2004ApJ...613L..97M} {613, L97}

\bibitem[\protect\citeauthoryear{{Morrissey} et~al.,}{{Morrissey}
  et~al.}{2012}]{morrissey2012}
{Morrissey} P.,  et~al., 2012, in \procspie. p. 844613,
  \mn@doi{10.1117/12.924729}

\bibitem[\protect\citeauthoryear{{Nestor}, {Hamann}  \& {Rodriguez
  Hidalgo}}{{Nestor} et~al.}{2008}]{Nestor2008}
{Nestor} D.,  {Hamann} F.,   {Rodriguez Hidalgo} P.,  2008, \mn@doi [\mnras]
  {10.1111/j.1365-2966.2008.13156.x}, \href
  {https://ui.adsabs.harvard.edu/abs/2008MNRAS.386.2055N} {386, 2055}

\bibitem[\protect\citeauthoryear{{Noterdaeme}, {Balashev}, {Krogager},
  {Srianand}, {Fathivavsari}, {Petitjean}  \& {Ledoux}}{{Noterdaeme}
  et~al.}{2019}]{Noterdaeme2019}
{Noterdaeme} P.,  {Balashev} S.,  {Krogager} J.~K.,  {Srianand} R.,
  {Fathivavsari} H.,  {Petitjean} P.,   {Ledoux} C.,  2019, \mn@doi [\aap]
  {10.1051/0004-6361/201935371}, \href
  {https://ui.adsabs.harvard.edu/abs/2019A&A...627A..32N} {627, A32}

\bibitem[\protect\citeauthoryear{{O'Dea} \& {Saikia}}{{O'Dea} \&
  {Saikia}}{2021}]{Odea21}
{O'Dea} C.~P.,  {Saikia} D.~J.,  2021, \mn@doi [\aapr]
  {10.1007/s00159-021-00131-w}, \href
  {https://ui.adsabs.harvard.edu/abs/2021A&ARv..29....3O} {29, 3}

\bibitem[\protect\citeauthoryear{{O'Sullivan}, {Martin}, {Matuszewski},
  {Hoadley}, {Hamden}, {Neill}, {Lin}  \& {Parihar}}{{O'Sullivan}
  et~al.}{2020}]{Osullivan2020}
{O'Sullivan} D.~B.,  {Martin} C.,  {Matuszewski} M.,  {Hoadley} K.,  {Hamden}
  E.,  {Neill} J.~D.,  {Lin} Z.,   {Parihar} P.,  2020, \mn@doi [\apj]
  {10.3847/1538-4357/ab838c}, \href
  {https://ui.adsabs.harvard.edu/abs/2020ApJ...894....3O} {894, 3}

\bibitem[\protect\citeauthoryear{{Overzier}, {Nesvadba}, {Dijkstra}, {Hatch},
  {Lehnert}, {Villar-Mart{\'\i}n}, {Wilman}  \& {Zirm}}{{Overzier}
  et~al.}{2013}]{overzier2013}
{Overzier} R.~A.,  {Nesvadba} N.~P.~H.,  {Dijkstra} M.,  {Hatch} N.~A.,
  {Lehnert} M.~D.,  {Villar-Mart{\'\i}n} M.,  {Wilman} R.~J.,   {Zirm} A.~W.,
  2013, \mn@doi [\apj] {10.1088/0004-637X/771/2/89}, \href
  {https://ui.adsabs.harvard.edu/abs/2013ApJ...771...89O} {771, 89}

\bibitem[\protect\citeauthoryear{{P{\^a}ris} et~al.,}{{P{\^a}ris}
  et~al.}{2012}]{Paris2012}
{P{\^a}ris} I.,  et~al., 2012, \mn@doi [\aap] {10.1051/0004-6361/201220142},
  \href {https://ui.adsabs.harvard.edu/abs/2012A&A...548A..66P} {548, A66}

\bibitem[\protect\citeauthoryear{{Park}, {Woo}, {Denney}  \& {Shin}}{{Park}
  et~al.}{2013}]{park2013}
{Park} D.,  {Woo} J.-H.,  {Denney} K.~D.,   {Shin} J.,  2013, \mn@doi [\apj]
  {10.1088/0004-637X/770/2/87}, \href
  {https://ui.adsabs.harvard.edu/abs/2013ApJ...770...87P} {770, 87}

\bibitem[\protect\citeauthoryear{{Peacock}, {Hambly}, {Bilicki},
  {MacGillivray}, {Miller}, {Read}  \& {Tritton}}{{Peacock}
  et~al.}{2016}]{Peacock2016}
{Peacock} J.~A.,  {Hambly} N.~C.,  {Bilicki} M.,  {MacGillivray} H.~T.,
  {Miller} L.,  {Read} M.~A.,   {Tritton} S.~B.,  2016, \mn@doi [\mnras]
  {10.1093/mnras/stw1818}, \href
  {https://ui.adsabs.harvard.edu/abs/2016MNRAS.462.2085P} {462, 2085}

\bibitem[\protect\citeauthoryear{P\'erez \& Granger}{P\'erez \&
  Granger}{2007}]{ipython}
P\'erez F.,  Granger B.~E.,  2007, \mn@doi [Computing in Science and
  Engineering] {10.1109/MCSE.2007.53}, 9, 21

\bibitem[\protect\citeauthoryear{{Perrotta} et~al.,}{{Perrotta}
  et~al.}{2016}]{Perrotta2016}
{Perrotta} S.,  et~al., 2016, \mn@doi [\mnras] {10.1093/mnras/stw1703}, \href
  {https://ui.adsabs.harvard.edu/abs/2016MNRAS.462.3285P} {462, 3285}

\bibitem[\protect\citeauthoryear{{Planck Collaboration} et~al.,}{{Planck
  Collaboration} et~al.}{2018}]{planck2018}
{Planck Collaboration} et~al., 2018, arXiv e-prints, \href
  {https://ui.adsabs.harvard.edu/abs/2018arXiv180706209P} {p. arXiv:1807.06209}

\bibitem[\protect\citeauthoryear{{Prescott}, {Martin}  \& {Dey}}{{Prescott}
  et~al.}{2015}]{Prescott2015b}
{Prescott} M. K.~M.,  {Martin} C.~L.,   {Dey} A.,  2015, \mn@doi [\apj]
  {10.1088/0004-637X/799/1/62}, \href
  {https://ui.adsabs.harvard.edu/abs/2015ApJ...799...62P} {799, 62}

\bibitem[\protect\citeauthoryear{{Privon} et~al.,}{{Privon}
  et~al.}{2020}]{Privon2020}
{Privon} G.~C.,  et~al., 2020, \mn@doi [\apj] {10.3847/1538-4357/ab8015}, \href
  {https://ui.adsabs.harvard.edu/abs/2020ApJ...893..149P} {893, 149}

\bibitem[\protect\citeauthoryear{{Reuland} et~al.,}{{Reuland}
  et~al.}{2003}]{Reuland2003}
{Reuland} M.,  et~al., 2003, \mn@doi [\apj] {10.1086/375619}, \href
  {https://ui.adsabs.harvard.edu/abs/2003ApJ...592..755R} {592, 755}

\bibitem[\protect\citeauthoryear{{Richards} et~al.,}{{Richards}
  et~al.}{2006}]{richards2006a}
{Richards} G.~T.,  et~al., 2006, \mn@doi [\apjs] {10.1086/506525}, \href
  {https://ui.adsabs.harvard.edu/abs/2006ApJS..166..470R} {166, 470}

\bibitem[\protect\citeauthoryear{{Richards} et~al.,}{{Richards}
  et~al.}{2011}]{Richards2011}
{Richards} G.~T.,  et~al., 2011, \mn@doi [\aj] {10.1088/0004-6256/141/5/167},
  \href {https://ui.adsabs.harvard.edu/abs/2011AJ....141..167R} {141, 167}

\bibitem[\protect\citeauthoryear{{Roche}, {Humphrey}  \& {Binette}}{{Roche}
  et~al.}{2014}]{Roche2014}
{Roche} N.,  {Humphrey} A.,   {Binette} L.,  2014, \mn@doi [\mnras]
  {10.1093/mnras/stu1430}, \href
  {http://adsabs.harvard.edu/abs/2014MNRAS.443.3795R} {443, 3795}

\bibitem[\protect\citeauthoryear{{Roettgering}, {van Ojik}, {Miley},
  {Chambers}, {van Breugel}  \& {de Koff}}{{Roettgering}
  et~al.}{1997}]{roettgering1997}
{Roettgering} H.~J.~A.,  {van Ojik} R.,  {Miley} G.~K.,  {Chambers} K.~C.,
  {van Breugel} W.~J.~M.,   {de Koff} S.,  1997, \aap, \href
  {https://ui.adsabs.harvard.edu/abs/1997A&A...326..505R} {326, 505}

\bibitem[\protect\citeauthoryear{{Rosdahl} \& {Blaizot}}{{Rosdahl} \&
  {Blaizot}}{2012}]{Rosdahl2012}
{Rosdahl} J.,  {Blaizot} J.,  2012, \mn@doi [\mnras]
  {10.1111/j.1365-2966.2012.20883.x}, \href
  {https://ui.adsabs.harvard.edu/abs/2012MNRAS.423..344R} {423, 344}

\bibitem[\protect\citeauthoryear{{Ross} et~al.,}{{Ross}
  et~al.}{2013}]{Ross2013}
{Ross} N.~P.,  et~al., 2013, \mn@doi [\apj] {10.1088/0004-637X/773/1/14}, \href
  {https://ui.adsabs.harvard.edu/abs/2013ApJ...773...14R} {773, 14}

\bibitem[\protect\citeauthoryear{{Rybicki} \& {Lightman}}{{Rybicki} \&
  {Lightman}}{1986}]{Rybicki1986}
{Rybicki} G.~B.,  {Lightman} A.~P.,  1986, {Radiative Processes in
  Astrophysics}

\bibitem[\protect\citeauthoryear{{Saikia} \& {Gupta}}{{Saikia} \&
  {Gupta}}{2003}]{Saikia03}
{Saikia} D.~J.,  {Gupta} N.,  2003, \mn@doi [\aap]
  {10.1051/0004-6361:20030635}, \href
  {https://ui.adsabs.harvard.edu/abs/2003A&A...405..499S} {405, 499}

\bibitem[\protect\citeauthoryear{{Shen} \& {Liu}}{{Shen} \&
  {Liu}}{2012}]{Shen2012}
{Shen} Y.,  {Liu} X.,  2012, \mn@doi [\apj] {10.1088/0004-637X/753/2/125},
  \href {https://ui.adsabs.harvard.edu/abs/2012ApJ...753..125S} {753, 125}

\bibitem[\protect\citeauthoryear{{Shen}, {Greene}, {Strauss}, {Richards}  \&
  {Schneider}}{{Shen} et~al.}{2008}]{shen2008}
{Shen} Y.,  {Greene} J.~E.,  {Strauss} M.~A.,  {Richards} G.~T.,   {Schneider}
  D.~P.,  2008, \mn@doi [\apj] {10.1086/587475}, \href
  {https://ui.adsabs.harvard.edu/abs/2008ApJ...680..169S} {680, 169}

\bibitem[\protect\citeauthoryear{{Shen} et~al.,}{{Shen}
  et~al.}{2011}]{shen2011}
{Shen} Y.,  et~al., 2011, \mn@doi [\apjs] {10.1088/0067-0049/194/2/45}, \href
  {https://ui.adsabs.harvard.edu/abs/2011ApJS..194...45S} {194, 45}

\bibitem[\protect\citeauthoryear{{Shen} et~al.,}{{Shen}
  et~al.}{2016}]{shen2016}
{Shen} Y.,  et~al., 2016, \mn@doi [\apj] {10.3847/0004-637X/831/1/7}, \href
  {https://ui.adsabs.harvard.edu/abs/2016ApJ...831....7S} {831, 7}

\bibitem[\protect\citeauthoryear{{Shukla}, {Srianand}, {Gupta}, {Petitjean},
  {Baker}, {Krogager}  \& {Noterdaeme}}{{Shukla} et~al.}{2021}]{Shukla2021}
{Shukla} G.,  {Srianand} R.,  {Gupta} N.,  {Petitjean} P.,  {Baker} A.~J.,
  {Krogager} J.-K.,   {Noterdaeme} P.,  2021, \mn@doi [\mnras]
  {10.1093/mnras/staa3977}, \href
  {https://ui.adsabs.harvard.edu/abs/2021MNRAS.501.5362S} {501, 5362}

\bibitem[\protect\citeauthoryear{{Srianand}, {Gupta}, {Momjian}  \&
  {Vivek}}{{Srianand} et~al.}{2015}]{Srianand15}
{Srianand} R.,  {Gupta} N.,  {Momjian} E.,   {Vivek} M.,  2015, \mn@doi
  [\mnras] {10.1093/mnras/stv1004}, \href
  {http://adsabs.harvard.edu/abs/2015MNRAS.451..917S} {451, 917}

\bibitem[\protect\citeauthoryear{{Timlin}, {Brandt}  \& {Laor}}{{Timlin}
  et~al.}{2021}]{Timlin2021a}
{Timlin} John~D. I.,  {Brandt} W.~N.,   {Laor} A.,  2021, \mn@doi [\mnras]
  {10.1093/mnras/stab1217}, \href
  {https://ui.adsabs.harvard.edu/abs/2021MNRAS.504.5556T} {504, 5556}

\bibitem[\protect\citeauthoryear{{Tonry} et~al.,}{{Tonry}
  et~al.}{2012}]{Tonry2012}
{Tonry} J.~L.,  et~al., 2012, \mn@doi [\apj] {10.1088/0004-637X/750/2/99},
  \href {https://ui.adsabs.harvard.edu/abs/2012ApJ...750...99T} {750, 99}

\bibitem[\protect\citeauthoryear{{Trakhtenbrot} \& {Netzer}}{{Trakhtenbrot} \&
  {Netzer}}{2012}]{Trakhtenbrot2012}
{Trakhtenbrot} B.,  {Netzer} H.,  2012, \mn@doi [\mnras]
  {10.1111/j.1365-2966.2012.22056.x}, \href
  {https://ui.adsabs.harvard.edu/abs/2012MNRAS.427.3081T} {427, 3081}

\bibitem[\protect\citeauthoryear{Vestergaard \& Peterson}{Vestergaard \&
  Peterson}{2006}]{Vestergaard_2006}
Vestergaard M.,  Peterson B.~M.,  2006, \mn@doi [The Astrophysical Journal]
  {10.1086/500572}, 641, 689

\bibitem[\protect\citeauthoryear{{Villar-Martin}, {Binette}  \&
  {Fosbury}}{{Villar-Martin} et~al.}{1996}]{villar1996}
{Villar-Martin} M.,  {Binette} L.,   {Fosbury} R.~A.~E.,  1996, \aap, \href
  {https://ui.adsabs.harvard.edu/abs/1996A&A...312..751V} {312, 751}

\bibitem[\protect\citeauthoryear{{Villar-Mart{\'\i}n}, {Humphrey}, {De Breuck},
  {Fosbury}, {Binette}  \& {Vernet}}{{Villar-Mart{\'\i}n}
  et~al.}{2007a}]{villar2007a}
{Villar-Mart{\'\i}n} M.,  {Humphrey} A.,  {De Breuck} C.,  {Fosbury} R.,
  {Binette} L.,   {Vernet} J.,  2007a, \mn@doi [\mnras]
  {10.1111/j.1365-2966.2006.11371.x}, \href
  {https://ui.adsabs.harvard.edu/abs/2007MNRAS.375.1299V} {375, 1299}

\bibitem[\protect\citeauthoryear{{Villar-Mart{\'\i}n}, {S{\'a}nchez},
  {Humphrey}, {Dijkstra}, {di Serego Alighieri}, {De Breuck}  \& {Gonz{\'a}lez
  Delgado}}{{Villar-Mart{\'\i}n} et~al.}{2007b}]{villar2007}
{Villar-Mart{\'\i}n} M.,  {S{\'a}nchez} S.~F.,  {Humphrey} A.,  {Dijkstra} M.,
  {di Serego Alighieri} S.,  {De Breuck} C.,   {Gonz{\'a}lez Delgado} R.,
  2007b, \mn@doi [\mnras] {10.1111/j.1365-2966.2007.11811.x}, \href
  {https://ui.adsabs.harvard.edu/abs/2007MNRAS.378..416V} {378, 416}

\bibitem[\protect\citeauthoryear{{Virtanen} et~al.,}{{Virtanen}
  et~al.}{2020}]{scipy}
{Virtanen} P.,  et~al., 2020, \mn@doi [Nature Methods]
  {https://doi.org/10.1038/s41592-019-0686-2}, \href {https://rdcu.be/b08Wh} {}

\bibitem[\protect\citeauthoryear{{Wild} et~al.,}{{Wild}
  et~al.}{2008}]{Wild2008}
{Wild} V.,  et~al., 2008, \mn@doi [\mnras] {10.1111/j.1365-2966.2008.13375.x},
  \href {https://ui.adsabs.harvard.edu/abs/2008MNRAS.388..227W} {388, 227}

\bibitem[\protect\citeauthoryear{{Wisotzki} et~al.,}{{Wisotzki}
  et~al.}{2016}]{Wisotzki2016}
{Wisotzki} L.,  et~al., 2016, \mn@doi [\aap] {10.1051/0004-6361/201527384},
  \href {https://ui.adsabs.harvard.edu/abs/2016A&A...587A..98W} {587, A98}

\bibitem[\protect\citeauthoryear{{Wolfe}, {Gawiser}  \& {Prochaska}}{{Wolfe}
  et~al.}{2005}]{Wolfe2005}
{Wolfe} A.~M.,  {Gawiser} E.,   {Prochaska} J.~X.,  2005, \mn@doi [\araa]
  {10.1146/annurev.astro.42.053102.133950}, \href
  {https://ui.adsabs.harvard.edu/abs/2005ARA&A..43..861W} {43, 861}

\bibitem[\protect\citeauthoryear{{Wylezalek} et~al.,}{{Wylezalek}
  et~al.}{2013}]{Wylezalek2013}
{Wylezalek} D.,  et~al., 2013, \mn@doi [\apj] {10.1088/0004-637X/769/1/79},
  \href {https://ui.adsabs.harvard.edu/abs/2013ApJ...769...79W} {769, 79}

\bibitem[\protect\citeauthoryear{{Zheng} \& {Malkan}}{{Zheng} \&
  {Malkan}}{1993}]{Zheng1993}
{Zheng} W.,  {Malkan} M.~A.,  1993, \mn@doi [\apj] {10.1086/173182}, \href
  {https://ui.adsabs.harvard.edu/abs/1993ApJ...415..517Z} {415, 517}

\bibitem[\protect\citeauthoryear{{van Ojik}, {Roettgering}, {Miley}  \&
  {Hunstead}}{{van Ojik} et~al.}{1997}]{vanojik1997}
{van Ojik} R.,  {Roettgering} H.~J.~A.,  {Miley} G.~K.,   {Hunstead} R.~W.,
  1997, \aap, \href {https://ui.adsabs.harvard.edu/abs/1997A&A...317..358V}
  {317, 358}

\bibitem[\protect\citeauthoryear{{van der Walt}, {Colbert}  \&
  {Varoquaux}}{{van der Walt} et~al.}{2011}]{numpy}
{van der Walt} S.,  {Colbert} S.~C.,   {Varoquaux} G.,  2011, \mn@doi
  [Computing in Science and Engineering] {10.1109/MCSE.2011.37}, \href
  {https://ui.adsabs.harvard.edu/abs/2011CSE....13b..22V} {13, 22}

\makeatother
\end{thebibliography}

\appendix
\section{APPENDIX}
\label{sec_appendix}

\begin{figure*} 
\centerline{\vbox{
\centerline{\hbox{ 
\includegraphics[height=0.95\textheight,angle=0]{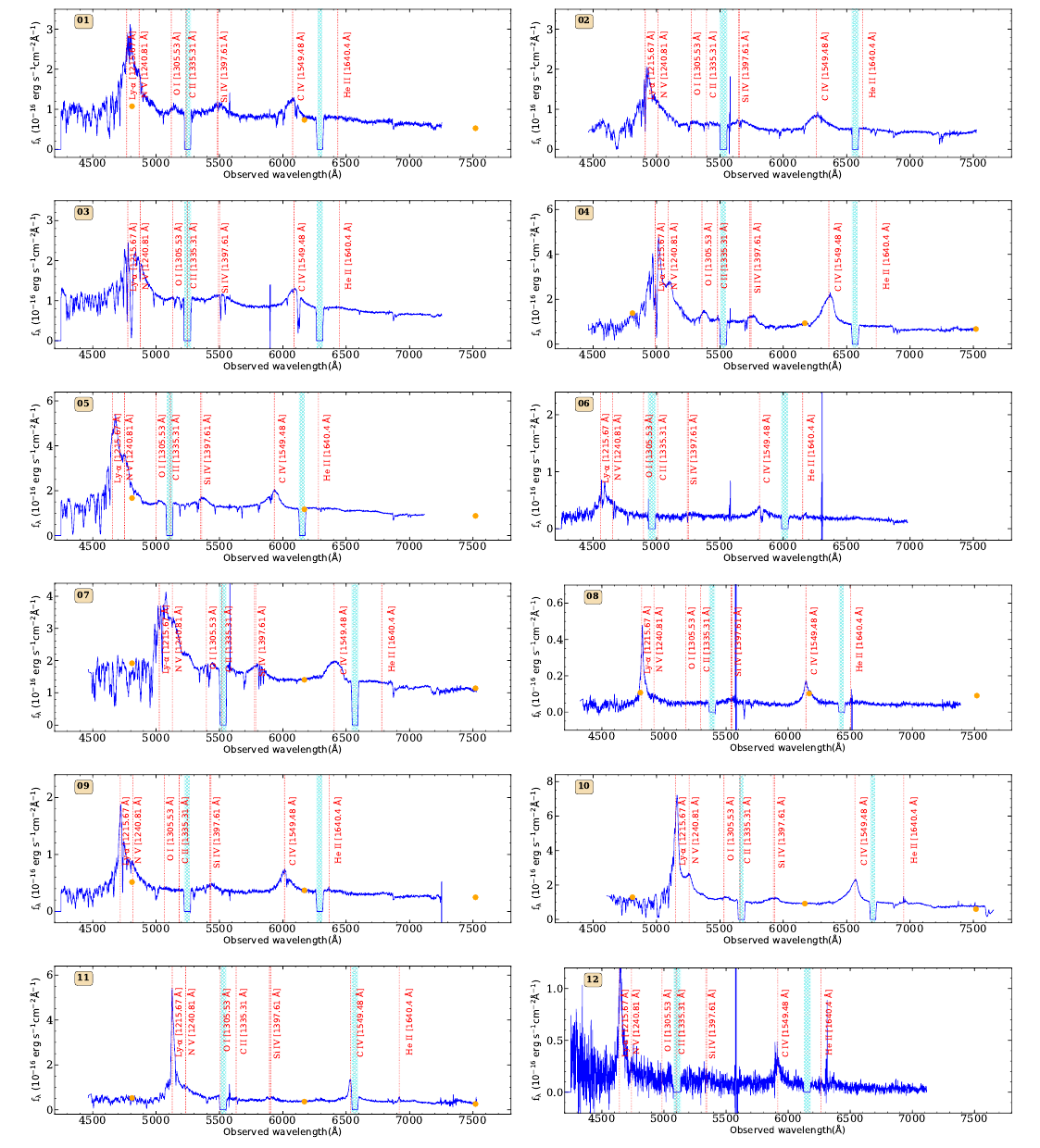}
}} 
}}  
\vskip+0.0cm  
\caption{1D spectrum of the sources in our sample. Vertical lines mark the locations of different emission lines and the orange points are photometry from Pan-Starrs1. The cyan shaded regions indicate the two ccd gap ranges as mentioned in Section~\ref{sub_longslit}. The source IDs are shown in the top-left corner. Note that the spectra are combination of all available PAs for each source.
} 
\label{fig_1dspectra}   
\end{figure*}

\begin{figure*} 
\ContinuedFloat
\centerline{\vbox{
\centerline{\hbox{ 
\includegraphics[height=0.95\textheight,angle=0]{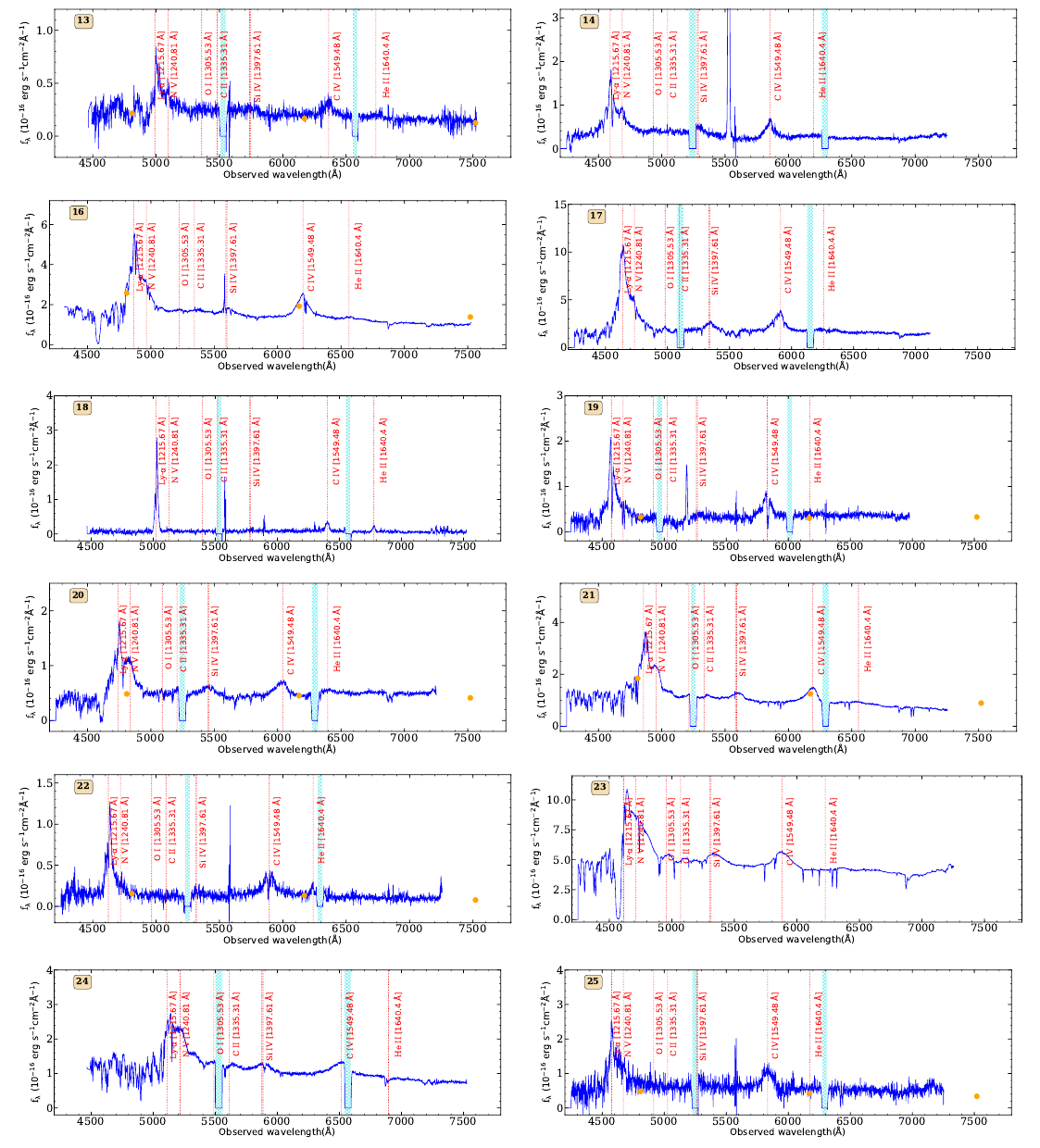}
}} 
}}  
\vskip+0.0cm  
\caption{Continued.
} 
\label{fig_1dspectra2}   
\end{figure*}


\begin{figure*} 
\centerline{\vbox{
\centerline{\hbox{ 
\includegraphics[height=0.8\textheight,angle=0]{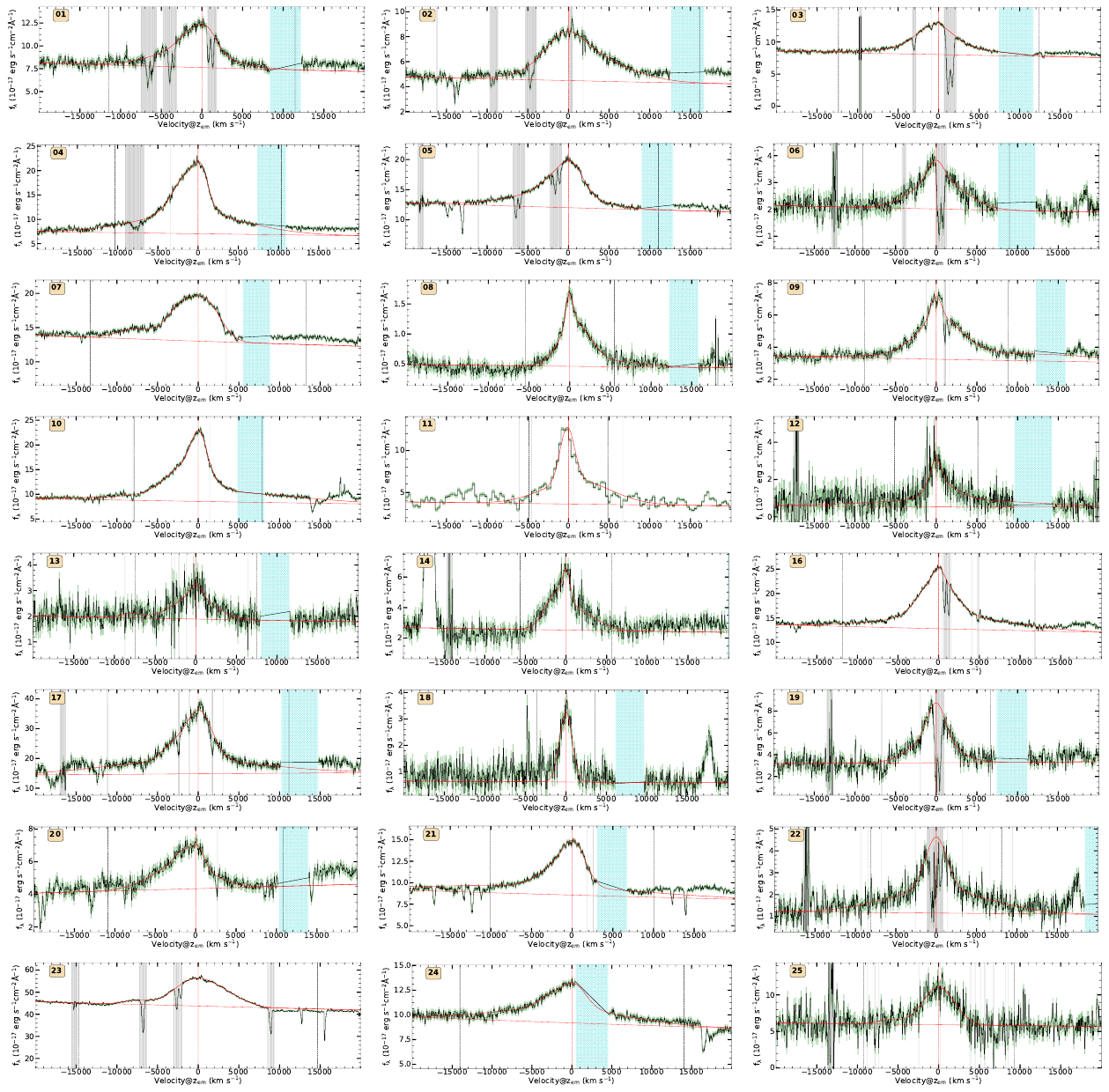}
}} 
}}  
\caption{C~{\sc iv} (black) emission line profiles of the quasars from our sample. The green shaded region is 1$\sigma$ error bar, the red curve is the Gaussian+power law fit and the dashed horizontal red line is the power law fit to the local continuum region. The grey and cyan shaded regions were masked while fitting the Gaussian to the emission line profile and linearly interpolated to measure the line fluxes. The cyan regions indicated ccdgap ranges. For M121514.42-062803.50, the C~{\sc iv} line falls in the CCD gap, so we have used our NOT spectrum to measure the C~{\sc iv} line properties.} 
\label{fig_civfit}   
\end{figure*} 
\begin{figure*} 
\centerline{\vbox{
\centerline{\hbox{ 
\includegraphics[height=0.8\textheight,angle=0]{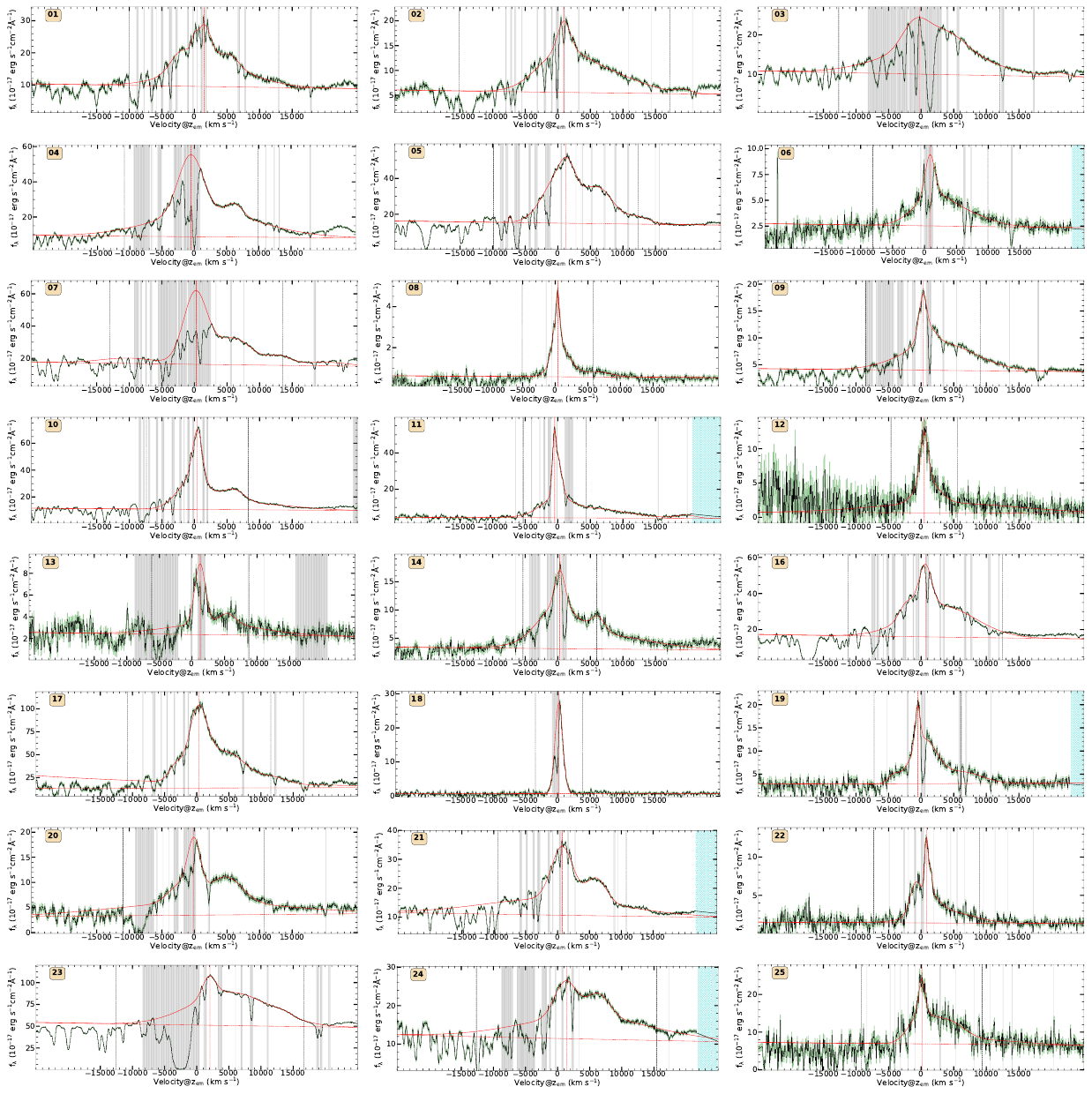}
}} 
}}  
\vskip+0.0cm  
\caption{Same as Fig. \ref{fig_civfit} for \lya\ emission line.
} 

\label{fig_lyafit}   
\end{figure*} 

\begin{figure*} 
\centerline{\vbox{
\centerline{\hbox{ 
\includegraphics[height=0.6\textheight,angle=0]{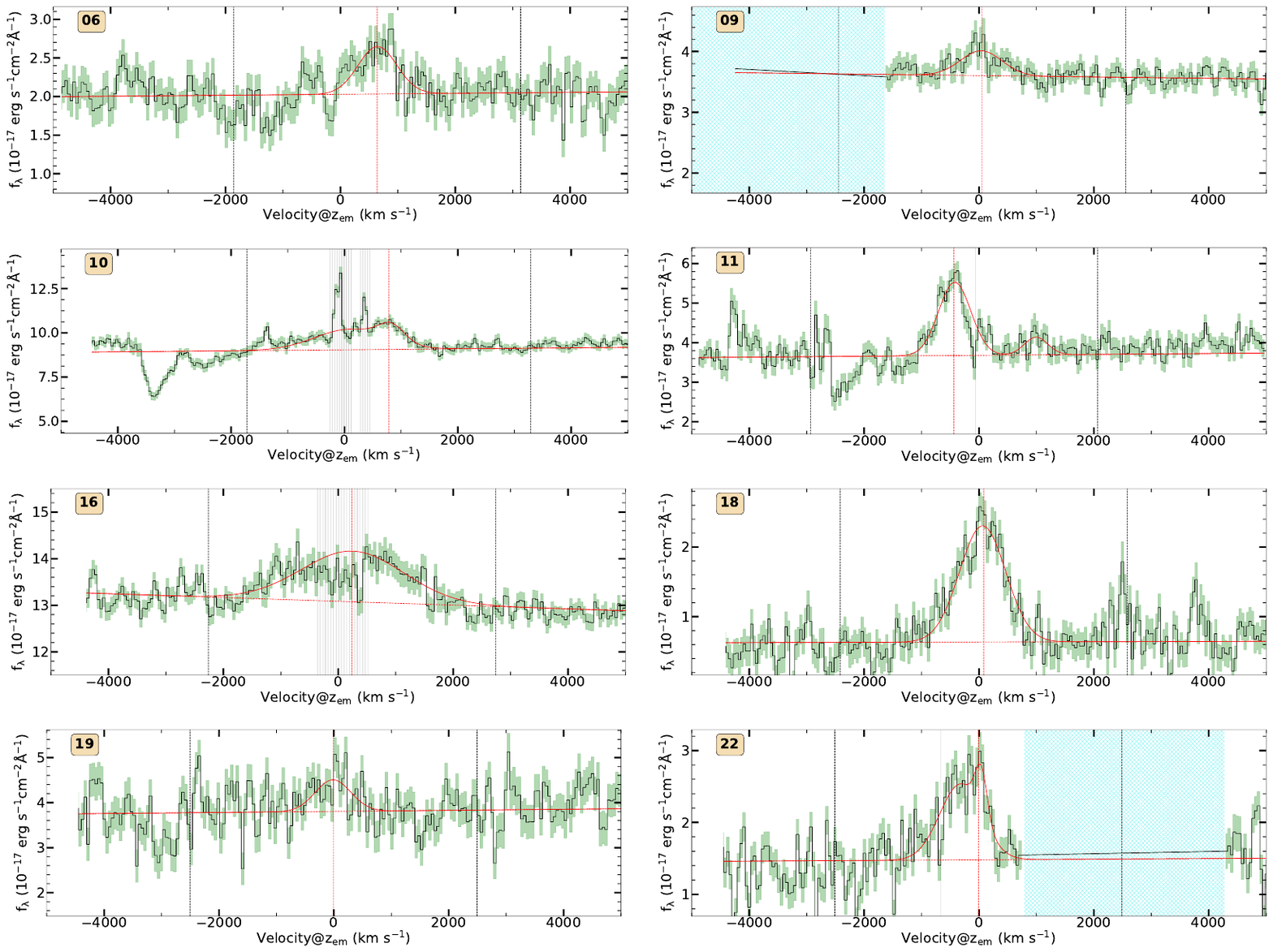}
}} 
}}  
\vskip+0.0cm  
\caption{Same as Fig. \ref{fig_civfit} for \heii\ emission line for sources in our sample with clear detection with 4\sig\ significance.
} 
\label{fig_heiifit}   
\end{figure*} 

\begin{figure*} 
\centerline{\vbox{
\centerline{\hbox{ 
\includegraphics[height=0.75\textheight,angle=0]{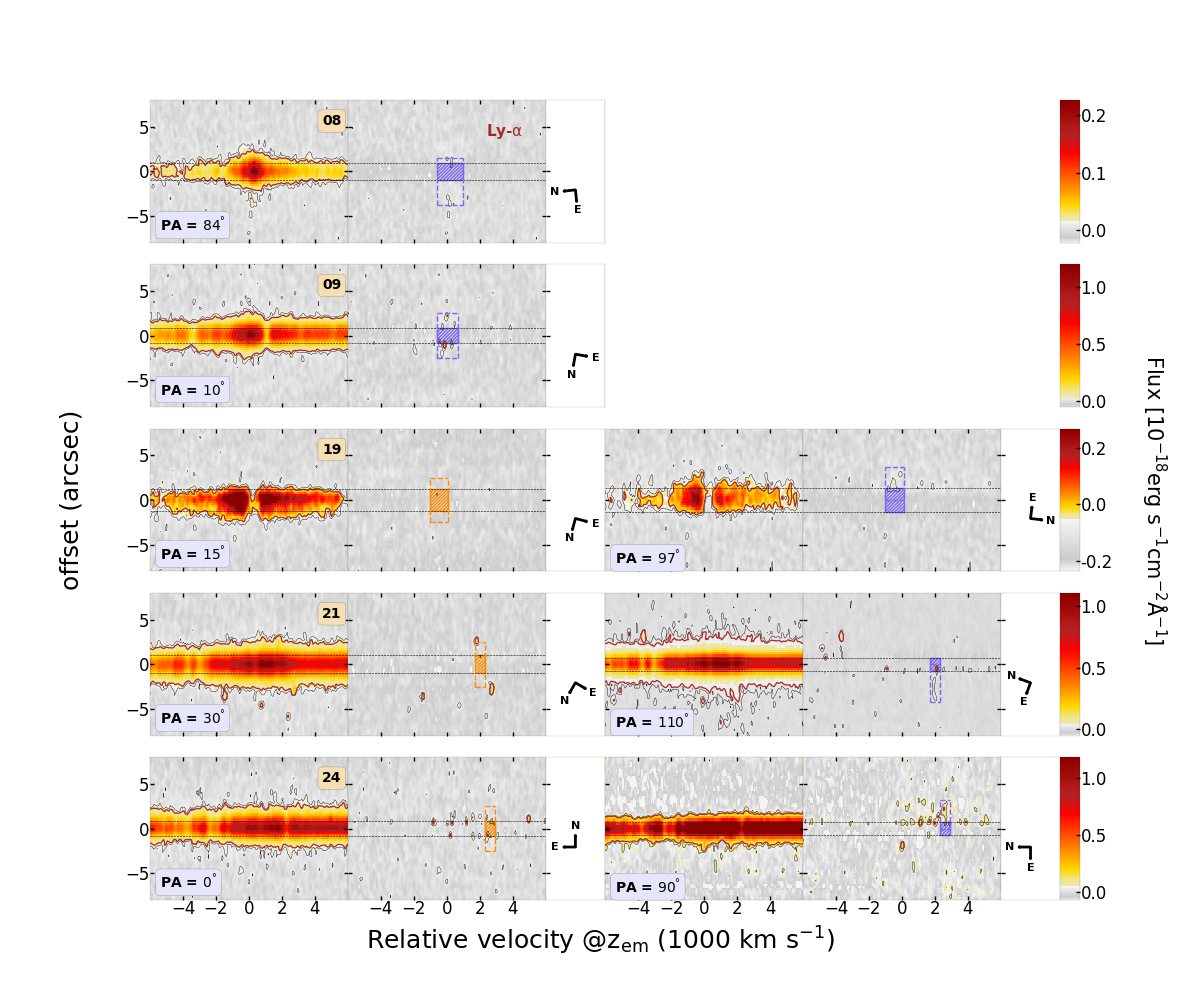}
}} 
}}  
\vskip+0.0cm  
\caption{Same as Fig. \ref{fig_lya_samp1} for five sources with tentative \lya\ halo detection. For \# 21 along PA=30\degree\ the two tiny blobs in the SPSF subtracted image are residuals of cosmic ray. In \#24, we see significant residual emission from SPSF subtraction within the FWHM especially at the top. 
} 
\label{fig_lya_samp2}   
\end{figure*} 

\begin{figure*} 
\centerline{\vbox{
\centerline{\hbox{ 
\includegraphics[
height=0.75\textheight,angle=0]{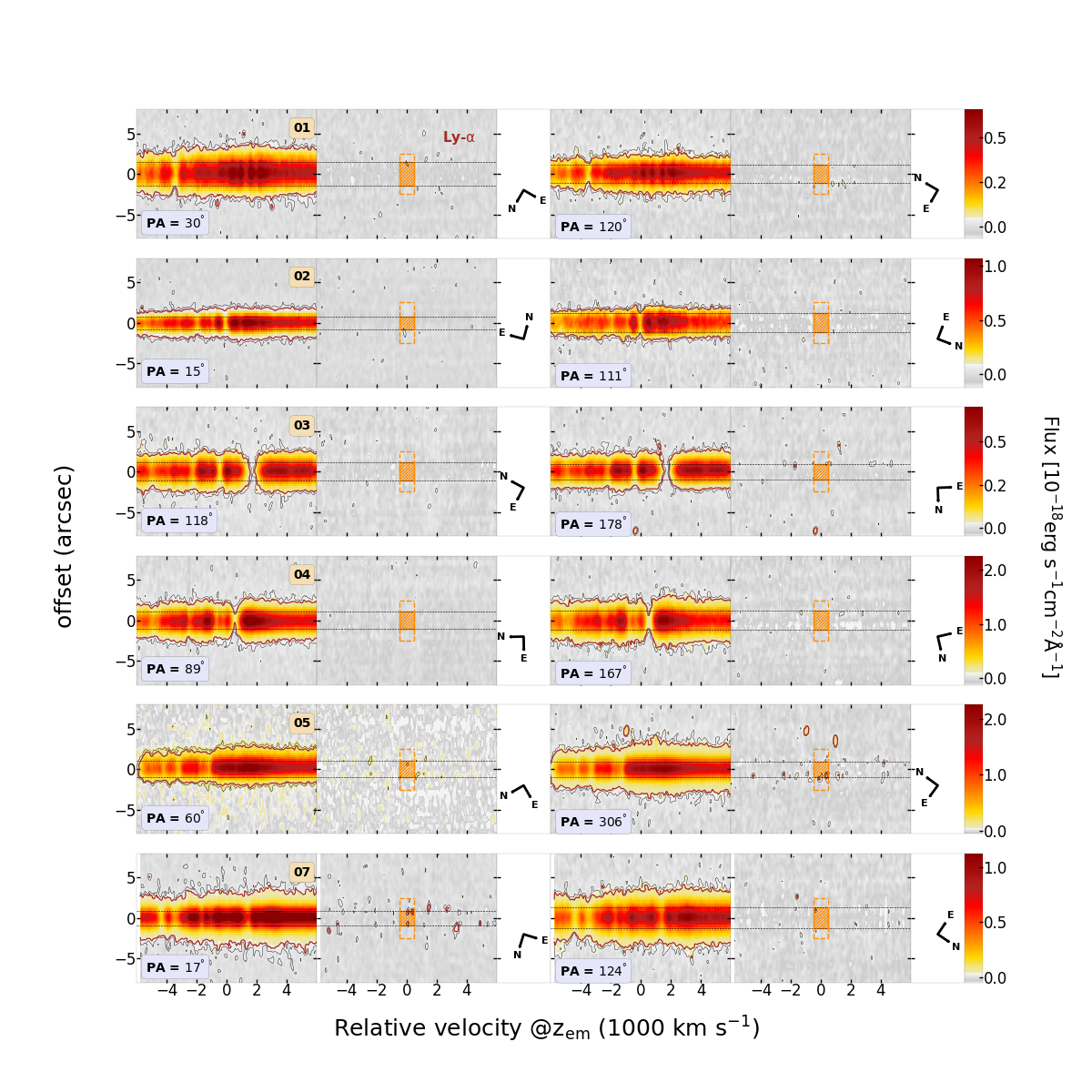}
}} 
}}  
\vskip+0.0cm  
\caption{Same as Fig. \ref{fig_lya_samp1} for rest of the sources with no clear \lya\ emission beyond SPSF FWHM at 3\sig\ level (see column 7 of Table \ref{tab_halo_props}).
} 
\label{fig_lya_samp3}   
\end{figure*}

\begin{figure*} 
\ContinuedFloat
\centerline{\vbox{
\centerline{\hbox{ 
\includegraphics[
height=0.75\textheight,angle=0]{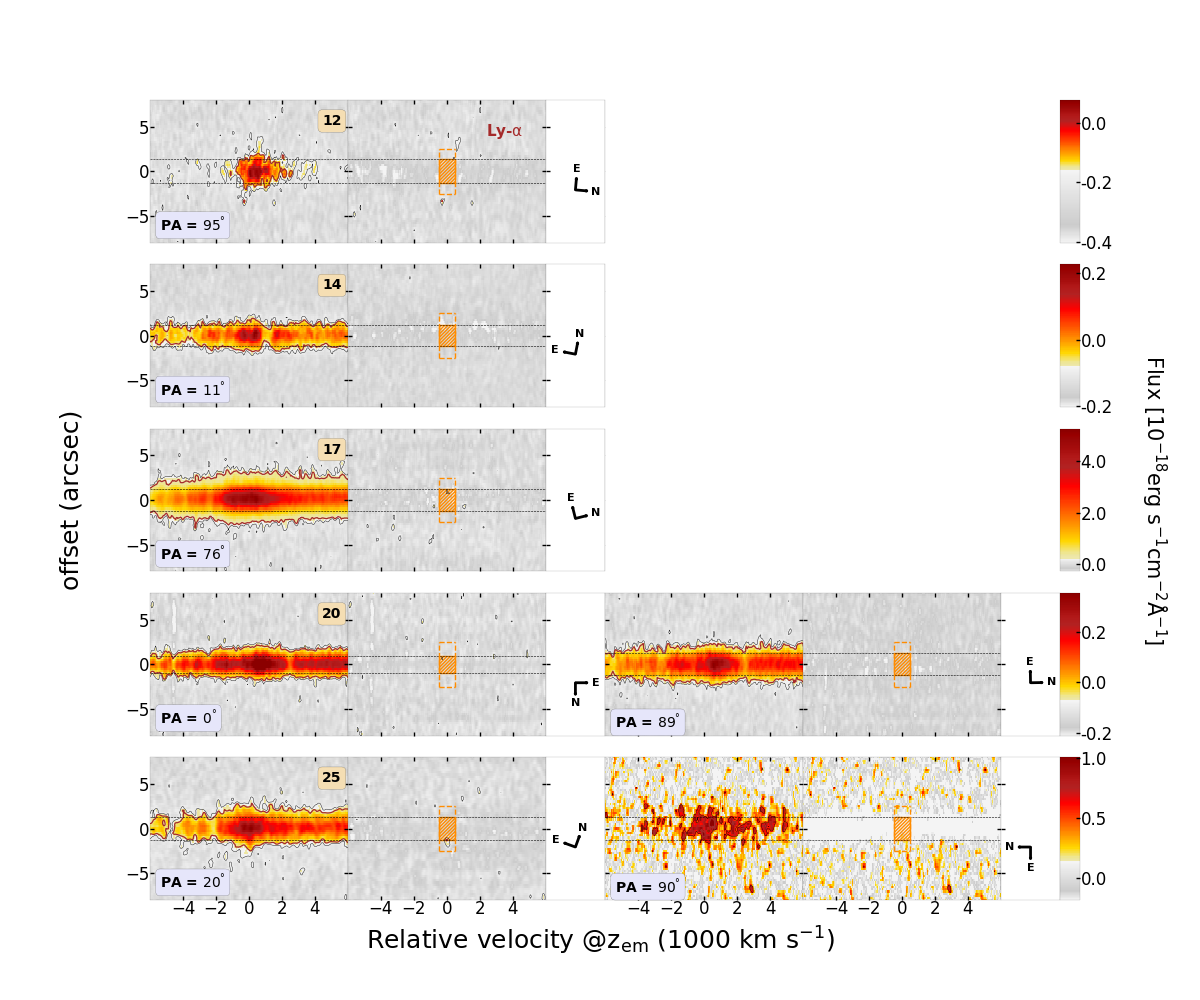}
}} 
}}  
\vskip+0.0cm  
\caption{Continued.
} 
\label{fig_lya_samp4}   
\end{figure*}

\begin{figure} 
\centerline{\vbox{
\centerline{\hbox{ 
\includegraphics[width=0.5\textwidth,angle=0]{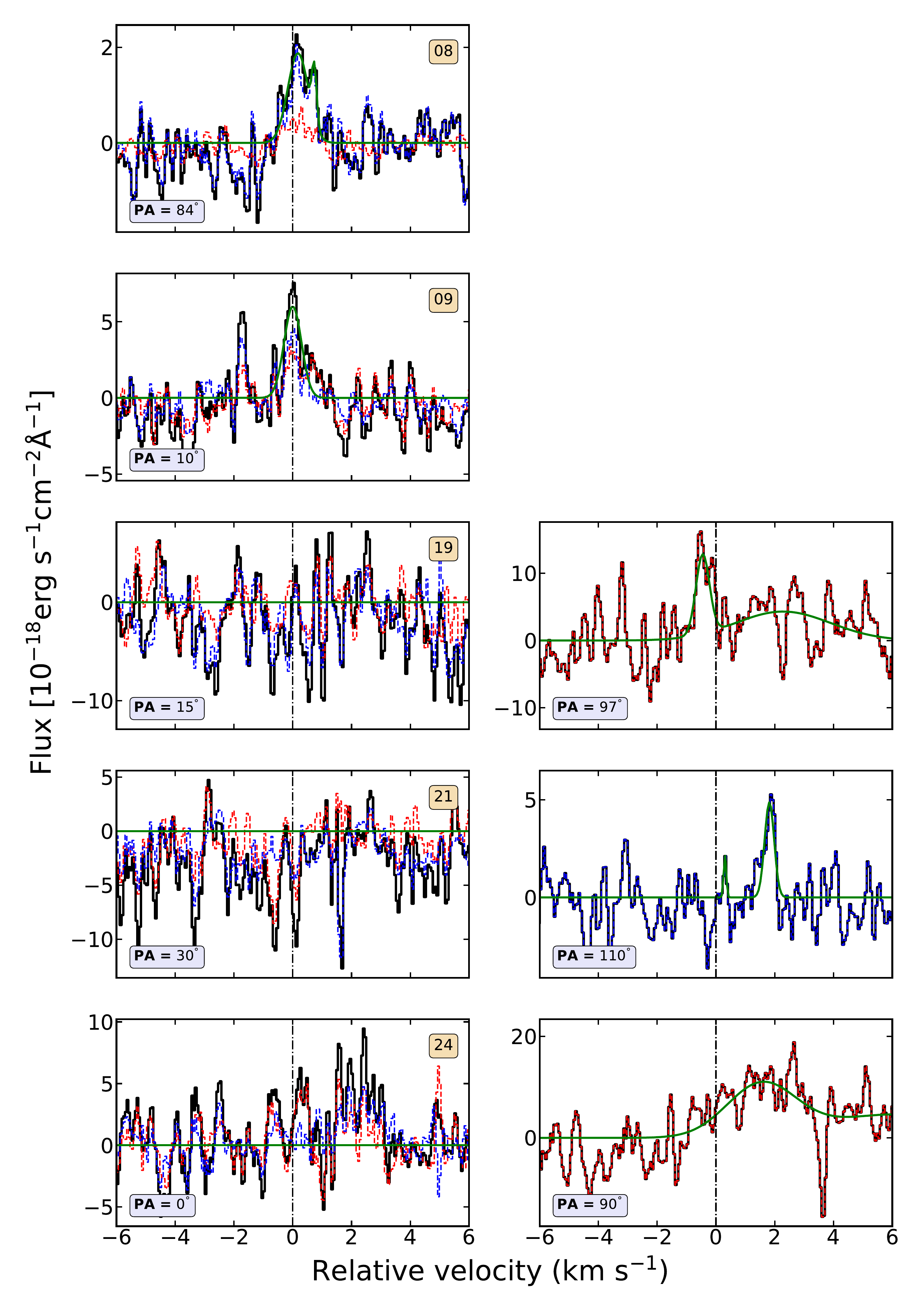}
}} 
}}  
\vskip+0.0cm  
\caption{Same as Fig. \ref{fig_halo_velprof} for five sources with tentative \lya\ halo detection.} 
\label{fig_halo_velprof_tentative}   
\end{figure} 

\subsubsection{M135131.98-101932.90 (\#16)} 
\label{sub_M1351}
This quasar is among top three in our sample in terms of radio size, radio power and $L_{bol}$ (see Table~\ref{tab_luminosities}) and with \heii\ line clearly detected in its spectra (see Fig.~\ref{fig_heiifit}). We have obtained spectra along PA = 0\degree\ and 157\degree, where PA = 157\degree\ is aligned with the axis of radio emission (see Fig.~\ref{fig_ps1}). We detect the \lya\ halo in both these spectra (see Fig.~\ref{fig_lya_samp1}). 

To better understand the correlation between radio and \lya\ halo morphology, we measure the extent of the \lya\ as a distance between quasar trace and outermost spatial locations of 3\sig\ contour for the emission on either side of the trace separately. For PA=157\degree, the extents are \til\ 40 kpc, 44 kpc and the \lya\ luminosities are $1.30\times 10^{43}$, $1.54\times 10^{43}$ \ergs\ for the South-East and North-West direction, respectively (refer to the slits shown in Fig.~\ref{fig_ps1}). Therefore, total extent of the \lya\ halo is 84 kpc (distance between the outer most contours on either sides of the quasar trace) along PA=157\degree\ and has a total luminosity of $2.84\times 10^{43}$\ergs.

As mentioned above the radio emission is extended with a double lobe structure  (see Fig.~\ref{fig_ps1}). The separation between the WISE location and peak of the South-East and North-West lobes are \til\ 28 kpc and 58 kpc with peak flux densities of \til\ 0.41 and 0.09 mJy $\mathrm{b^{-1}}$, respectively.  Clearly the observed \lya\ emission is more symmetric than the radio emission with respect to optical source.
The radio emission is extended beyond the \lya\ halo in the North-west direction whereas it is well within the \lya\ halo (in projection) in the South-East direction. In the case of radio galaxies, \citet{vanojik1997} have found the inner parts of the \lya\ halo within the extent of the radio emission to show perturbed kinematics (FWHM $>$ 1000 \kms)  due  to  jet–gas  interaction. The fact that the radio emission is within the \lya\ halo along PA=157\degree\ and the halo velocity widths are \til\ 600 \kms (see Table~\ref{tab_halo_props}) indicates that the velocity field of the gas associated with the extended \lya\ emission is not heavily influenced by the radio source.

In the case of spectra taken with PA = 0\degree, the \lya\ is extended \til\ 36 kpc both along North and South direction with total \lya\ luminosities of $1.0\times 10^{43}$ and $2.2\times 10^{43}$ \ergs, respectively.  The line widths are also similar to what we find along the PA = 157\degree. This also confirms that the effect of turbulence introduced by any possible interaction of the radio source with the ambient medium, if at all present, is not that strong in this quasar.

As can be seen from column 9 of the Table~\ref{tab_halo_props} and Fig.~\ref{fig_halo_velprof} the peak of the \lya\ emission from the halo is shifted with respect to the systemic redshift by 975 and 1027 \kms\ for PA = 157\degree\ and 0\degree, respectively. As we discussed before the systemic redshifts are determined based on the \civ\ emission and therefore underestimated in our study. We detect redshifted \civ\ associated absorption system (See Table~\ref{tab:associated_abs} and Fig.~\ref{fig_civfit} where \civ\ absorption is seen in the red wing of the \civ\ emission line) in its spectrum.
We clearly detect \lya\ also from this absorber that is redshift by 990 \kms\ with respect to the systemic redshift. This is within 15 \kms\ to the redshift of the \lya\ halo. Thus it is possible that the \lya\ halo is associated with an infalling gas and with minimal interaction with the radio emission. To substantiate this interpretation, it is important to measure the systemic redshift of this quasar using rest frame optical lines.

\subsubsection{M210143.29-174759.20 (\#22)} 
As can be seen from Table~\ref{tab_radio_props}, this sources has the largest radio luminosity in our sample.  However, it has the fourth lowest bolometric luminosity  and inferred \lya\ continuum luminosity in our sample (see Table~\ref{tab_luminosities}). Thus naively one expects the photoionization to be less efficient compared to other sources in the list. However, we do see a strong nuclear He~{\sc ii} emission  in this quasar (see Fig.~\ref{fig_heiifit}).

This is another radio source for which we have obtained optical spectra along (PA = 112\degree) and perpendicular (i.e PA = 30\degree) to the extended radio emission (see Fig~\ref{fig_spsf}). The radio source shows double component with the optical counterpart coinciding with one of the radio peaks (which we believe is the core of the radio emission). The other radio peak (probably related to an one sided jet) is along the South-East direction with respect to the optical source. Extended \lya\ halo is detected along both the PAs, with maximum extension of 53 kpc and luminosity 2.35$\times 10^{44}$ \ergs\ along PA = 112\degree.  In this case the \lya\ halo is found to be asymmetric both in terms of size and luminosity.  It is much brighter (i.e by more than 40 times) and extended (i.e up to 30 kpc from the quasar trace) in the North-West direction (see Fig.~\ref{fig_lya_samp1}) compared to that in the South-East direction.
The FWHM of this extended \lya\ emission is found to be 1259 \kms. This suggests a perturbed kinematics \citep[as per the definitions used in][]{vanojik1997} and possible jet-gas interaction for the \lya\ emission detected in the North-West direction.  

In the  spectra obtained with PA = 30\degree, we detect \lya\ emission both in the North-East and South-West directions (i.e PA = 112\degree). The measured luminosities are at least a factor 1.5 less than what has been found for North-west direction. Also we notice that, the velocity width of the extended \lya\ emission measured along North-East and South-West direction are smaller than what has been measured along PA = 112\degree. This once again confirms the asymmetric nature of the \lya\ halo. {The fact that the radio source has a one sided jet structure with the strongest \lya\ emission (also having large velocity width) is found in the opposite direction may favor jet cloud interaction leading to the observed \lya\ asymmetry both in terms of \lya\ luminosity and velocity field.} Higher resolution radio images with IFU spectroscopy will be very important to explore this possibility.

As can be seen from Table~\ref{tab:associated_abs}, we detect two \civ\ associated absorption systems with relative velocities of $-$316 and 275 \kms with respect to the systemic redshift. These are well within the measurement uncertainties of the systemic redshift. Thus we do not have any clear signatures of infall, with velocity beyond that is allowed by errors in the systemic redshift, in this case.

\subsubsection{M114226.58-263313.70 (\#10)}
\label{sub_M1142}

As can be seen from Table~\ref{tab_radio_props}, this sources has the second largest linear size for radio emission in our sample.  This also has the fifth highest bolometric luminosity and Lyman continuum luminosity in our sample (see Table~\ref{tab_luminosities}). The nuclear He~{\sc ii} emission is clearly detected in this quasar (see Fig.~\ref{fig_heiifit}).

The radio sources shows a compact component coinciding well with the optical source and an extended component towards south with respect to the optical position. The radio emission extend up to 92 \kpc. Thus the radio morphology is consistent with the one-sided jet. We have long-slit observations taken alone only one PA (=100\degree) for this source, which is roughly perpendicular to the the axis of radio emission (see Fig.~\ref{fig_ps1}). We have also taken narrow band observations using SALT centered around the \lya\ emission,  to understand overall distribution of the gas. But unfortunately the data quality (in particular the image quality) is not good for the analysis. Therefore, a direct comparison of radio morphology with \lya\ halo is difficult. 

The \lya\ halo is clearly detected showing asymmetry both in the size and the \lya\ luminosity. The \lya\ halo luminosity (\llya) of this source is one of the highest in our sample, with luminosity of 1.7$\times 10^{44}$\ergs\ and halos size of 68 kpc. The emission is asymmetric being more extended and brightest along South-East direction. The emitting gas is perturbed with FHWM \til\ 1007 \kms, indicating interaction with the radio source. As discussed before, this is the only quasar in our sample where we also detect the extended \civ\ emission. There is no extended radio emission along the direction of our slit (see Fig.~\ref{fig_ps1}). Therefore, if at all there is any unresolved radio structure it has to be well inside the observed \lya\ halo. Thus with the present radio images in hand we are not in a position to associate the observed asymmetry we have seen in the radio emission with the jet-gas interaction.

Unlike the other two cases discussed above we do not detect any associated \civ\ absorption (see Table \ref{tab:associated_abs}) for this source. However, the peak of the \lya\ emission is shifted by \til\ 713 \kms\ with respect to the systemic redshift of the quasar (see Fig~\ref{fig_halo_velprof}). Presence of detectable C~{\sc iv} and He~{\sc ii} extended emission, strong asymmetric \lya\ emission and one side jet radio morphology clearly make this target ideal for VLBA imaging and IFU based spectroscopic followup study to probe the radio jet interactions with the ambient medium.

\bsp
\label{lastpage}
\end{document}